\documentclass[11pt,a4paper]{article}
\pdfoutput=1  
\usepackage{epsfig}
\usepackage{graphicx,psfrag}
\usepackage{multirow}
\usepackage{slashed}
\usepackage{pstricks}
\usepackage{caption}
\usepackage{subcaption}
\usepackage{url}
\usepackage{braket}
\usepackage{jheppub}
\usepackage{enumerate}
\usepackage{wasysym} 
\usepackage{mathrsfs} 
\usepackage{amsfonts} 
\usepackage{amsbsy} 
\usepackage{amscd}
\usepackage{color}
\usepackage{changepage}
\usepackage{floatrow}

\makeatletter

\def\eeq{\end{equation}}
\def\beq{\begin{equation}}

\newcommand{\Rmnum}[1]{\expandafter\@slowromancap\romannumeral #1@}

\newcommand{\bea} {\begin{eqnarray}}
\newcommand{\eea} {\end{eqnarray}}

\newcommand{\gsim}{\raisebox{-0.13cm}{~\shortstack{$>$ \\[-0.07cm]
      $\sim$}}~}
\newcommand{\lsim}{\raisebox{-0.13cm}{~\shortstack{$<$ \\[-0.07cm]
      $\sim$}}~}
\makeatother

\title{Muon g-2 and a type-X two Higgs doublet scenario: some studies in high-scale validity }

\author[a]{Atri Dey,}
  \affiliation[a]{Regional Centre for Accelerator-based Particle Physics,
Harish-Chandra Research Institute, HBNI,
Chhatnag Road, Jhunsi, Allahabad - 211 019, India}

\author[b]{Jayita Lahiri,}
   \affiliation[b]{Department of Physics, Indian Institute of Technology Guwahati,
North Guwahati, Assam - 781039, India}

\author[c]{Biswarup Mukhopadhyaya,} 
  \affiliation[c]{Department of Physical Sciences, Indian Institute of Science Education and Research Kolkata, Mohanpur - 741246, India}

\emailAdd{atridey@hri.res.in}
\emailAdd{jayitalahiri@rnd.iitg.ac.in}
\emailAdd{biswarup@iiserkol.ac.in}

\abstract{
We study the high-scale validity of a Type-X two Higgs doublet scenario which
 provides an explanation of the observed value of muon $(g-2)$. This region admits
 of a pseudoscalar physical state, which is well below the observed 125-GeV scalar 
 in mass. A second neutral scalar particle can be both above and below 125 GeV in
 such a scenario. Admissible regions in the parameter space are obtained
 by using the most recent data on muon $(g-2)$, theoretical constraints such as
 low-scale perturbativity and vacuum stability, and also all experimental constraints,
 including the available LHC results. Among other things, both the aforesaid orders of
 CP-even neutral scalar masses are included in our benchmark studies.
 Two-loop renormalisation group equations 
 are used to predict the values of various couplings at high scales, and the regions
 in the space spanned by low-scale parameters, which retain perturbative unitarity as
 well as vacuum stability upto various scales are identified. We thus conclude that 
 such a scenario, while successfully explaining the observed muon $(g-2)$, can be
 valid upto energy scales ranging from $10^{4}$ GeV to the Planck scale, thus
 opening up directions of thought on its ultraviolet completion.

}

\preprint{HRI-RECAPP-2021-005\\$\textrm{}$}

\DeclareUnicodeCharacter{2212}{-}
\begin{document}

\maketitle

\newpage

\section{Introduction}
\label{sec1}

It is often speculated that the spontaneous symmetry-breaking sector of the electroweak theory may include additional ingredients, over and above the single complex Higgs doublet postulated in the `minimal' original framework. The simplest and most obvious extension is a scenario with two complex scalar doublets. The spectrum of physical fields in such a case, after the absorption of three Goldstone bosons, consists of two CP-odd neutral scalars, one CP-even neutral scalar and a pair of mutually conjugate charged scalar bosons. While the doublets can both acquire vacuum expectation values (VEV) in such a scenario, the Yukawa couplings to fermions are more model-specific, depending on the various possibilities restricted  by the principle of natural flavour conservation. Various phenomenological features of a two Higgs doublet model (2HDM) are accordingly decided.

A scenario of particular interest is the Type-X 2HDM where one of the scalar doublets couples only to quarks, and the other, to leptons. The doublet that couples to quarks dominates the mass eigenstate corresponding to the 125-GeV scalar that has been experimentally discovered. We are concerned with this kind of a theory in the present work.

A rather striking consequence of a Type-X 2HDM is that it admits of scalar physical states
considerably lighter than 125 GeV, consistently with all experimental observations so far. In particular, the neutral pseudoscalar here can be well below 100 GeV. This is worthy of special mention because such a light pseudoscalar can mediate contributions to the muon anomalous magnetic moment, leading to a closer agreement with the experimental observation~\cite{Bennett:2006fi,Abi:2021gix,Albahri:2021ixb}. Keeping this in mind, the region of the parameter space answering to such a light pseudoscalar has been investigated from various angles in recent times, including its implications for the LHC~\cite{Arhrib:2011wc,Liu:2015oaa,Chun:2017yob,Chun:2018vsn,Chakrabarty:2018qtt,Iguro:2019sly,Chun:2019oix,Bandyopadhyay:2019xfb,Frank:2020smf,Chun:2020uzw,Jueid:2021avn}.

For experimental detectability, one largely depends on the pair-production of the light pseudoscalar in the decay of the 125-GeV scalar, a process whose branching ratio (and hence its parameter space) is  restricted by the four-lepton branching ratio. This is because the pseudoscalar in the alignment limit is dominated by the doublet giving masses to leptons.  Also, the 125-GeV particle can be either the lighter or the heavier of the two CP-even neutral scalars. A deciding factor here is
$\sin  (\beta - \alpha)$, where $\alpha$ is the mixing angle between the real parts of the two doublets, and $\tan\beta$ is the ratio of the two VEVs. In addition, wrong-sign Yukawa coupling is an allowed feature of a Type-X 2HDM of the kind we are interested in. The sign of $\sin(\beta-\alpha)$ and the function $\tan \beta$ are of primary importance in this context. The parameter $\tan \beta$ plays crucial role throughout our analysis. This is especially because the closest agreement with the experimental value of muon $g-2$ (henceforth to be called $g_{\mu}-2$) can be achieved for large (\gsim 20) $\tan \beta$. The viability of such $\tan \beta$ in this scenario, from both phenomenological and  theoretical angles, is therefore worth studying. 

The question we ask here is: can the aforesaid  aspects of low-energy phenomenology provide any hint of the UV completion of this scenario? If so, then not only does the muon anomalous magnetic moment get related to high-scale physics, but we also build towards some insights into features such as the signs of Yukawa interactions at low energy. With this in view, we have undertaken a detailed study of the high-scale behavior of the various quartic couplings in the scalar potential of the Type-X 2HDM. Limits on its high-scale validity then arise from vacuum stability, perturbativity and unitarity of the couplings. This exercise has been carried out across various regions of the parameter space, including both right-and wrong-sign Yukawa interaction regions, where the muon anomalous magnetic moment is better explained, and all other theoretical and experimental constraints are satisfied. We make use of two-loop renormalisation group (RG) equations. However, it is demonstrated in some illustrative cases that the difference in the results is not qualitative, as compared to those obtained with one-loop RGs. Therefore, the explanation of the allowed regions of the parameter space corresponding to various levels of high-scale validity has been often given by referring to the one-loop RGs where the effects of different parameters of the theory are more transparent.

The plan of this work is as follows. In Section~\ref{sec2}, we discuss the Type-X two Higgs doublet model and its various aspects relevant for our analysis. Section~\ref{sec3} summarises the implications of the observed $g_{\mu}-2$ for this specific model. We discuss various theoretical and experimental constraints on this model and the allowed parameter space in Section~\ref{sec4}. In Section~\ref{sec5}, we study the renormalization group evolution of various couplings for a few benchmarks. We next identify in Section~\ref{sec6}, the regions of parameter space which are valid upto various high scales and are also interesting from the perpective of the anomalous magnetic moment of muon as well as relevant collider searches. This way we try to explore the validity of Type-X 2HDM as a UV-complete theory. Finally, we conclude our analysis in Section~\ref{sec7}.  

\section{Type-X two Higgs Doublet Model}
\label{sec2}

The most general scalar potential involving two scalar doublets with hypercharge $Y = 1$, under the assumption of a softly broken discrete $Z_2$ symmetry, is given by
\cite{Branco:2011iw}
\begin{eqnarray} \label{V2HDM} \mathrm{V} &=& m_{11}^2
(\Phi_1^{\dagger} \Phi_1) + m_{22}^2 (\Phi_2^{\dagger}
\Phi_2) - \left[m_{12}^2 (\Phi_1^{\dagger} \Phi_2 + \rm h.c.)\right]\nonumber \\
&&+ \frac{\lambda_1}{2}  (\Phi_1^{\dagger} \Phi_1)^2 +
\frac{\lambda_2}{2} (\Phi_2^{\dagger} \Phi_2)^2 + \lambda_3
(\Phi_1^{\dagger} \Phi_1)(\Phi_2^{\dagger} \Phi_2) + \lambda_4
(\Phi_1^{\dagger}
\Phi_2)(\Phi_2^{\dagger} \Phi_1) \nonumber \\
&&+ \left[\frac{\lambda_5}{2} (\Phi_1^{\dagger} \Phi_2)^2 + \rm
h.c.\right].
\end{eqnarray}
We assume CP-conservation, in which case all $\lambda_i$'s and
$m_{12}^2$ are real.

The two complex Higgs doublets with hypercharge $Y = 1$ can be written as
\begin{equation}
\Phi_1=\left(\begin{array}{c} \phi_1^+ \\
\frac{1}{\sqrt{2}}\,(v_1+\phi_1^0+ia_1)
\end{array}\right)\,, \ \ \
\Phi_2=\left(\begin{array}{c} \phi_2^+ \\
\frac{1}{\sqrt{2}}\,(v_2+\phi_2^0+ia_2)
\end{array}\right).
\end{equation}
Where $v_1$ and $v_2$ are the vacuum expectation values with $v^2 = v^2_1 + v^2_2 = (246~\rm GeV)^2$ and $\tan\beta=v_2 /v_1$. After electroweak
symmetry breaking, we obtain five physical states, two neutral CP-even scalars, the lighter of which will be called $h$,  and the heavier $H$, one neutral pseudoscalar $A$, and a pair of charged scalars $H^{\pm}$.

In Type-X 2HDM the Yukawa interactions can be given as
 \bea
- {\cal L}_{Yukawa} &=&Y_{u2}\,\overline{Q}_L \, \tilde{{ \Phi}}_2 \,u_R
+\,Y_{d2}\,
\overline{Q}_L\,{\Phi}_2 \, d_R\, + \, Y_{\ell 1}\,\overline{L}_L \, {\Phi}_1\,e_R+\, \mbox{h.c.}\, \eea in which
$Q_L^T=(u_L\,,d_L)$, $L_L^T=(\nu_L\,,l_L)$, and
$\widetilde\Phi_{1,2}=i\tau_2 \Phi_{1,2}^*$. $Y_{u2}$,
$Y_{d2}$ and $Y_{\ell 1}$ are the couplings of the up, down quarks and leptons with the two doublets, family indices are suppressed.

The factors by which the Standard Model(SM) Higgs interaction strengths need to be scaled to obtain the neutral scalar Yukawa couplings, are
\begin{eqnarray}\label{hffcoupling} &&
y_{h}^{f_i}=\left[\sin(\beta-\alpha)+\cos(\beta-\alpha)\kappa_f\right], \nonumber\\
&&y_{H}^{f_i}=\left[\cos(\beta-\alpha)-\sin(\beta-\alpha)\kappa_f\right], \nonumber\\
&&y_{A}^{f_i}=-i\kappa_f~{\rm (for~u)},~~~~y_{A}^{f_i}=i \kappa_f~{\rm (for~d,~\ell)},\nonumber\\
&&{\rm with}~\kappa_\ell\equiv-\tan\beta,~~~\kappa_u=\kappa_d\equiv 1/\tan\beta.\end{eqnarray}

The corresponding charged Higgs Yukawa couplings are:
\begin{align} \label{eq:Yukawa2}
 \mathcal{L}_Y & = - \frac{\sqrt{2}}{v}\, H^+\, \Big\{\bar{u}_i \left[\kappa_d\,(V_{CKM})_{ij}~ m_{dj} P_R
 - \kappa_u\,m_{ui}~ (V_{CKM})_{ij} ~P_L\right] d_j + \kappa_\ell\,\bar{\nu} m_\ell P_R \ell
 \Big\}+h.c.,
 \end{align}
in which $i,j=1,2,3$.

The couplings of gauge boson pairs with the neutral scalars are given by 
\begin{equation}
y^{V}_h=\sin(\beta-\alpha)\times g^{V}_{SM},~~~
y^{V}_H=\cos(\beta-\alpha)\times g^{V}_{SM},\label{hvvcoupling}
\end{equation}
Where $V$ denotes $W$ or $Z$ and $g^{V}_{SM}$ is the coupling strength of the SM Higgs with a gauge boson pair.

Furthermore, Yukawa couplings here may or may not have the same sign as in the SM case~\cite{Han:2020zqg},
\bea
&&y_h^{f_i}~\times~y^{V}_h > 0~{\rm for~SM-like~coupling~or~right-sign(RS)},~~~\nonumber\\
&&y_h^{f_i}~\times~y^{V}_h < 0~{\rm for~wrong-sign(WS)}.\label{wrongsign}
\eea

\noindent
This can happen, for example, for down-type Yukawa couplings in Type II 2HDM~\cite{Han:2020zqg} as well. However, in Type-X 2HDM the wrong-sign Yukawa coupling can arise in the lepton Yukawa sector alone, unless one allows $\tan \beta < 1$. 
In case of the SM-like coupling, the 125-GeV Higgs couplings are very close to those in the SM, which is the so-called alignment limit. Now in the wrong-sign regime,
the absolute values of $y_h^{\ell}$ and $y^{V}_h$ should still be close to unity because of the restrictions of 125-GeV Higgs signal data~\cite{Sirunyan:2018koj,Aad:2019mbh}.
Moreover, there are two scenarios, a) The lightest CP-even scalar $h$ is SM-like ie. $m_h = m_{h_{SM}} = 125$ GeV, we call this {\bf Scenario 1} and b) when the heavier CP-even scalar $H$ is SM-like, ie. $m_H = m_{h_{SM}} = 125$ GeV, we call this {\bf Scenario 2}. Both scenario 1 and 2 can in principle lead to right-sign or wrong-sign of Yukawa coupling depending on the conditions stated in Equation~\ref{wrongsign}.

\noindent
Let us first consider Scenario 1 in the right- and wrong-sign regions. In scenario 1, the 125-GeV Higgs couplings are:
\begin{equation}  
y_h^{\ell}=\sin(\beta - \alpha) - \cos(\beta -\alpha)\tan \beta,~~y^{V}_h\simeq \sin(\beta-\alpha) \nonumber\\
\end{equation}

\noindent
In the alignment limit $|\sin(\beta - \alpha)| \approx 1$. The following possibilities emerge depending on the sign of $\sin(\beta -\alpha)$ and range of $\tan \beta$.

\medskip
\noindent
$\bullet$ For $\sin(\beta-\alpha) < 0$, $\cos(\beta-\alpha) > 0$, $y_h^{\ell}$ takes the form $-(1+\epsilon)$. $y_h^{\ell}~\times~y^{V}_h > 0$ and it corresponds to right-sign region.  

\medskip
\noindent
$\bullet$ On the other hand, for $\sin(\beta-\alpha) > 0$, $\cos(\beta-\alpha) > 0$, $y_h^{\ell}$ takes the form $(1-\epsilon)$. This case also corresponds to the right-sign region. 

\medskip
\noindent
$\bullet$ When $\sin(\beta-\alpha) > 0$ and $\cos(\beta-\alpha) > 0$ and $\tan \beta \gsim 10$, $y_h^{\ell}$ becomes negative and $y_h^{\ell}~\times~y^{V}_h < 0$. This scenario gives rise to wrong-sign lepton-Yukawa coupling.

Having discussed the coupling structure in Scenario 1, we will now explore the same for Scenario 2. In this case, the heavier CP-even Higgs is the observed 125 GeV Higgs ie. $m_H = 125$ GeV. Here the couplings of $H$ with the leptons and gauge bosons take the following forms.

\begin{equation}  
y_H^{\ell}=\cos(\beta - \alpha) + \sin(\beta -\alpha)\tan \beta,~~y^{V}_h\simeq \cos(\beta-\alpha) \nonumber\\
\end{equation}
In the alignment limit, $|\sin(\beta - \alpha)| << 1$. The sign of $\sin(\beta-\alpha)$ and ranges of $\tan \beta$ in this case will give rise to the following conditions.

\medskip
\noindent
$\bullet$ For $\sin(\beta-\alpha) > 0$, $\cos(\beta-\alpha) > 0$, $y_H^{\ell}$ takes the form $(1+\epsilon)$ and $y_H^{\ell}~\times~y^{V}_H > 0$. Therefore this case corresponds to the right-sign region. 

\medskip
\noindent
$\bullet$ On the other hand, for $\sin(\beta-\alpha) < 0$, $\cos(\beta-\alpha) > 0$, $y_H^{\ell}$ takes the form $(1-\epsilon)$, $y_H^{\ell}~\times~y^{V}_H > 0$. Hence this region also gives rise to right-sign lepton-Yukawa coupling. 

\medskip
\noindent
$\bullet$ When $\sin(\beta-\alpha) < 0$ and $\cos(\beta-\alpha) > 0$ and $\tan \beta \gsim 10$, $y_H^{\ell}$ becomes negative and $y_H^{\ell}~\times~y^{V}_H < 0$. In this scenario, wrong-sign condition is satisfied.

Throughout the discussion concerning Scenario 1 and Scenario 2, $\epsilon$ is assumed to be an extremely small positive quantity. One should note that $\tan \beta (\gsim 10)$ in the right-sign region will give rise to Yukawa scale factors widely differing from unity (therefore disfavored by the Higgs signal strength data), unless $|\sin(\beta-\alpha)|$(Scenario 1) or $|\cos(\beta-\alpha)|$(Scenario 2) is very close to 1. Notably, $\cos(\beta-\alpha$) is kept positive in all the above cases, since the sign of the Yukawa interactions are unambiguously decided by ($\beta-\alpha$) lying in two of the four quadrants. The required ranges of $\tan \beta$ are not altered by such quadrant choice.


The main motivation of the present study is to explore the possibility of having a light ($\lsim 100$ GeV) pseudoscalar in Type-X 2HDM, which makes it easier to match the observed value of $g_{\mu}-2$. We will see in the following section that large $\tan \beta$ regions will be favored from this particular requirement. There will be further overlap or tension between various theoretical and experimental constraints on the model parameter space. These are decisive in understanding the high-scale validity of the scenario, which is our ultimate purpose here.

\section{Explanation of $g_{\mu}-2$}
\label{sec3}

The anomalous magnetic moment of muon is an early triumph of quantum field theory. In today's context, the long-standing discrepancy between SM prediction and experimental observation~\cite{Blum:2013xva} hints towards new physics. The recent result from Fermilab~\cite{Abi:2021gix,Albahri:2021ixb} has strengthened this disagreement further. The future E34 experiment at J-PARC~\cite{Iinuma:2011zz} may shed new light on this tension between theory and experiment.

The effect of loop corrections are usually parameterized in terms of $a_{\mu} = \frac{g_{\mu}-2}{2}$. The SM contributions to $a_{\mu} = \frac{g_{\mu}-2}{2}$ have been extensively studied~\cite{Davier:2010nc,Hagiwara:2011af,Davier:2017zfy,Davier:2019can,Aoyama:2020ynm,Keshavarzi:2018mgv,Colangelo:2018mtw,Hoferichter:2019mqg,Keshavarzi:2019abf,Kurz:2014wya,Melnikov:2003xd,Masjuan:2017tvw,Colangelo:2017fiz,Hoferichter:2018kwz,Gerardin:2019vio,Bijnens:2019ghy,Colangelo:2019uex,Colangelo:2014qya,Blum:2019ugy,Aoyama:2012wk,Czarnecki:2002nt,Aoyama:2019ryr,Gnendiger:2013pva}, the most recent estimate~\cite{Zyla:2020zbs} being
\begin{equation}
a_{\mu}^{SM}= 116591810(43)  \times 10^{-11}
\end{equation}
While the most recent experimental bound is obtained by combining the Fermilab data(2021)~\cite{Abi:2021gix,Albahri:2021ixb} and earlier BNL(2006) data~\cite{Bennett:2006fi}. 
\begin{equation}
a_{\mu}^{exp} = 116592040(54)  \times 10^{-11}
\label{gm2combined}
\end{equation}
This may be contrasted with the earlier limits from BNL data~\cite{Bennett:2006fi}.
\begin{equation}
a_{\mu}^{exp-BNL} = 116592089(63)  \times 10^{-11}
\label{gm2bnl}
\end{equation} 

\noindent
Thus there is approximately $4.2\sigma$ discrepancy when one uses the combined experimental result (Equation~\ref{gm2combined}). 

\begin{equation}
\Delta a_{\mu}=  a_{\mu}^{exp} - a_{\mu}^{SM} = 251(59)  \times 10^{-11}
\label{mgm2}
\end{equation}

\noindent
On the other hand, a discrepancy at the level of 3.7$\sigma$ is seen, if one uses only the BNL data (Equation~\ref{gm2bnl}). 
\begin{equation}
\Delta a_{\mu}^{BNL}=  a_{\mu}^{exp-BNL} - a_{\mu}^{SM} = 279(76)  \times 10^{-11}
\label{mgm2}
\end{equation}

\begin{figure}[!hptb]
	\centering
	\includegraphics[width=14cm,height=5.0cm]{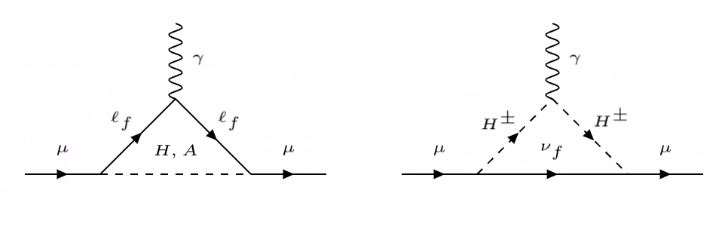}

	\caption{\it Non-standard contribution to $\Delta a_{\mu}$ at one-loop. }
	
	\label{mgm21loop}
\end{figure}

\begin{figure}[!h]
	\centering
	\includegraphics[width=12cm,height=11cm]{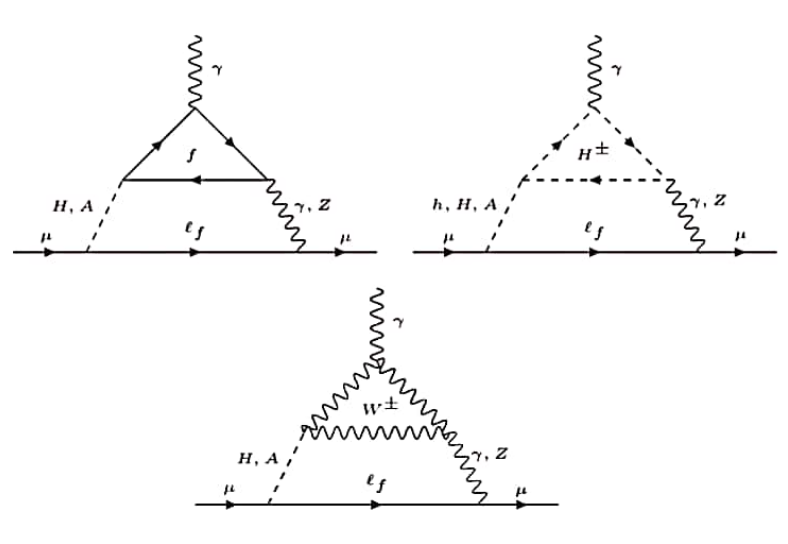}

	\caption{\it Non-standard contribution to $\Delta a_{\mu}$ from two-loop Bar-Zee diagrams with internal $\gamma/Z$. }
	
	\label{intgamma}
\end{figure}

\begin{figure}[!hptb]
	\centering
	\includegraphics[width=14cm,height=11cm]{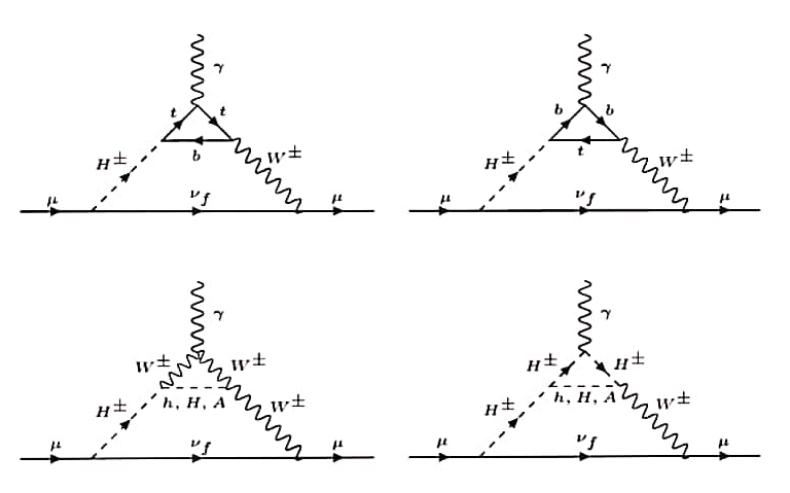}
	\caption{\it Non-standard contribution to $\Delta a_{\mu}$ from two-loop Bar-Zee diagrams with internal $W^{\pm}$ and $H^{\pm}$. Cross-diagrams with $H^{\pm}$ and $W^{\pm}$ interchanged are also considered.}
	
	\label{barzeewhp}
\end{figure}

\begin{figure}[!hptb]
	\centering
	\includegraphics[width=15cm,height=5cm]{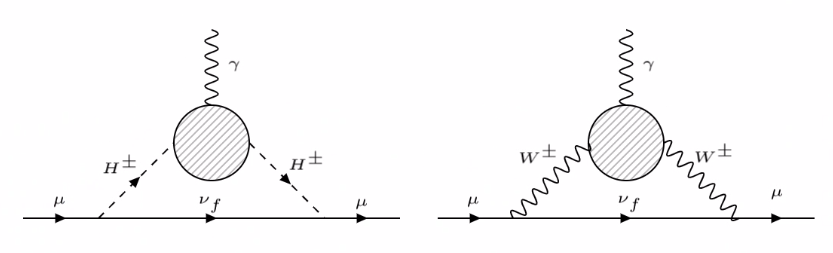}
	\caption{\it Same as in Figure~\ref{barzeewhp}, but with both internal lines $H^{\pm}H^{\pm}$ and $W^{\pm}W^{\pm}$.}
	\label{barzeeall}
\end{figure}

\noindent
We consider one loop as well as two loop Bar-Zee type contribution to $\Delta a_{\mu}$ in Type-X 2HDM. It has been shown in earlier works~\cite{Queiroz:2014zfa,Ilisie:2015tra}, that the two-loop Bar-Zee diagrams dominate over the one-loop contributions, both of which are shown in figures~\ref{mgm21loop}-\ref{barzeeall}. Although the two loop diagrams suffer from a loop suppression factor, they also have an enhancement factor of $\frac{M^2}{m_{\mu}^2}$, where $M$ is the mass of the heavy particle running in the loop namely, $t, b, \tau, H^{\pm}, W^{\pm}$ (see Figure~\ref{intgamma}). One should note that in Type-X 2HDM, the contribution from the $\tau$ loop gets additional enhancement factor from the $\tau$ coupling with pseudoscalar($A$) in the large $\tan \beta$ region. The enhancement factor in general dominates over the aforementioned loop suppression. The diagram involving $W^{\pm}$ in the loop (Figure~\ref{intgamma} bottom), will have negligible contribution due to suppression in the coupling between $W^{\pm}$ bosons and the non-standard CP-even Higgs in the alignment limit. We also consider the Bar-Zee diagrams where charged Higgs replaces the neutral Higgs and also $W^{\pm}$ substitutes the internal $\gamma$ (see Figure~\ref{barzeewhp} and~\ref{barzeeall}). The contribution from these diagrams can be sizable in some regions of the parameter space~\cite{Ilisie:2015tra}.

\begin{figure}[!hptb]
\centering
\includegraphics[width=75mm,height=59.0mm]{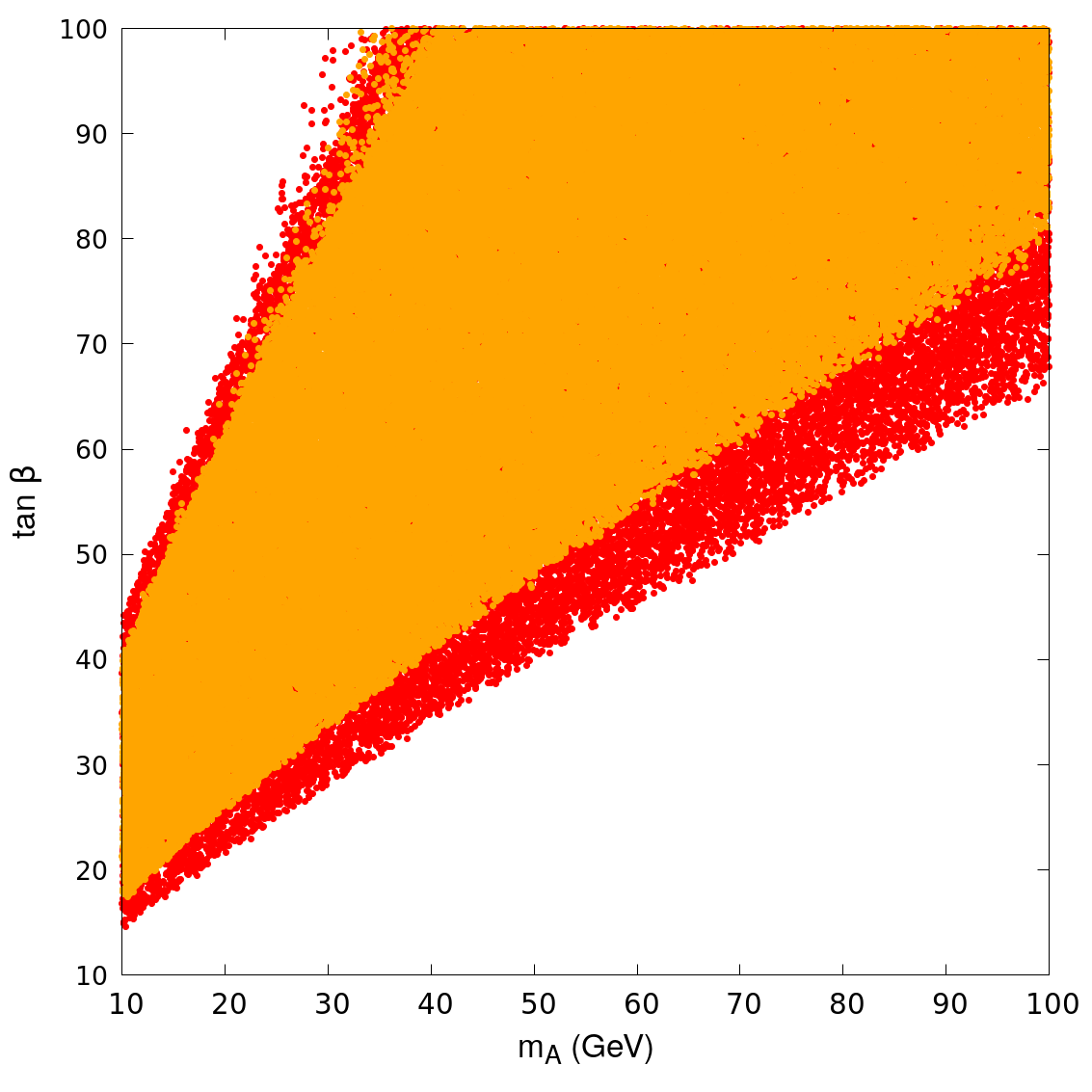}
\caption{\it The allowed region in $m_A - \tan \beta$ plane from $g_{\mu}-2$ data at $3 \sigma$ level. The limits have been obtained by marginalizing over all other parameters of the 2HDM, except the mass of one of the neutral CP-even scalars is set at 125 GeV. The yellowish interior corresponds to the combined constraints from older~\cite{Bennett:2006fi} and recent data~\cite{Abi:2021gix,Albahri:2021ixb}, while the red regions on both sides show the additional regions allowed when only the older data are used.}
\label{muon_anomaly}
\end{figure}

In order to obtain updated constraints on $\Delta a_{\mu}$, we have calculated afresh the contributions from all the aforementioned diagrams following~\cite{Queiroz:2014zfa,Ilisie:2015tra}. The resulting constraints on the $m_A - \tan \beta$ plane is shown in Figure~\ref{muon_anomaly}. 
$3\sigma$ upper and lower bound on the experimentally observed central value of $\Delta a_{\mu}$ have been used in the scan.
While such scans have been carried out earlier~\cite{Broggio:2014mna,Cherchiglia:2017uwv}, we have (a) used the most recent constraints and (b) have used all new physics diagrams exhaustively in our analysis.

In Figure~\ref{muon_anomaly}, the yellowish interior corresponds to the region that satisfies constraints coming from a combination of the BNL and Fermilab data. The red bands on both sides of this region, denote the additional regions which are allowed at the $3\sigma$ level before Fermilab data came into existence. The red band on the lower side is consistent even when the new data are included, so long as one allows experimental values to be undersaturated by Type-X 2HDM. On the other hand, points in the upper red band overshoots the 3$\sigma$ limit arising from the combined data, and therefore, may be taken to be in conflict with the most recent experiments.   

It is clear that a low mass pseudoscalar with an enhanced coupling to the $\tau$ leptons will give significant contribution to $\Delta a_{\mu}$(see Figure~\ref{intgamma}(top left)), especially for large $\tan \beta$. Overall, low $m_A$ and large $\tan\beta$ region is favored in the light of $g_{\mu}-2$ data in our model. In this work we are interested to know the high-scale behavior of this particular region of parameter space. Before exploring the high-scale validity of this region of the parameter space, we would like to consider the other important theoretical as well as experimental constraints on such a scenario.

\section{Other constraints on model parameters }
\label{sec4}

\subsection{Constraints from electroweak precision observables}

\begin{figure}[!hptb]
\centering

\includegraphics[width=7.5cm, height=6.5cm]{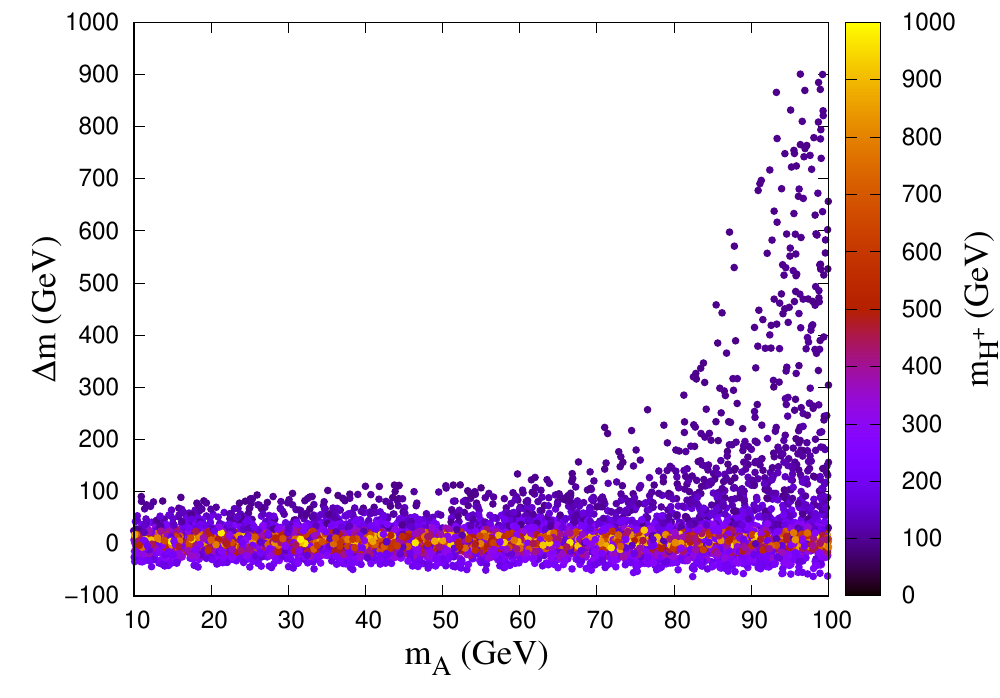}



\caption{\it Allowed parameter space in $m_A - \Delta m$ plane consistent with the observed values of S,T,U parameters at 3$\sigma$. The charged Higgs mass $m_H^{\pm}$ has been shown here as the third axis.}
\label{stu}
\end{figure}

The custodial SU(2) is a symmetry of the tree-level 2HDM potential and can be broken at the loop level due to corrections to weak boson masses as well as weak couplings by extra scalars in 2HDM. Electroweak precision measurements of the oblique parameters, have been performed by the Gfitter group~\cite{BAAK:2014gga}. This restricts the mass difference between the charged scalar and the non-standard CP-even scalar $|\Delta m|= |m_{h/H} - m_{H^\pm}|$ , depending on $m_A$ and values of $m_{H^\pm}$~\cite{Broggio:2014mna}. The status of two Higgs doublet models in the light of global electroweak data has been presented in~\cite{Haller:2018nnx}. The allowed parameter space in $m_A - \Delta m$ plane is shown in Figure~\ref{stu} with color-coded representation of $m_{H^{\pm}}$. We mention here that we have considered the elliptic contour in the $S-T$ plane computed with $U$ as a free parameter. This choice leaves us with a less constrained parameter space than the scenario when $U$ is fixed at 0.

It can be seen from Figure~\ref{stu} that, in the pseudoscalar mass range of our interest ($m_A \lsim 100$ GeV), when $m_{h/H} < m_{H^{\pm}}$, it is possible to get upto $|\Delta m| \lsim 50$ GeV, only in the limit of low $m_H^{\pm}(\lsim 200$ GeV). Notably, large positive $\Delta m$ (upto a TeV or so) can be allowed when $A$ and $H^{\pm}$ are closely degenerate.

\subsection{Theoretical constraints}
Theoretical constraints include  perturbativity, unitarity and vacuum stability conditions at the electroweak scale . Effects of these constraints on various 2HDMs have been studied in detail in earlier works~\cite{Crivellin:2013wna,Arbey:2017gmh,Hussain:2017tdf}. It has been pointed out that large separation between $m_A$ and $m_{H^{\pm}}$ is disfavored from the requirement of vacuum stability and perturbativity.
We concentrate on the low $m_A$ region and therefore it is crucial to look at the allowed upper limit on $m_H^{\pm}$ in this scenario.

\medskip
\noindent
$\bullet$ {\bf perturbativity and unitarity:} If 2HDM is a perturbative quantum field theory at a given scale, it would imply, all quartic couplings $C_{H_iH_jH_kH_l} < 4\pi$ and all Yukawa couplings $Y_j < \sqrt{4\pi}$. Further, unitarity bound on the tree level scattering
amplitude of the Higgses and longitudinal parts of EW gauge bosons put an upper bound on the eigenvalues $|a_i|\leq 8\pi$ of the $2 \rightarrow 2$ scattering matrices~\cite{Dicus:1992vj,Lee:1977yc}.

The physical masses can be written as a function of the quartic couplings in the following manner. 

\begin{eqnarray}
m_A^2 &=& \frac{m_{12}^2}{\sin\beta \cos\beta} - \lambda_5 v^2  \\
m_{H^{\pm}}^2 &\approx& m_A^2 + \frac{1}{2} v^2 (\lambda_5 - \lambda_4)
\label{massdiff}
\end{eqnarray}

\noindent
It is clear from Equation~\ref{massdiff} that $m_{H^{\pm}}^2 - m_A^2$ is proportional to $\lambda_5 - \lambda_4$ which should be less than $ \lambda_3 + \sqrt{\lambda_1 \lambda_2}$ from the requirement of vacuum stability (see Equation~\ref{stability}). Therefore these conditions along with the requirement of perturbativity ie. $C_{H_iH_jH_kH_l} < 4\pi$ puts an upper limit on the mass square difference  $m_{H^{\pm}}^2 - m_A^2$.

In what follows, we translate these constraints into those of the parameter space for both right- and wrong-sign Yukawa couplings. With this in view, we first express the quartic couplings in terms of physical masses and mixing angles.

\begin{eqnarray}
\nonumber && \lambda_1 = \frac{m_H^2\cos^2\alpha+m_h^2\sin^2\alpha-m_{12}^2\tan\beta}{v^2\cos^2\beta},\\
\nonumber && \lambda_2 = \frac{m_H^2\sin^2\alpha+m_h^2\cos^2\alpha-m_{12}^2\cot\beta}{v^2\sin^2\beta},\\
\nonumber && \lambda_3 = \frac{(m_H^2-m_h^2)\cos\alpha \sin\alpha+2m_{H^\pm}^2\sin\beta \cos\beta-m_{12}^2}{v^2\sin\beta \cos\beta},\\
\nonumber && \lambda_4 = \frac{(m_A^2-2m_{H^\pm}^2)\sin\beta \cos\beta+m_{12}^2}{v^2\sin\beta \cos\beta},\\
&& \lambda_5 = \frac{m_{12}^2 - m_A^2\sin\beta \cos\beta}{v^2\sin\beta \cos\beta}.
\label{eq:paratran}
\end{eqnarray}

\noindent
It is clear from the expression of $\lambda_1$ in Equation~\ref{eq:paratran} that, to have it in the perturbative limit, the soft $Z_2$ breaking parameter $m_{12}^2 \approx \frac{m_H^2}{\tan \beta}$.
Also the perturbativity condition of the quartic couplings $\lambda_4$ and $\lambda_5$ implies $m_{H^{\pm}}^2 - m_A^2 < 4\pi v^2$ which translates to the limit $m_{H^{\pm}} \lsim 870$ GeV for very low $m_A$. 

\medskip
\noindent
$\bullet$ {\bf Vacuum stability:} Vacuum stability demands there can exist no direction in the field space in which $\cal V \rightarrow -\infty$. This implies the following conditions on the quartic couplings of the Higgs potential~\cite{Deshpande:1977rw,Nie:1998yn,Gunion:2002zf}.
\begin{eqnarray}
\label{vs1}
\lambda_{1,2} > 0 \,, \\
\label{vs2}
\lambda_3  > -\sqrt{\lambda_1 \lambda_2} \,\\
\label{vs3}
|\lambda_5| < \lambda_3 + \lambda_4 + \sqrt{\lambda_1 \lambda_2}\,
\label{stability}
\end{eqnarray} 

\begin{figure}
\floatsetup[subfigure]{captionskip=10pt}
    \begin{subfigure}{.44\linewidth}
    \centering
    \includegraphics[width=7.0cm, height=5.5cm]{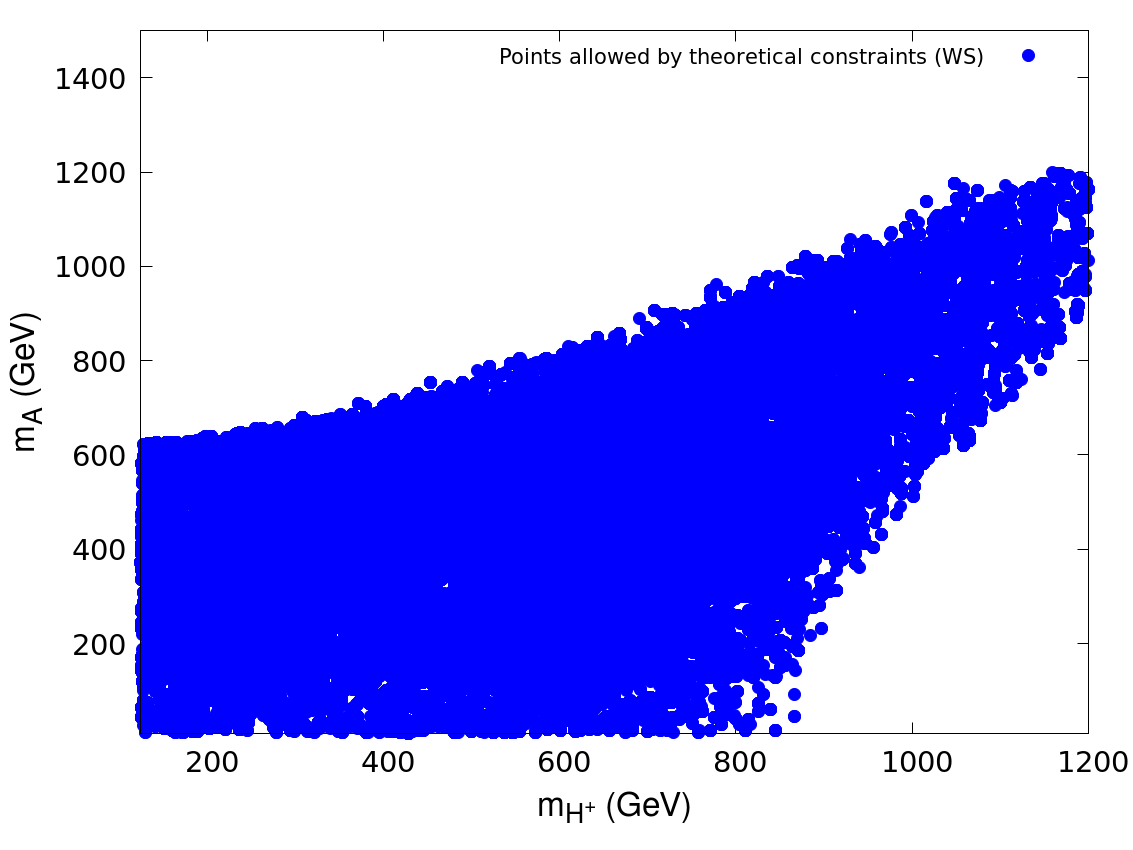}
    \caption{}\label{fig:image1}
    \end{subfigure} %
    \qquad
    \begin{subfigure}{.44\linewidth}
    \centering
    \includegraphics[width=7.0cm, height=5.5cm]{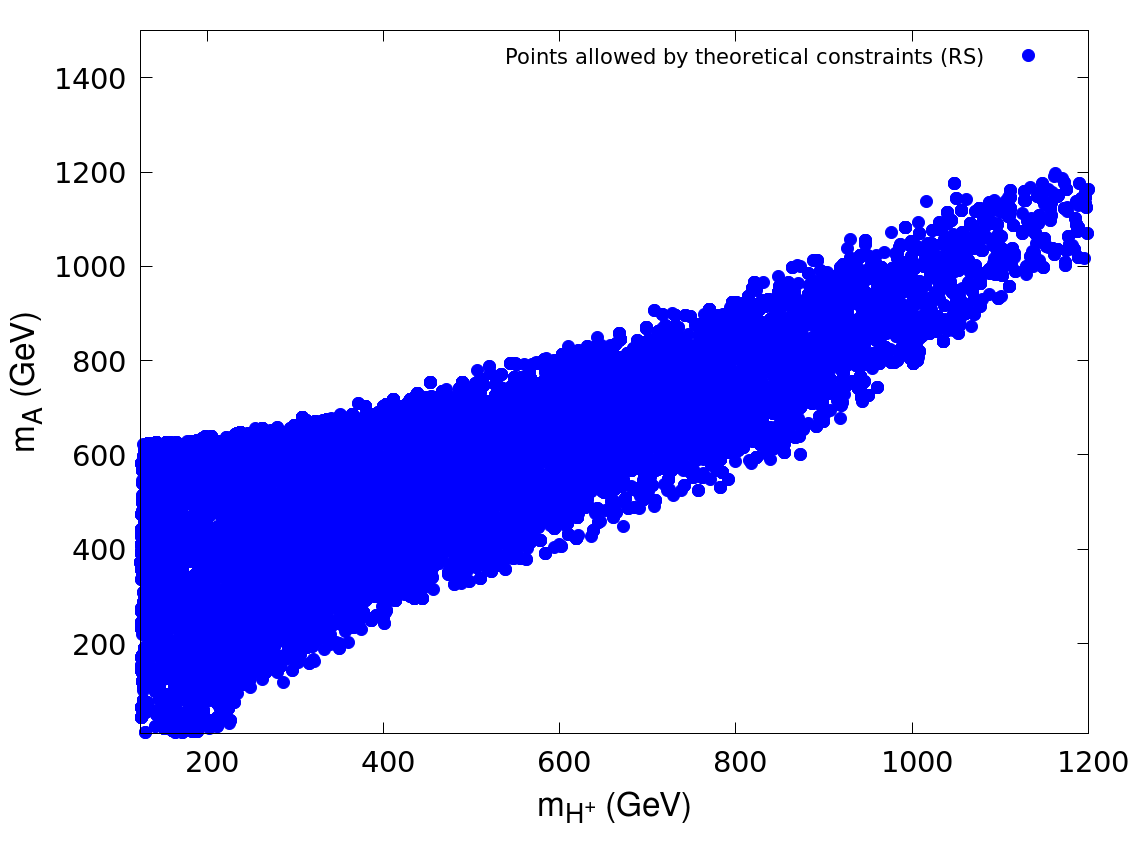}
    \caption{}\label{fig:image12}
   \end{subfigure}
\\[2ex]
\RawCaption{\caption{\it Allowed parameter space in $m_{H^\pm}-m_A$ plane consistent with theoretical bounds in the (a) WS and (b) RS case.}
\label{pert_bound_rs}}
\end{figure}

\noindent
The last condition in Equation~\ref{stability} can be rewritten as $\lambda_{3}+\lambda_4-\lambda_5>-\sqrt{\lambda_1\lambda_2}$ ~for $m_H>m_A$.
One of the key features to note is that, the upper limits on the heavy Higgs masses show quite different behaviors in the right-sign and wrong-sign limit of the Yukawa couplings \cite{Ferreira:2014naa}. The light-Higgs Yukawa couplings for leptons $y_{h}^{\ell}$ in Type-X 2HDM can be expressed as
\begin{equation} \label{xitau}
y_{h}^{\ell} = -{\sin\alpha\over \cos\beta} \equiv \sin(\beta-\alpha)-\tan\beta \cos(\beta-\alpha).
\end{equation}
The 125 GeV-Higgs boson couplings are experimentally found to be very much SM-like, implying, in particular,  $|\sin(\beta-\alpha)| \simeq 1$ and $|y^\ell_h|\approx 1$.  This can be achieved
when $\tan\beta \cos(\beta-\alpha) \approx 0$ (leading to the right-sign lepton coupling 
$y_{h}^{\ell} \approx +1$),
or in the large $\tan \beta$ limit with  $\tan\beta \cos(\beta-\alpha)\approx 2$ (leading to the wrong-sign coupling $y_{h}^{\ell} \approx -1$). Using the Equations~\ref{xitau} and \ref{eq:paratran}, one finds
\begin{equation}
 \lambda_3 + \lambda_4 -\lambda_5 =
 { 2 m_A^2 + y_{h}^{\ell} \sin(\beta-\alpha) m_h^2 - (\sin^2(\beta-\alpha) + y_{h}^{\ell} \sin(\beta-\alpha)) m_H^2
 \over v^2} +{\cal O}({1\over \tan^2\beta})
 \label{lam345}
 \end{equation}
 in the large $\tan\beta$ limit.
Now, in the right-sign case ($y_{h}^{\ell} \sin(\beta-\alpha) \to +1$), we have
\begin{equation}
2 {m_H^2 \over v^2} < \sqrt{0.26 \times 4\pi} + {2 m_A^2 + m_h^2 \over v^2}
\end{equation}
which puts a strong upper bound, $m_H \lsim 250$ GeV  for low $m_A$, which is consistent with~\cite{Broggio:2014mna}.
On the other hand, in the wrong-sign
limit ($y_{h}^{\ell} \sin(\beta-\alpha) \to -1$), $m_H$ can be arbitrarily large with the condition 
$\sin^2(\beta-\alpha)+y_{h}^{\ell} \sin(\beta-\alpha) \approx 0$ being trivially satisfied in the alignment limit. 
These particular properties of wrong-sign and right-sign regions can be seen from Figures~\ref{pert_bound_rs}(a) and (b).

\subsection{Constraints from the direct search at colliders}
\label{constraints_collider}

\noindent
$\bullet$ {\bf LEP data:}

\medskip

\noindent
The earliest collider constraint on the masses of charged and neutral scalars came from the LEP experiments~\cite{Abbiendi:2013hk}. The charged Higgs has been searched for at LEP in the process $e^+ e^- \rightarrow \gamma/Z \rightarrow H^+ H^-$ with subsequent decay $H^{\pm} \rightarrow \tau^{\pm}\nu_{\tau}$. Direct search at LEP in this channel provides a lower limit on charged Higgs mass $m_{H^\pm}$ as a function of BR($H^{\pm} \rightarrow \tau^{\pm} \nu_{\tau})$. The strongest bound results in $m_{H^{\pm}} \gsim 90$ GeV~\cite{Abbiendi:2013hk} at $95\%$ CL., considering BR($H^{\pm} \rightarrow \tau \nu_\tau \approx 100\%$). However, the upper limit varies only mildly with the BR($H^{\pm} \rightarrow \tau \nu_\tau)$ and therefore is fairly model-independent. On the other hand, another LEP search in the channel $p p \rightarrow h A \rightarrow 4\tau$ also puts an upper limit on BR$(h \rightarrow \tau^+\tau^-)\times$ BR$(A \rightarrow \tau^+\tau^-) \times R_{hA}$(mixing between two doublets) for $m_A+m_h$ upto 200 GeV~\cite{Abdallah:2004wy}. 
 
\bigskip

\noindent
$\bullet$ {\bf LHC data on the SM-like Higgs:}
\medskip

\noindent
An important constraint comes from the direct search for 125-GeV Higgs decaying into two light pseudoscalar final states when it is kinematically allowed. The upper bound on this branching ratio puts severe constraint on the parameter space of this model. As $g_{\mu}-2$ constraint pushes us to a region tilted towards large $\tan \beta$ with small $m_A$, it can lead to substantial branching fraction in the decay mode $h_{SM} \rightarrow A A$, when this particular decay is kinematically allowed ie. $m_A \lsim \frac{m_h}{2}$. At large $\tan \beta$, pseudoscalar $A$ decays to $\tau^+ \tau^-$ pair with $\gsim 99\%$ branching fraction, leaving a small branching fraction ($\sim 0.35\%$) in the $\mu^+ \mu^-$ final state ~\cite{Kanemura:2011kx,Kanemura:2014bqa,Chun:2017yob,Chun:2018vsn}. LHC searches for $h_{SM} \rightarrow AA$ in the $4\tau$ or $2\tau+2\mu$ final state disfavors a large BR($h_{SM} \rightarrow A A$). We impose the most stringent upper limit BR$(h_{SM} \rightarrow AA) \lsim 0.04$, consistent with the upper bounds provided by the experimental results~\cite{Sirunyan:2018mbx,Sirunyan:2018mbx}\footnote{The limit is taken on the strongest side in our analysis. It may become slightly relaxed with varying $m_A$. Thus our study is conservative.}.

First we consider Scenario 1 ie. $m_h = 125$ GeV. The partial decay width of Higgs decaying to a pair of pseudoscalars is given by
\begin{equation}
\Gamma(h\rightarrow AA) = \frac{1}{32\pi}\frac{g_{hAA}^2}{m_h}\sqrt{1-4m_A^2/m_h^2}
\label{widthhaa}
\end{equation}

\noindent
Using the relations between the quartic couplings $\lambda$'s and the physical masses and Higgs mixing parameter $m_{12}^2$, in the alignment limit $|\sin(\beta - \alpha)| \approx 1$, $hAA$ coupling~\cite{Gunion:2002zf} takes the following form.

\begin{equation}
g_{hAA} \propto (\lambda_3 + \lambda_4 - \lambda_5 )v \approx  \frac{\sin(\beta - \alpha){y_h^{\ell}}({m^2_h} - {m^2_H}) + 2{m^2_A} - m_{12}^2/\sin\beta\cos\beta}{v} 
\end{equation}
Expressing the quantity $y_h^{\ell}\sin(\beta - \alpha)$ in terms of $g_{hAA}$ and mass parameters we get
\begin{equation}
{y_h^{\ell}}\sin(\beta - \alpha) = \frac{g_{hAA}v + m_{12}^2/\sin\beta\cos\beta - 2m_A^2}{m_h^2 - m_H^2}
\label{ylhwrong}
\end{equation}

\noindent
We can see from Equation~\ref{widthhaa} that when $m_A \lsim \frac{m_h}{2}$, the only way a small branching ratio for BR($h \rightarrow AA)$ can be achieved is when the coupling $g_{hAA}$ is extremely small.
We should also remember from our discussion of perturbativity that, in this scenario $m_{12}^2 \approx \frac{m_H^2}{\tan \beta}$, in order to ensure perturbativity of the quartic couplings. If we demand perturbativity as well as the condition, $g_{hAA} \approx 0$, Equation~\ref{ylhwrong} implies ${y_h^{\ell}}\sin(\beta - \alpha) < 0$. In other words, wrong-sign lepton Yukawa coupling is more favored in Scenario 1, when one demands smallness of BR($h_{SM}\rightarrow AA)$ as well as perturbativity of the quartic couplings.


The other possibility is to consider the case when the heavier CP even scalar is the SM-like Higgs, ie $m_H = 125$ GeV, which is our Scenario 2. However, in this case the LEP limit implies either $m_A$ or $m_h$ can be less than $\frac{m_H}{2}$~\cite{Bernon:2014nxa}. We consider the low mass pseudoscalar and therefore $m_h > \frac{m_H}{2}$.  
Here the decay width of 125-GeV Higgs decaying to a pair of pseudoscalars is given by

\begin{equation}
\Gamma(H\rightarrow AA) = \frac{1}{32\pi}\frac{g_{HAA}^2}{m_h}\sqrt{1-4m_A^2/m_H^2}
\label{widthHaa}
\end{equation}

\noindent
Here too, like the previous scenario, the limit on BR($H \rightarrow AA$) will indicate extremely small value of the coupling $g_{HAA}$, whose expression in the alignment limit ie. $|\cos(\beta -\alpha)| \approx 1$ is given as follows: 

\begin{equation}
g_{HAA} \propto (\lambda_3 + \lambda_4 - \lambda_5 )v \approx  \frac{\cos(\beta - \alpha){y_H^{\ell}}({m^2_H} - {m^2_h}) + 2{m^2_A} - m_{12}^2/\sin\beta\cos\beta}{v} 
\end{equation}

\noindent
Expressing the quantity $y_H^{\ell}\cos(\beta - \alpha)$ in terms of $g_{HAA}$ and mass parameters we get
\begin{equation}
{y_H^{\ell}}\cos(\beta - \alpha) = \frac{g_{HAA}v + m_{12}^2/\sin\beta\cos\beta - 2m_A^2}{m_H^2 - m_h^2}
\label{ylHrght}
\end{equation}

\noindent
We can see that, as we are concerned with low pseudoscalar mass here($m_A \lsim \frac{m_H}{2}$), in the limit $g_{HAA} \approx 0$, $y_H^{\ell}\cos(\beta - \alpha)$ will be positive for the most part of our parameter space. Therefore we can conclude that the right-sign region will be favored in case of Scenario 2. We will see the implications of these in the next section.

\bigskip

\noindent
$\bullet$ {\bf Signal strengths of the 125-GeV scalar}

\medskip

\noindent
Important limits come from the signal strength measurements of the 125-GeV Higgs in various final states
including $\gamma \gamma$, $ZZ$, $WW$, $b \bar b$ and $\tau \tau$ final states~\cite{Sirunyan:2018koj,Aad:2019mbh}. The experimental data indicate that the gauge boson and Yukawa couplings of the 125-GeV scalar are very close to their SM value. Therefore in our analysis we confine ourselves to the alignment limit ie. $|y^V_{h/H}| \approx 1$ ($y^V_h = \sin(\beta - \alpha)$ for Scenario 1 and $y^V_H = \cos(\beta -\alpha)$ for Scenario 2) and $|y^{\ell}_{h/H}|$ is also close to unity. This in turn implies that $\tan \beta$ can not be very large in the RS region. However, in the WS region this condition gets slightly relaxed and $|y^V_{h/H}|$ can deviate slightly further from unity, within the allowed range and $\tan \beta$ can be large as long as $|y^{\ell}_{h/H}|$ is close to 1. Another important constraint comes from the direct search for 125-GeV Higgs decaying into two light pseudoscalar final states, when it is kinematically allowed. The upper bound on this branching ratio translates into severe constraint on the parameter space of this model. We have discussed the effect of this constraint on our parameter space in detail in a previous subsection~\ref{constraints_collider}.

\bigskip
\noindent
$\bullet$ {\bf Direct search for heavier (pseudo)scalars at the LHC:}

\medskip
\noindent
Collider searches for the non-standard neutral scalar states also put constraints on the parameter space of interest. Such searches are performed at the LHC, in various SM final states. As we are particularly interested in the low pseudoscalar mass region with its enhanced coupling to leptons, the limits which are crucial for our analyses, come from the search for low pseudoscalar produced in association with a pair of $b$ quarks and decaying into $\tau \tau$ final state~\cite{Khachatryan:2015baw,CMS:2019hvr}. Constraints from the search for low mass (pseudo)scalar produced in association with $b \bar b$ and decaying into $b \bar b$~\cite{Khachatryan:2015tra,CMS:2016ncz} has also been taken into account. 

We have also taken into account the upper limits from CP-even non-standard scalars( $h/H$) decaying to $\tau \tau$~\cite{Arbey:2017gmh} final state. CMS has also looked for decay involving two non-standard Higgs bosons such as $h/H \rightarrow AZ$~\cite{Khachatryan:2016are,CMS-PAS-HIG-16-010}, $H \rightarrow hh$~\cite{Sirunyan:2018zkk,Sirunyan:2018iwt,Aaboud:2018knk} and $h/H \rightarrow VV$~\cite{ATLAS:2016kjy,TheATLAScollaboration:2016gnu,Aad:2021yhv}.   

At the LHC, the charged Higgs search can be produced in several ways. When $m_H^{\pm} < m_t$, charged Higgs can be produced from the decay of top quark($t \rightarrow bH^{\pm})$. This decay has been searched for in $\tau \nu$~\cite{Aad:2014kga,Khachatryan:2015qxa} and $c\bar s$~\cite{Aad:2013hla,Khachatryan:2015uua} final state. These searches put an upper limit on BR($t \rightarrow bH^{\pm})\times(H^{\pm}\rightarrow \tau \nu/c \bar s)$. The other important search mode at the LHC is $(p p \rightarrow tbH^{\pm}$) in the final states $\tau \nu$~\cite{Khachatryan:2015qxa,Aaboud:2016dig} and $c \bar s$~\cite{ATLAS-CONF-2016-088,CMS-PAS-HIG-16-031} and $t \bar b$~\cite{ATLAS:2016qiq}.

The most stringent bounds in the context of direct search for non-standard scalars come from $p p \rightarrow h/H/A \rightarrow \tau \tau$ and $p p \rightarrow tbH^{\pm}(H^{\pm} \rightarrow \tau \nu)$. Although in Type-X model at large $\tan \beta$, the neutral non-standard scalars decay to $\tau \tau$ with almost 100\% BR and the charged Higgs decays to $\tau\nu$ final state almost exclusively, the production cross-section is suppressed at large $\tan \beta$ as the quark couplings scale as $1/\tan \beta$. This in turn puts an lower bound on $\tan \beta$~\cite{Arbey:2017gmh}.

\subsection{Allowed parameter space}

Having listed all these constraints, our next task is to use them to constrain a type-X 2HDM, for both the scenarios 1 and 2. We take this up in the present section. 

\medskip

\noindent
$\bullet$ {\bf Scenario 1:}

\noindent
This scenario corresponds to the case, where lighter CP-even scalar is the SM-like Higgs boson. One can further categorize this scenario with WS and RS regions depending on leptonic coupling of $h$, as discussed earlier. 
We scan our parameter space in the following ranges:

\medskip

\noindent
$m_{H} \in \left[125,870\right]$ GeV, $m_H^{\pm} \in\left[125,870\right]$ GeV, $m_A \in \left[20,100\right]$ GeV,
$\tan\beta \in \left[20,100\right]$,\\ $\sin(\beta-\alpha) \in \left[0.99,1\right]$,
$m_{12}^2 \in \left[\frac{{m^2_{H}}}{\tan\beta} - 200,\frac{{m^2_{H}}}{\tan\beta} + 200\right]$. 

\medskip

\noindent
We also mention here that $\lambda_6 = \lambda_7 = 0$, as we only consider soft $Z_2$ breaking terms. \\

\begin{figure}[!hptb]
\floatsetup[subfigure]{captionskip=10pt}
    \begin{subfigure}{.44\linewidth}
    \centering
    \includegraphics[width=7.0cm, height=5.5cm]{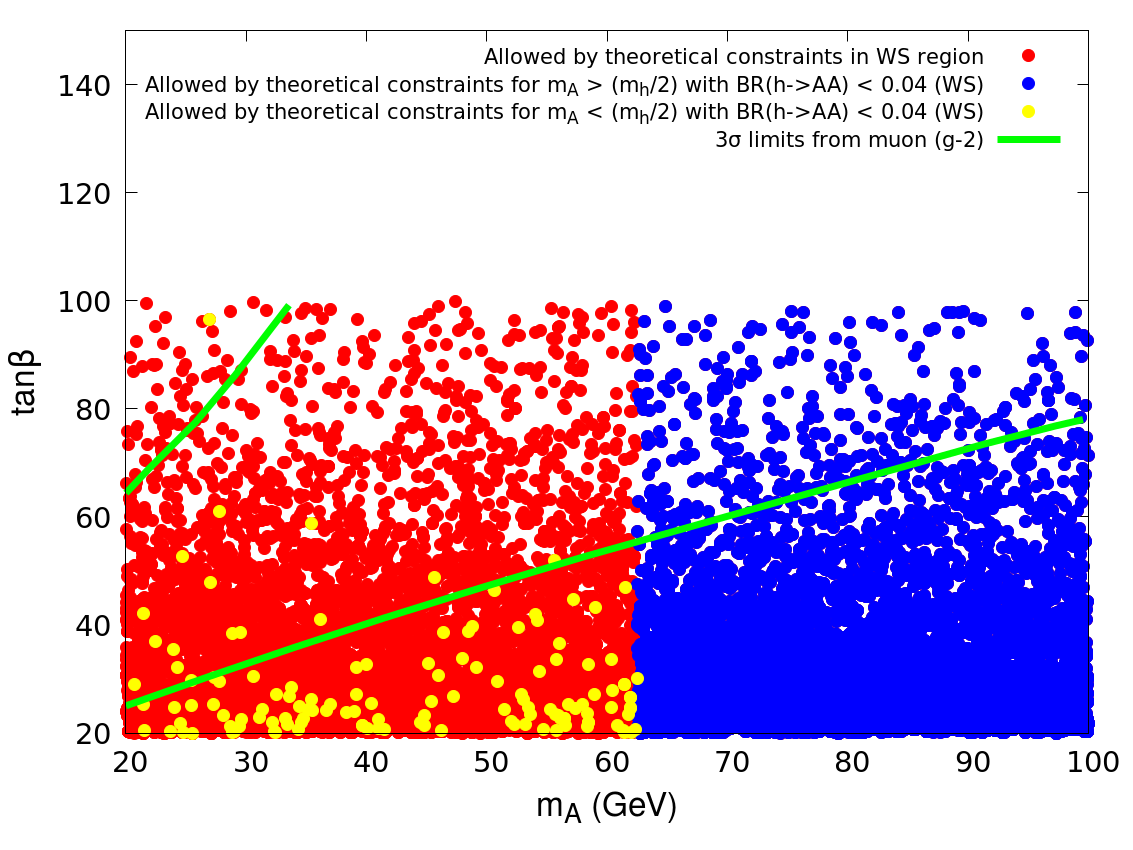}
    \caption{}\label{fig:image1}
    \end{subfigure} %
    \qquad
    \begin{subfigure}{.44\linewidth}
    \centering
    \includegraphics[width=7.0cm, height=5.5cm]{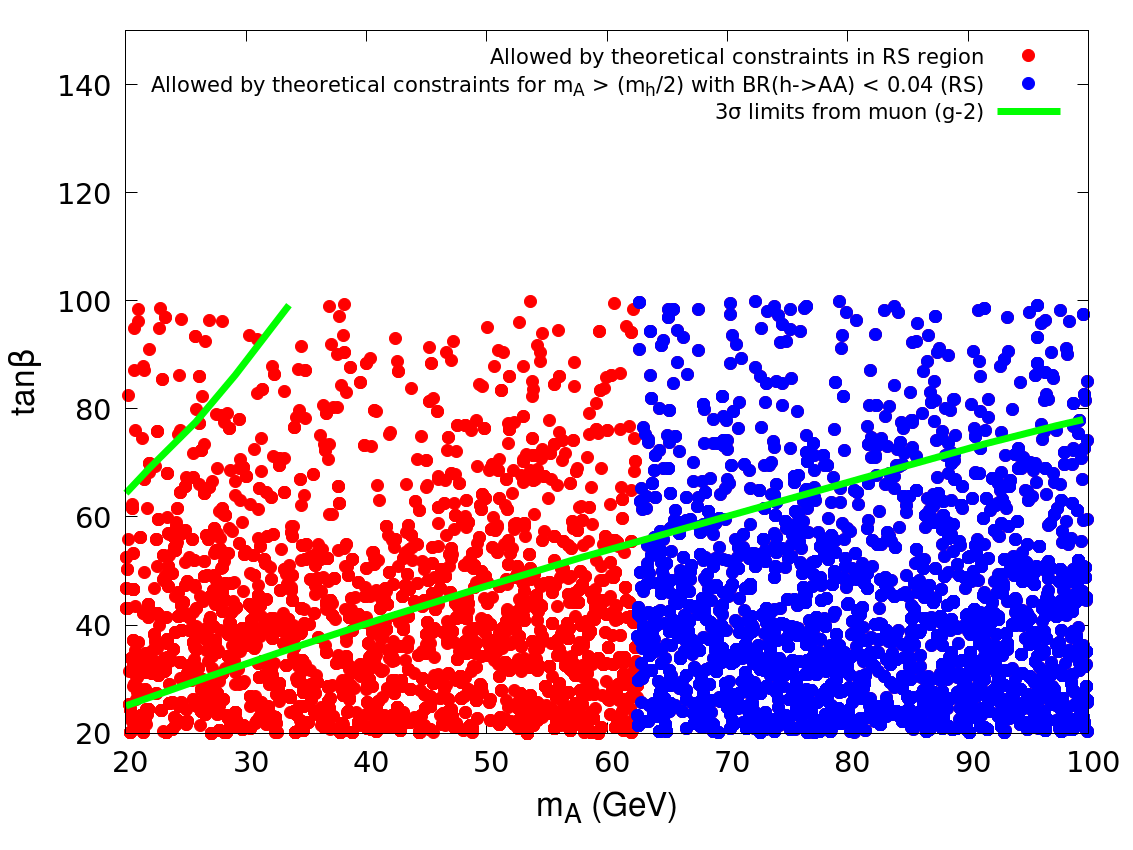}
    \caption{}\label{fig:image12}
   \end{subfigure}
\\[2ex]
\RawCaption{\caption{\it Allowed parameter spaces in $m_A - tan \beta$ plane for Scenario 1 with (a) WS Yukawa and (b) RS Yukawa. The green lines denote the upper and lower limits coming from the observed $g_{\mu}-2$ at 3$\sigma$ level.}
\label{mA_tb_ws_rs_g125_mAall}}
\end{figure}

\begin{figure}[!hptb]
\floatsetup[subfigure]{captionskip=10pt}
    \begin{subfigure}{.44\linewidth}
    \centering
    \includegraphics[width=7.0cm, height=5.5cm]{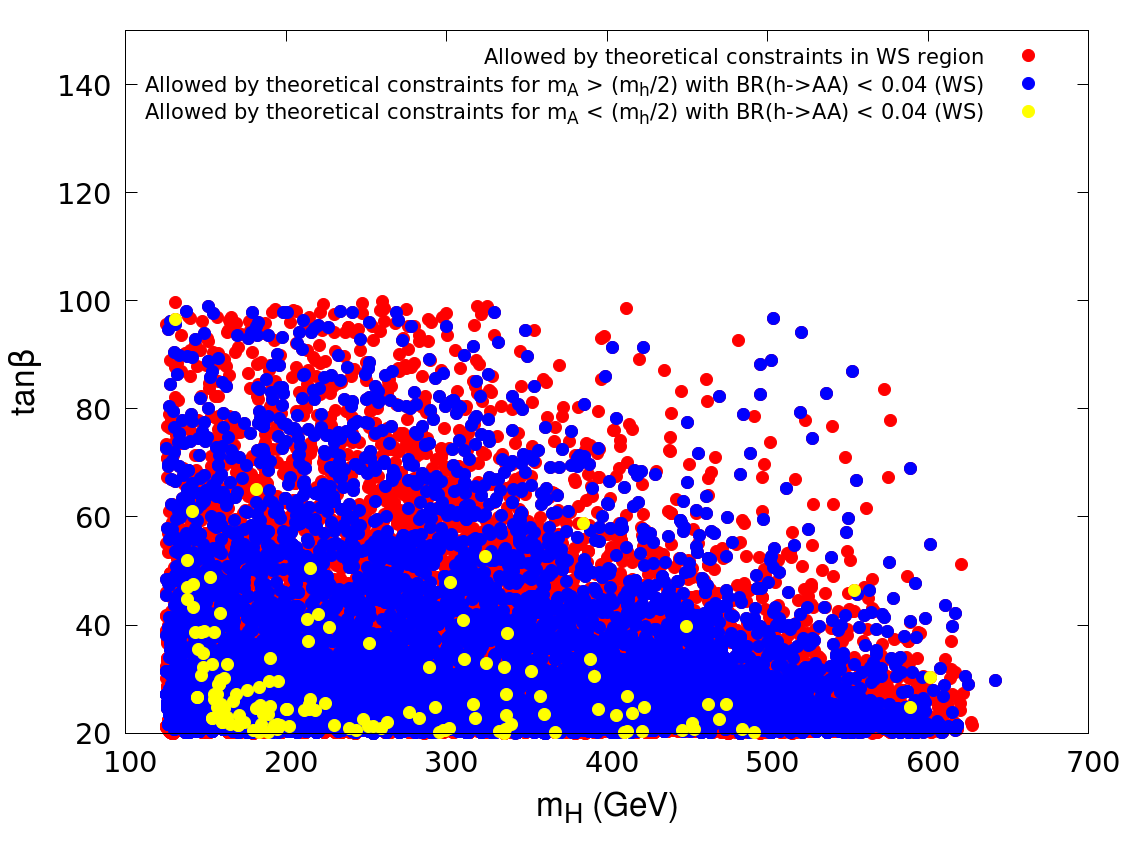}
    \caption{}\label{fig:image1}
    \end{subfigure} %
    \qquad
    \begin{subfigure}{.44\linewidth}
    \centering
    \includegraphics[width=7.0cm, height=5.5cm]{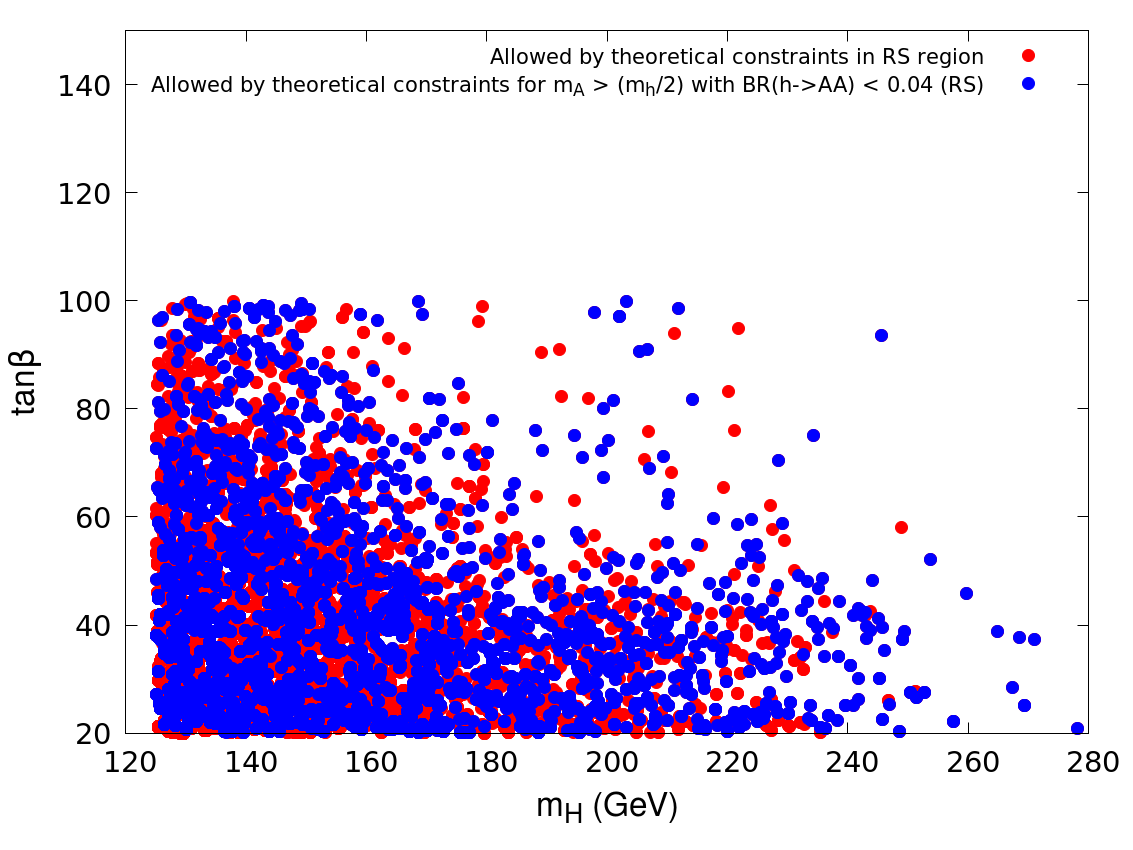}
    \caption{}\label{fig:image12}
   \end{subfigure}
\\[2ex]
    \begin{subfigure}{.44\linewidth}
    \centering
    \includegraphics[width=7.0cm, height=5.5cm]{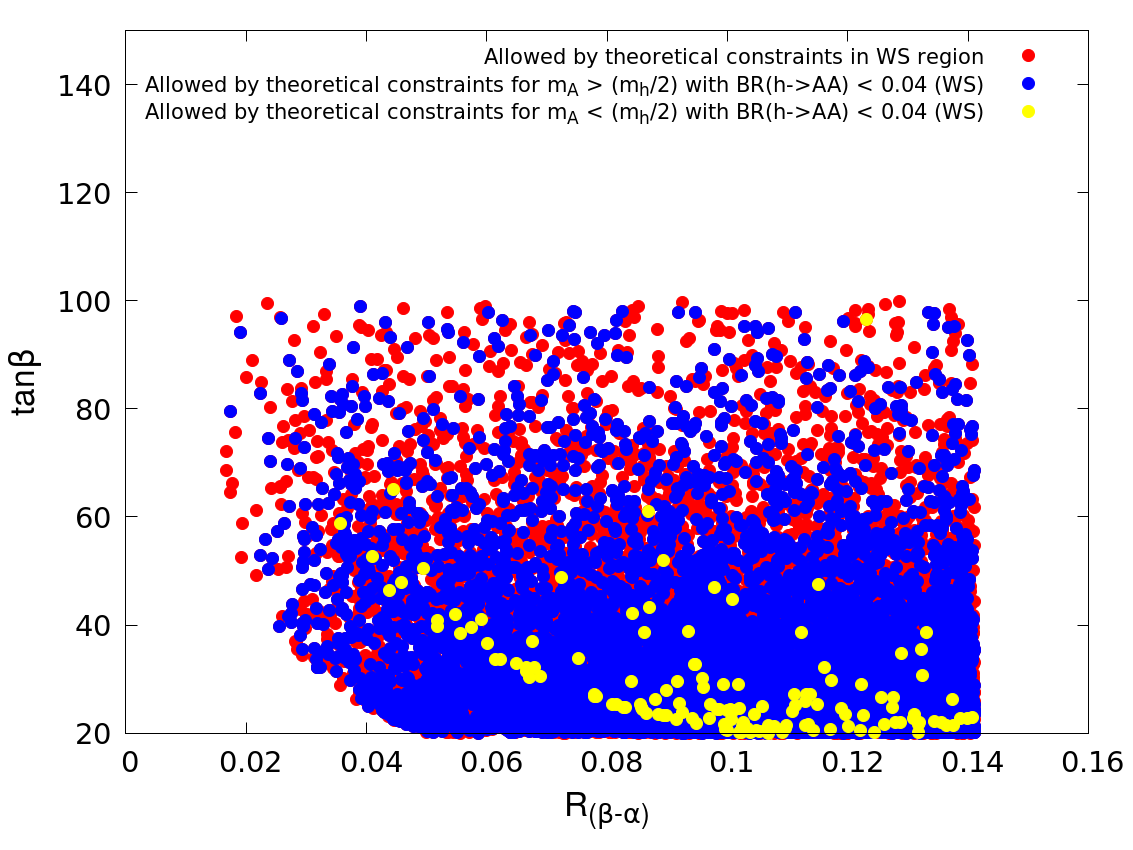}
    \caption{}\label{fig:image1}
    \end{subfigure} %
    \qquad
    \begin{subfigure}{.44\linewidth}
    \centering
    \includegraphics[width=7.0cm, height=5.5cm]{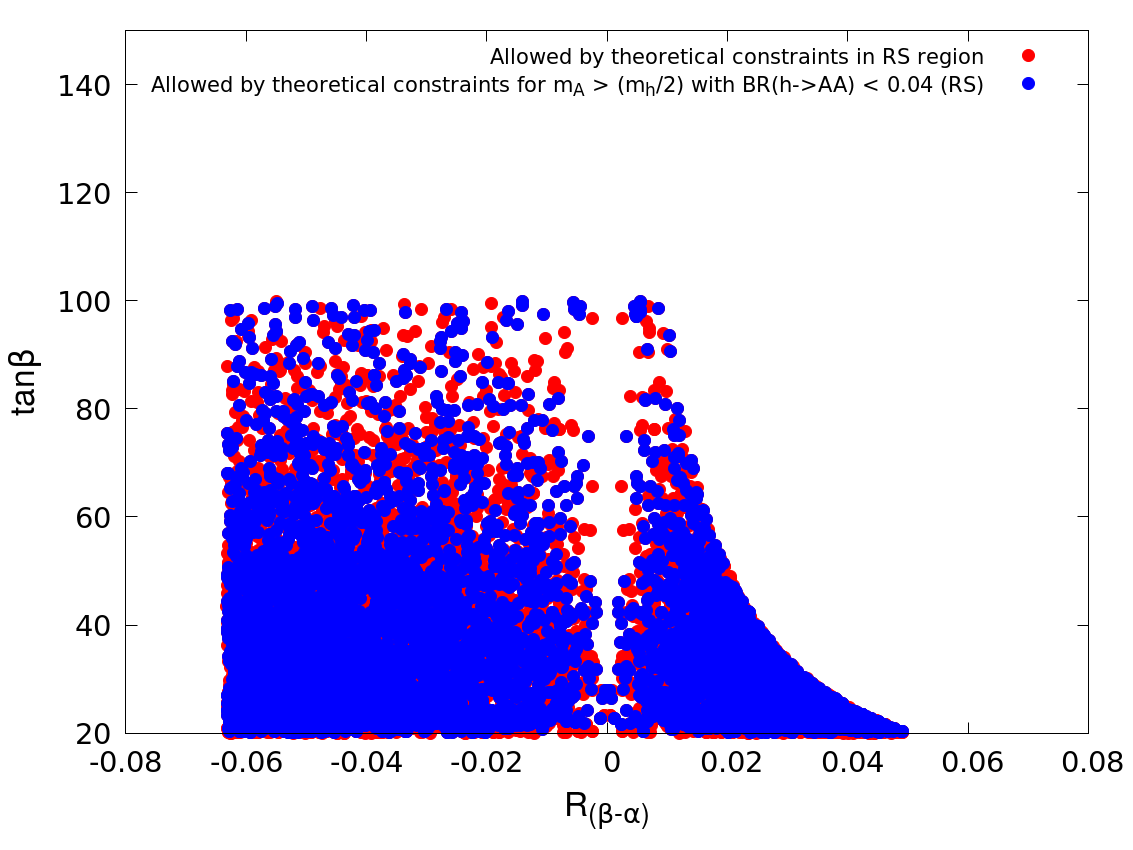}
    \caption{}\label{fig:image12}
   \end{subfigure}
\\[2ex]
    \begin{subfigure}{.44\linewidth}
    \centering
    \includegraphics[width=7.0cm, height=5.5cm]{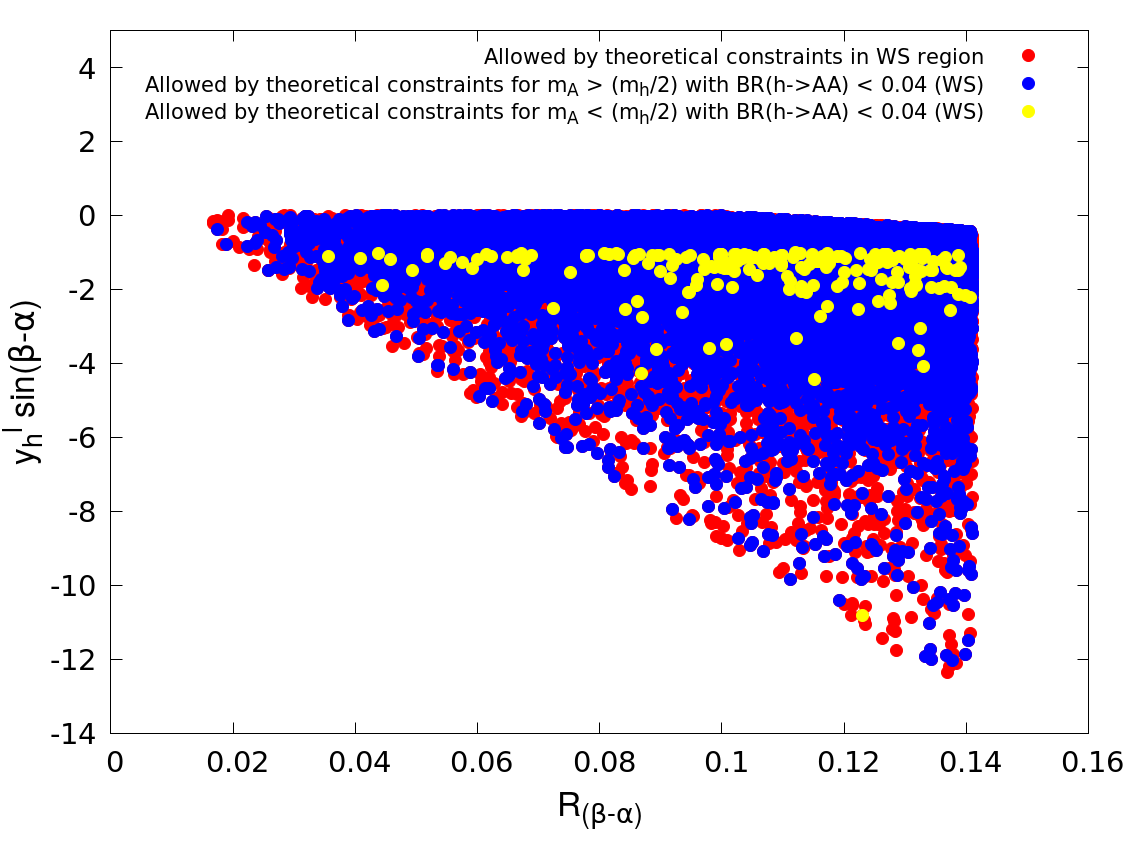}
    \caption{}\label{fig:image1}
    \end{subfigure} %
    \qquad
    \begin{subfigure}{.44\linewidth}
    \centering
    \includegraphics[width=7.0cm, height=5.5cm]{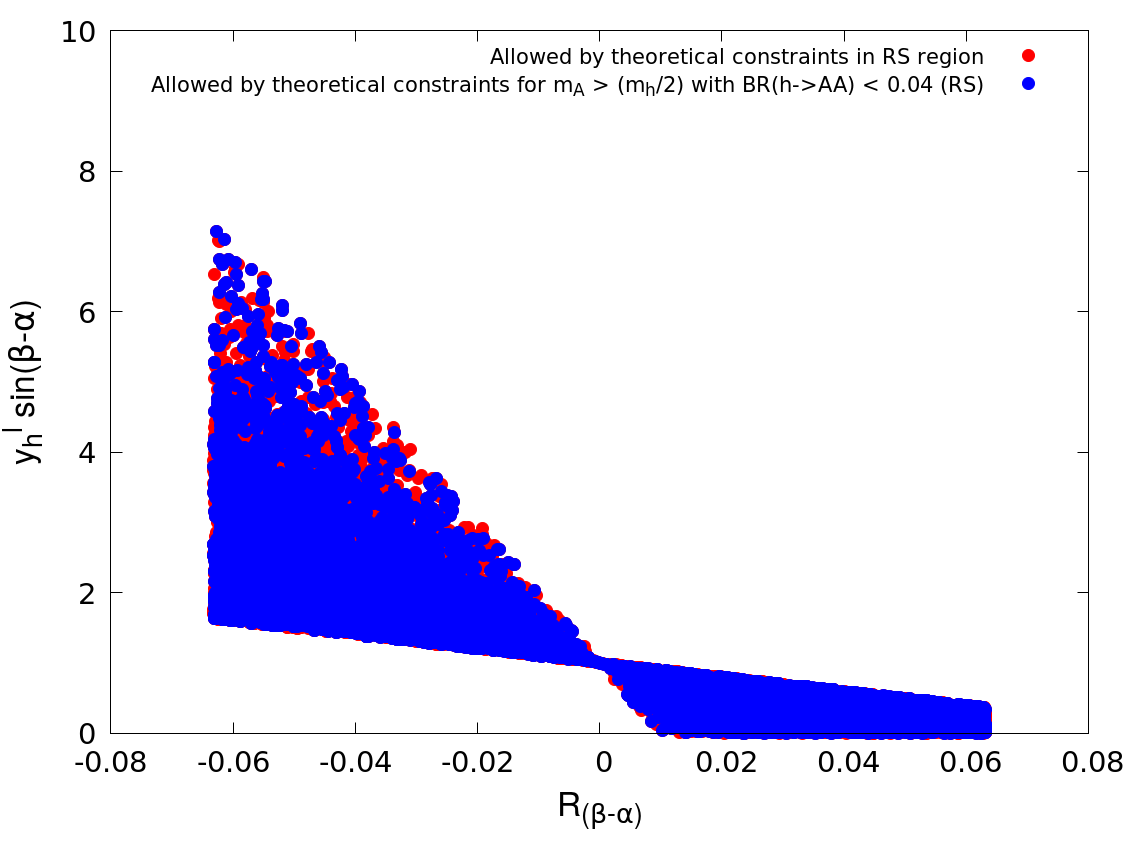}
    \caption{}\label{fig:image12}
   \end{subfigure}
\\[2ex]
\RawCaption{\caption{\it Allowed parameter spaces in scenario 1 in (a) and (b) $m_H - \tan \beta$, (c) and (d) $R_{(\beta - \alpha)} - \tan \beta$, (e) and (f) $R_{(\beta - \alpha)} - y_h^{\ell} \times \sin(\beta - \alpha)$ plane. (a),(c) and (e) correspond to WS Yukawa, (b),(d) and (f) correspond to RS Yukawa.}
\label{sba_yh_tb_1}}
\end{figure}

\noindent
In Figure~\ref{mA_tb_ws_rs_g125_mAall}, we see that though theoretical constraints (namely, perturbativity, unitarity and stability) prefer low to moderate $\tan \beta$, we can still get a large parameter space in WS domain which alleviates $g_{\mu}-2$ discrepancy. On the other hand, in the RS region, large $\tan \beta$ is less favored. 
As discussed earlier in Section~\ref{sec4}, we do not get a small enough BR($h \rightarrow A A)\lsim 4\%$ for $m_A \lsim m_h/2$, as long as we are in the RS domain. This is clear from Figure~\ref{mA_tb_ws_rs_g125_mAall}.

In Figure~\ref{sba_yh_tb_1} (a) and (b), we show the allowed region in $m_H-\tan \beta$ plane. We can see that moderate $\tan \beta$ regions are favored by the theoretical constraints, especially when $m_H$ is large. In addition to that, the upper bound from BR($h_{SM} \rightarrow AA$), pushes the allowed range of $\tan \beta$ to further lower side.

Figure~\ref{sba_yh_tb_1} (c) and (d) displays the allowed region in the $R_{(\beta-\alpha)}- \tan \beta$ plane, where $R_{(\beta-\alpha)}$ is defined as $sgn[\sin(\beta-\alpha)]\times \cos(\beta-\alpha)$. On the whole, while the RS case admits $\sin(\beta-\alpha$) with both signs, it is restricted to positive values only for WS. Furthermore, the WS picture disfavors large $\tan \beta$ from the limit on BR($h_{SM} \rightarrow AA$) so long as $m_A \lsim \frac{m_h}{2}$.

In Figure~\ref{sba_yh_tb_1} (e) and (f), we plot $y_h^{\ell}\times \sin(\beta - \alpha$) against $R_{(\beta-\alpha)}$. The limit on BR($h \rightarrow A A$) for $m_A \lsim m_h/2$ does not allow much deviation of $y^{\ell}_h\times \sin(\beta - \alpha$) from unity, which is also consistent with the alignment limit. In Figure~\ref{sba_yh_tb_1}(f), one can see that, both positive and negative signs for $\sin(\beta-\alpha)$ are equally consistent with the alignment limit($|y_h^{\ell}|\times  \sin(\beta - \alpha) \approx 1)$ in the RS region.


\bigskip

\noindent
$\bullet$ {\bf Scenario 2:}

\begin{figure}[!hptb]
\floatsetup[subfigure]{captionskip=10pt}
    \begin{subfigure}{.44\linewidth}
    \centering
    \includegraphics[width=7.0cm, height=5.5cm]{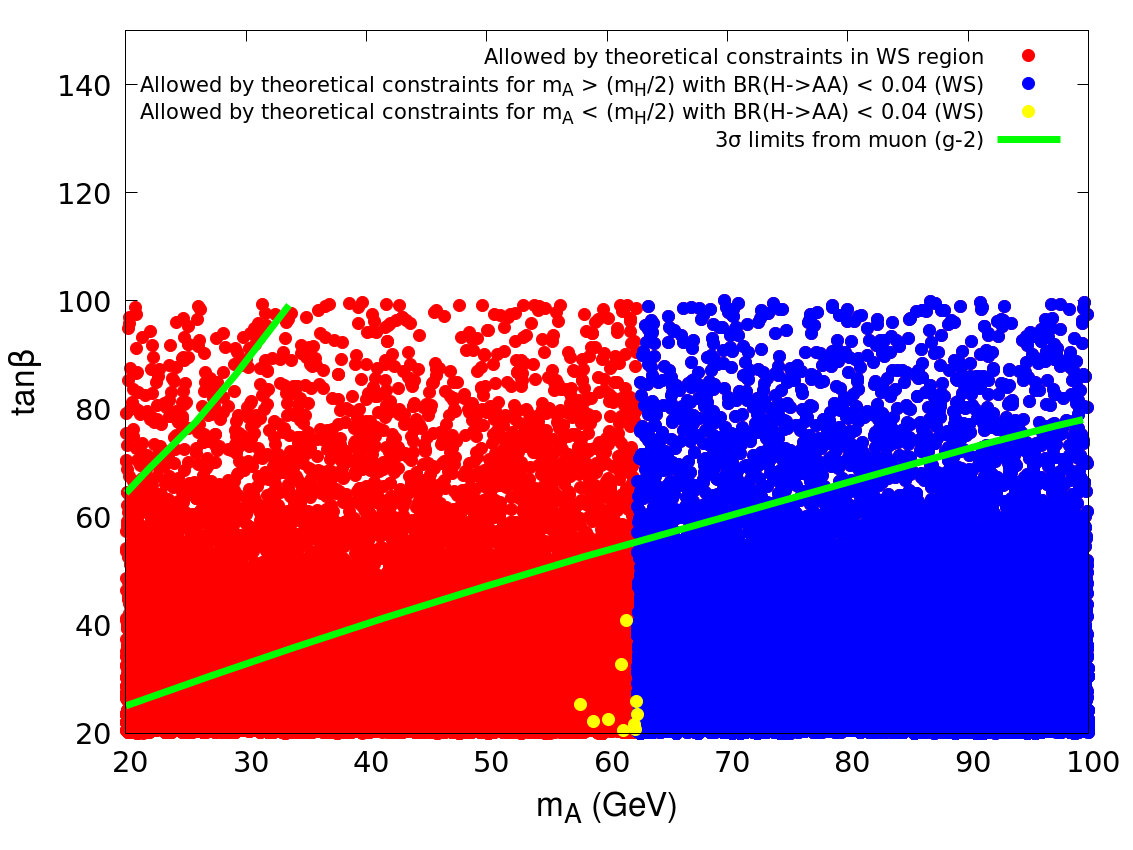}
    \caption{}\label{fig:image1}
    \end{subfigure} %
    \qquad
    \begin{subfigure}{.44\linewidth}
    \centering
    \includegraphics[width=7.0cm, height=5.5cm]{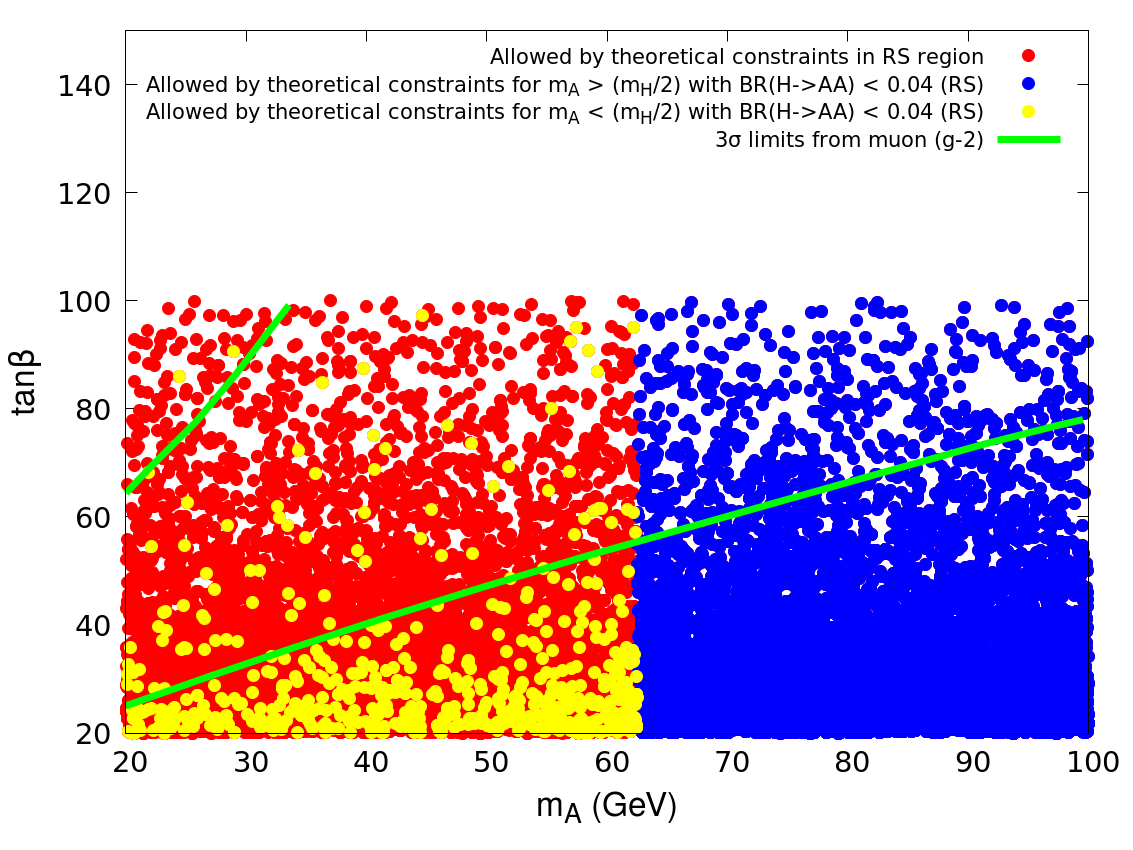}
    \caption{}\label{fig:image12}
   \end{subfigure}
\\[2ex]
\RawCaption{\caption{\it Allowed parameter spaces in $m_A - tan \beta$ plane for Scenario 2 with (a) WS Yukawa and (b) RS Yukawa. The green lines denote the upper and lower limits coming from the observed $g_{\mu} - 2$.}
\label{mA_tb_ws_rs_l125_mAall}}
\end{figure}

\begin{figure}[!hptb]
\floatsetup[subfigure]{captionskip=10pt}
    \begin{subfigure}{.44\linewidth}
    \centering
    \includegraphics[width=7.0cm, height=5.5cm]{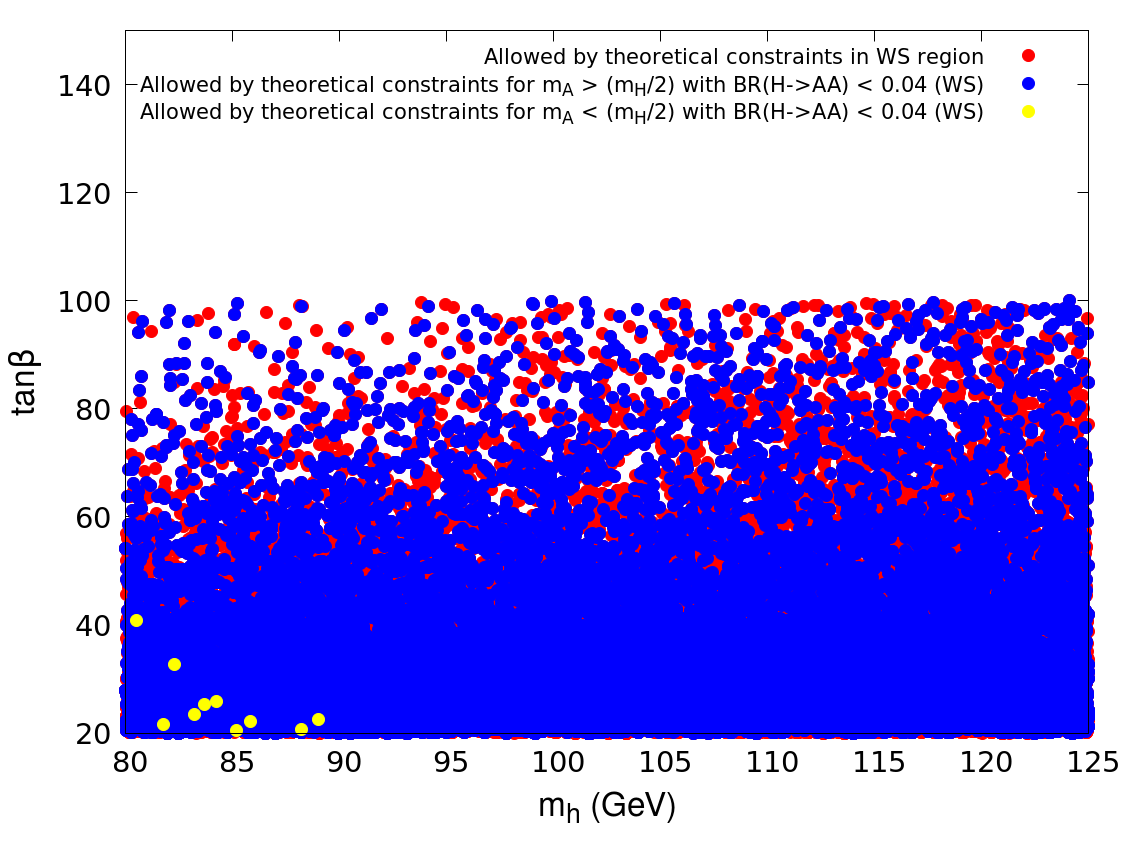}
    \caption{}\label{fig:image1}
    \end{subfigure} %
    \qquad
    \begin{subfigure}{.44\linewidth}
    \centering
    \includegraphics[width=7.0cm, height=5.5cm]{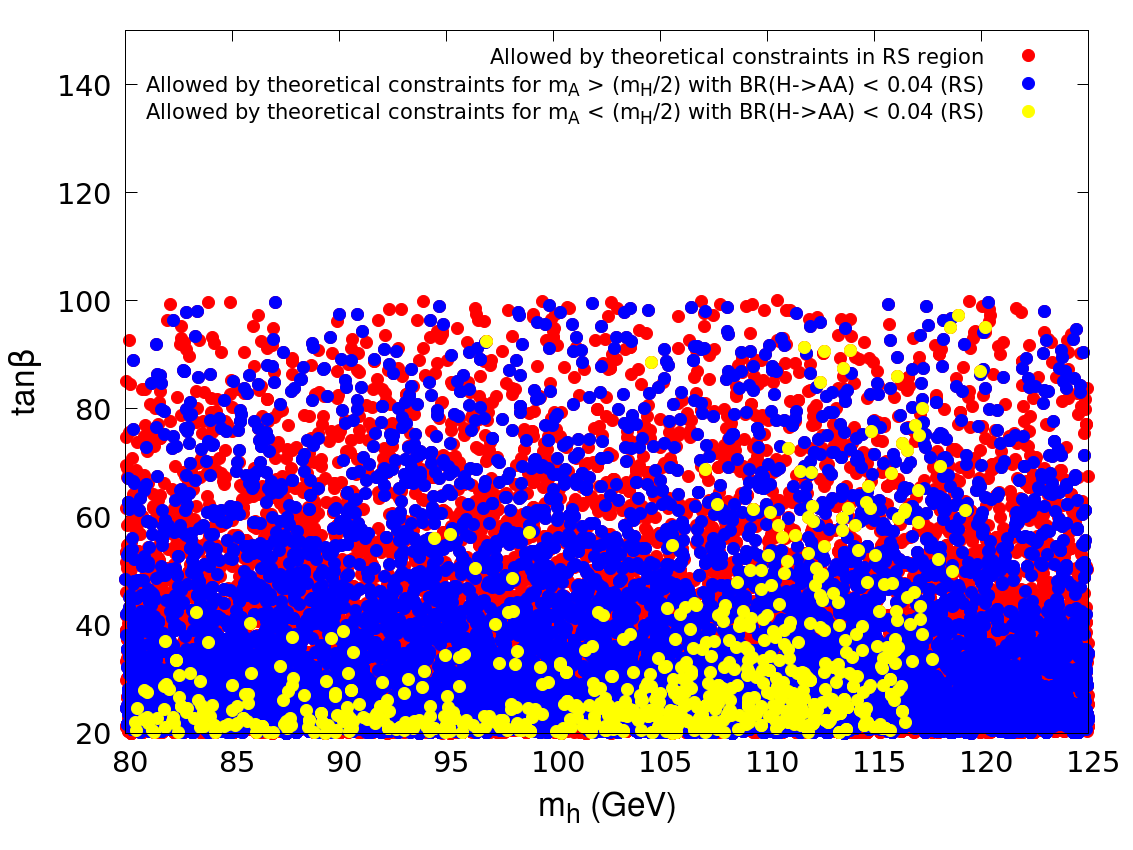}
    \caption{}\label{fig:image12}
   \end{subfigure}
\\[2ex]
    \begin{subfigure}{.44\linewidth}
    \centering
    \includegraphics[width=7.0cm, height=5.5cm]{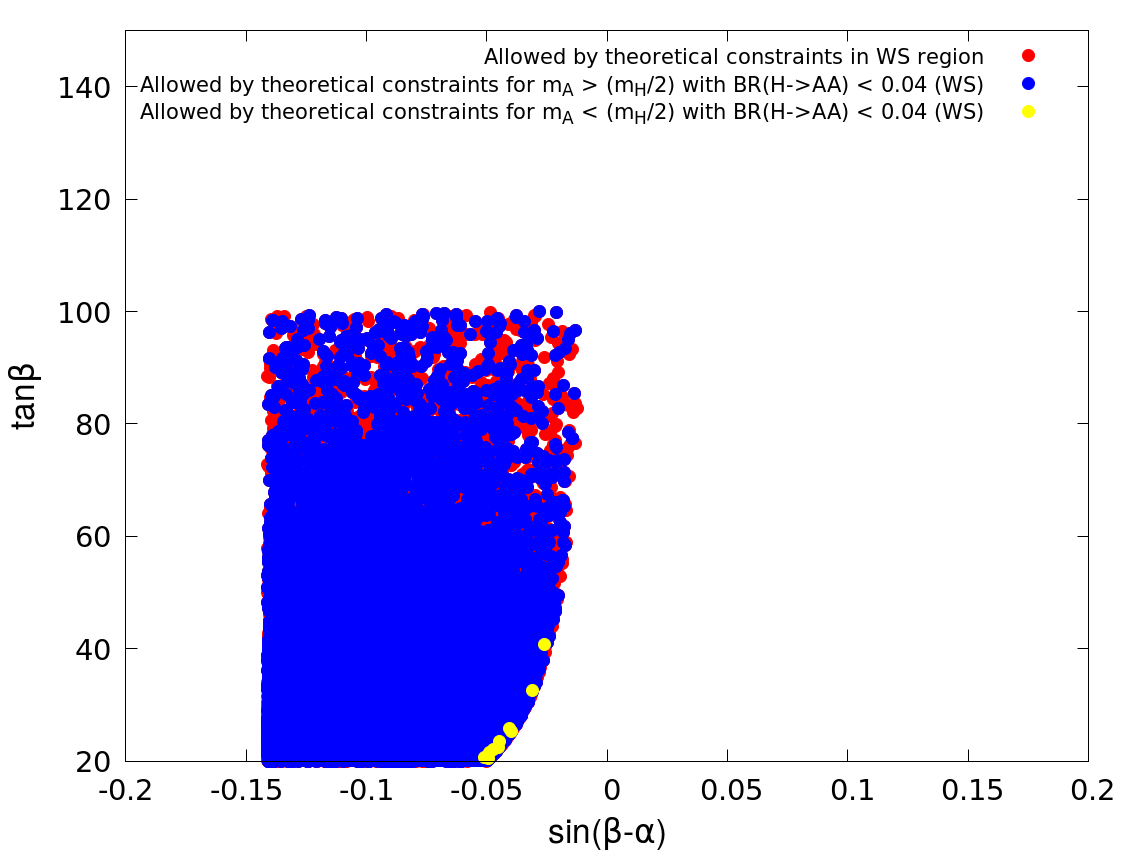}
    \caption{}\label{fig:image1}
    \end{subfigure} %
    \qquad
    \begin{subfigure}{.44\linewidth}
    \centering
    \includegraphics[width=7.0cm, height=5.5cm]{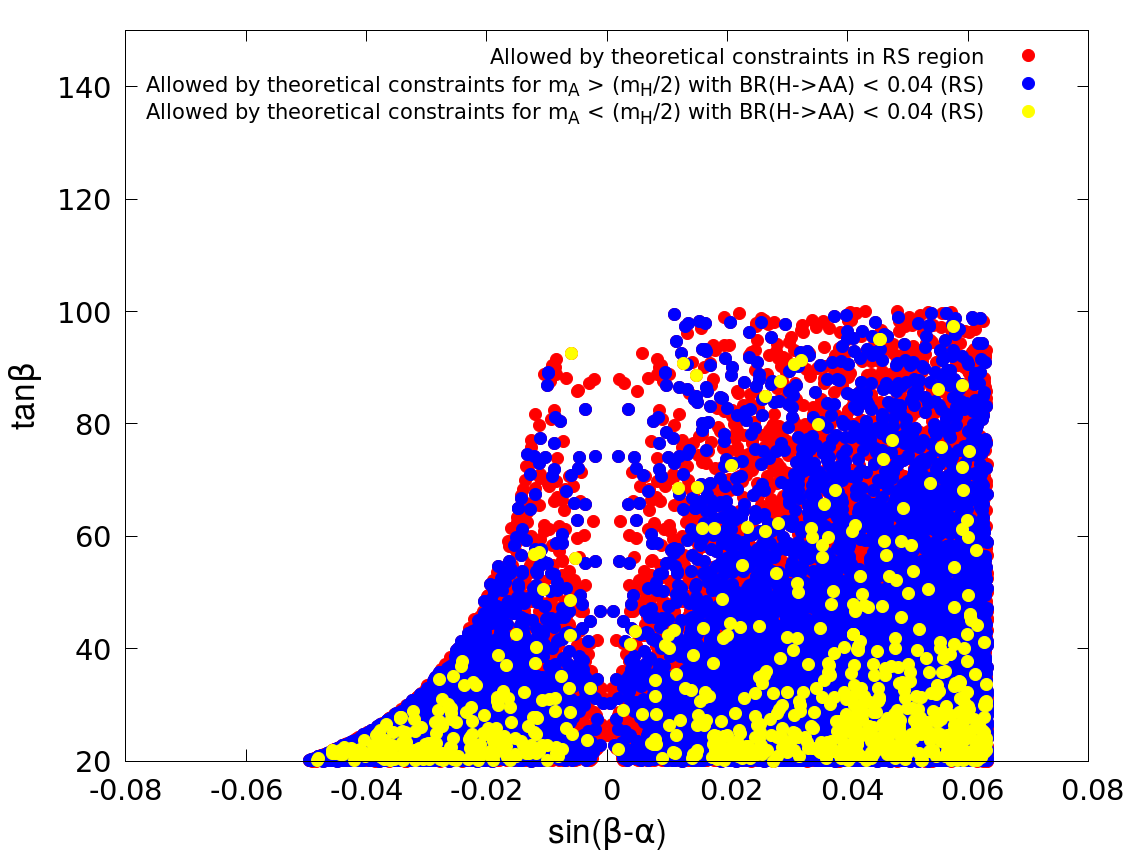}
    \caption{}\label{fig:image12}
   \end{subfigure}
\\[2ex]
    \begin{subfigure}{.44\linewidth}
    \centering
    \includegraphics[width=7.0cm, height=5.5cm]{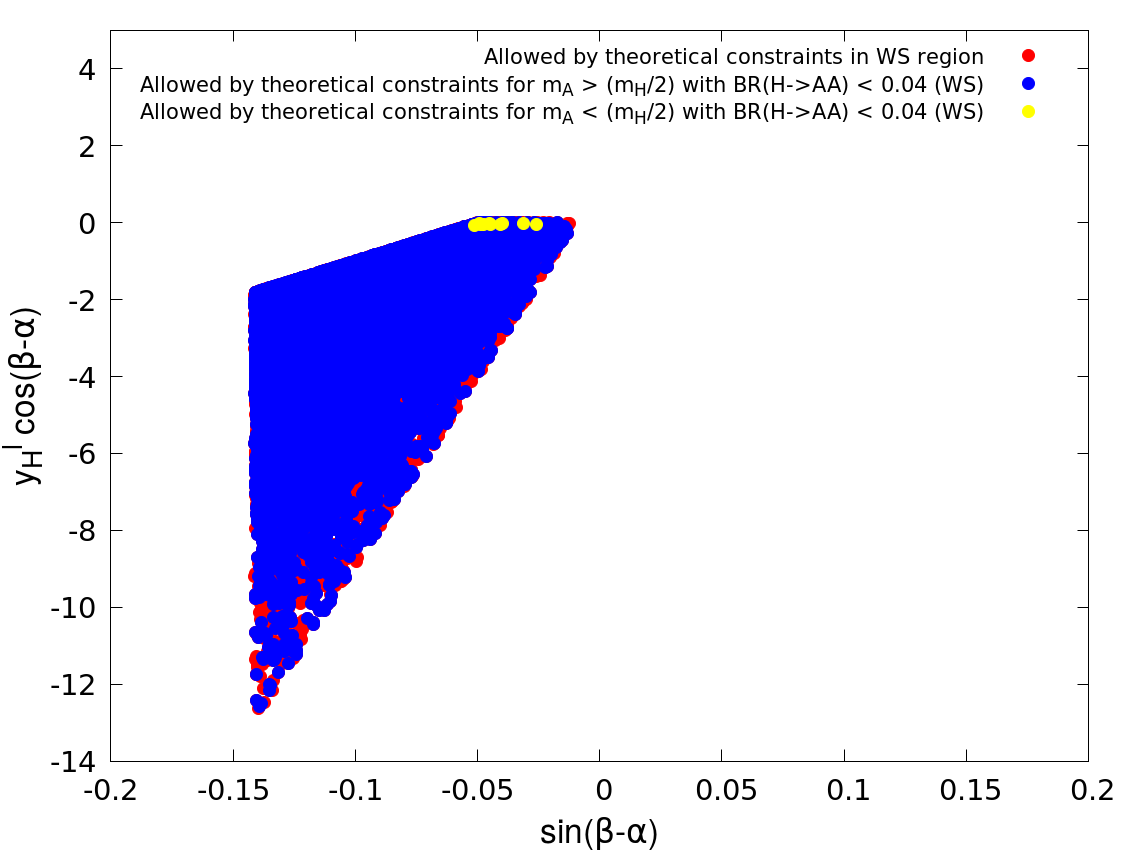}
    \caption{}\label{fig:image1}
    \end{subfigure} %
    \qquad
    \begin{subfigure}{.44\linewidth}
    \centering
    \includegraphics[width=7.0cm, height=5.5cm]{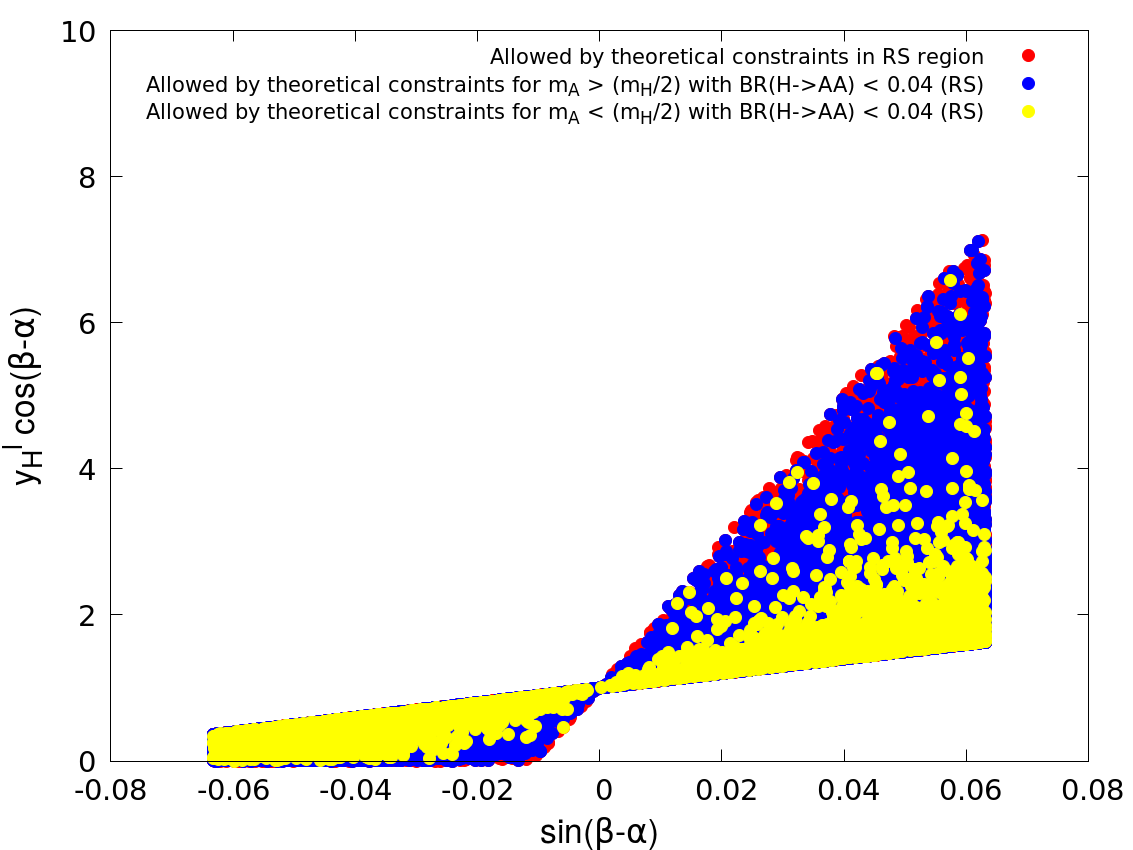}
    \caption{}\label{fig:image12}
   \end{subfigure}
\\[2ex]
\RawCaption{\caption{\it Allowed parameter spaces in scenario 2 in (a) and (b) $m_H - \tan \beta$, (c) and (d) $\sin(\beta - \alpha) - \tan \beta$, (e) and (f) $\sin(\beta - \alpha) - y_h^{\ell} \times \sin(\beta - \alpha)$ plane. (a),(c) and (e) correspond to WS Yukawa, (b),(d) and (f) correspond to RS Yukawa.}
\label{sba_yh_tb_2}}
\end{figure}

\noindent
In this scenario, the heavier CP-even scalar $H$ is the SM-like Higgs.
To get the allowed regions in this scenario, we scan our parameter space in the following range:

\medskip
\noindent
$m_{h} \in \left[80,125\right]$ GeV, $m_H^{\pm} \in\left[80,180\right]$ GeV, $m_A \in \left[20,100\right]$ GeV,
$\tan\beta \in \left[20,100\right]$,\\ $\cos(\beta-\alpha) \in \left[0.99,1\right]$, $m_{12}^2 \in \left[\frac{{m^2_{H}}}{\tan\beta} - 200,\frac{{m^2_{H}}}{\tan\beta} + 200\right]$.
\medskip

\noindent
In Figures~\ref{mA_tb_ws_rs_l125_mAall} and~\ref{sba_yh_tb_2}, we plot the points allowed by theoretical constraints as well as constraints on BR$(H \rightarrow AA)$, in two-dimensional planes of various model parameters.

If we focus on Figure~\ref{mA_tb_ws_rs_l125_mAall}(a), it becomes clear that for $m_A \lsim \frac{m_H}{2}$, the constraints on $HAA$ coupling can leave a very narrow region near resonance $m_A \approx m_H/2$, for WS cases, which is not quite compatible with the $g_{\mu}-2$ observation, within $3\sigma$. But the situation will be more relaxed in the RS domain for scenario 2 (yellow points in Figure~\ref{mA_tb_ws_rs_l125_mAall}(b)). On the other hand, we can get a large parameter space both in WS and RS region, which can solve $g_{\mu}-2$ discrepancy, for $m_A > \frac{m_H}{2}$.

In Figure~\ref{sba_yh_tb_2} (a) and (b), the allowed regions in $m_h-\tan \beta$ plane are shown for the WS and RS respectively. We can see that, in the RS case, low to moderate $\tan \beta$ will be favored from the requirement of low BR($h_{SM}\rightarrow AA)$. However, when the difference between the lighter and heavier CP-even scalar masses decreases, even larger $\tan \beta$ becomes allowed.  

In Figure~\ref{sba_yh_tb_2} (c) and (d), we show the allowed region in $\sin(\beta-\alpha)- \tan \beta$ plane where both positive and negative $\sin(\beta-\alpha)$ is allowed for RS cases, but WS is attained with only negative $\sin(\beta-\alpha)$.

In Figure~\ref{sba_yh_tb_2} (e) and (f), one can see similar behavior as scenario 1, where small BR($H \rightarrow AA)$ for $m_A \lsim \frac{m_H}{2}$ prefers lepton Yukawa coupling $y_h^{\ell}$ close to unity, consistent with the observed Higgs signals for both WS and RS cases.



\section{The running of various couplings}
\label{sec5}

\subsection{The Renormalization Group Equations(RGEs)}

The parameters constrained above are considered at the electroweak scale, set at the pole mass of top quark $(\sim 173.34$ GeV). We now investigate how they evolve at higher scales and thus obtain their domain of validity in the light of vacuum stability(following~\cite{Nie:1998yn}) and perturbative unitarity(following~\cite{Lee:1977yc}). This yields the cut-off scale $\Lambda^{cut-off}_{UV}$.

In this subsection we present the one-loop RG equations for the various quartic couplings as well as the gauge and third generation Yukawa couplings. For actual presentation of our results, we will take recourse to the two-loop renormalization group equations~\cite{Chowdhury:2015yja} for enhanced precision and rigor. However, we will soon see that, qualitatively the evolution trajectories at the one- and two-loop levels are very similar in our case and that the quantitative differences are rather minor, at least at energy scales well below the perturbative limits of couplings. Keeping this in view, we start by presenting the one-loop RGEs so that we can fall back on them to provide intuitive explanations of the trajectories. At the same time, the detailed results presented in the next subsection are all based on two-loop equations, although we take the liberty of explaining them in terms of one-loop equations, empowered by reasons summarized above.

First we present the one-loop RGEs for the gauge couplings. They form a stand-alone set, at one loop, as we can see from Equation~\ref{gauge_rge}, and therefore they remain unchanged for different types of 2HDMs. We mention here that in writing Equation~\ref{gauge_rge}, GUT normalization has not been used.

\begin{align}
\label{gauge_rge}
16 \pi ^2 \beta _{g_1} =&7 g_1^3 \nonumber \\
16 \pi ^2 \beta _{g_2} =&-3 g_2^3 \nonumber \\
16 \pi ^2 \beta _{g_3} =&-7 g_3^3
\end{align}

Next we focus on the RGE of the Yukawa couplings in Type-X 2HDM. The corresponding equations are as follows. Here $g$ and $Y$ in the superscripts, respectively, denote gauge and Yukawa interactions, contributing to the running of the Yukawa couplings(taken here as real).

\begin{align}
16 \pi ^2 \beta _{Y_t}^{g} &=-\left( \frac{17}{12} g_1^2 +\frac{9}{4} g_2^2 +8 g_3^2\right) Y_t \nonumber \\
16 \pi ^2 \beta _{Y_t}^{Y} &=\left( \frac{3}{2} Y_b^2 +\frac{9}{2} Y_t^2 \right) Y_t \nonumber\\ 
16 \pi ^2 \beta _{Y_b}^{g} &=-\left( \frac{5}{12} g_1^2 +\frac{9}{4} g_2^2 +8 g_3^2\right) Y_b \nonumber \\
16 \pi ^2 \beta _{Y_b}^{Y} &=\left( \frac{9}{2} Y_b^2 +\frac{3}{2} Y_t^2 \right) Y_b \nonumber\\
16 \pi ^2 \beta _{Y_\tau}^{g} &=-\left( \frac{15}{4}g_1^2 +\frac{9}{4}g_2^2\right) Y_\tau \nonumber \\
16 \pi ^2 \beta _{Y_\tau}^{Y} &=\frac{5}{2}Y_\tau^3
\label{yuk_rge}
\end{align}

\noindent
The resulting beta-function will be the sum of the gauge and Yukawa components. 

\begin{equation}
\beta _{Y}= \beta _{Y}^{g} + \beta _{Y}^{Y}
\end{equation}

\noindent
The Yukawa and gauge contributions show similar behavior for $Y_t$ and $Y_b$. It is clear from Equations.~\ref{yuk_rge} that the gauge contribution decreases with energy whereas the Yukawa part go up at higher energy. However, the terms involving the strong coupling constant $g_3$ dominates over the other terms and therefore the top and bottom Yukawa couplings monotonically decrease with energy. The $\tau$-Yukawa coupling on the other hand, unaffected by the strong interaction, remains almost constant. This behavior can be seen from Figure~\ref{RG_gauge_yukawa_2loop}.

The relevant equations for the running of quartic couplings are given below. Here, the superscripts $b$ and $Y$ denote, respectively, bosonic(gauge couplings and quartic couplings) and Yukawa interactions, contributing to the running of $\lambda'$s.

\begin{align*}
16 \pi ^2 \beta _{\lambda_1}^{b} =&\frac{3}{4} g_1^4 +\frac{3}{2} g_1^2 g_2^2 +\frac{9}{4} g_2^4 -3g_1^2 \lambda_1 -9 g_2^2 \lambda_1 +12 \lambda_1^2 +4 \lambda_3^2 +4 \lambda_3 \lambda_4 +2\lambda_4^2 +2\lambda_5^2 \nonumber \\
16 \pi ^2 \beta _{\lambda_1}^{Y} =& -4 Y_\tau^4 +4 Y_\tau^2 \lambda_1 \nonumber\\
16 \pi ^2 \beta _{\lambda_2}^{b} =& \frac{3}{4} g_1^4 +\frac{3}{2} g_1^2 g_2^2 +\frac{9}{4} g_2^4 -3g_1^2 \lambda_2 -9 g_2^2 \lambda_2 +12 \lambda_2^2 +4 \lambda_3^2 +4 \lambda_3 \lambda_4 +2\lambda_4^2 +2\lambda_5^2 \nonumber\\
16 \pi ^2 \beta _{\lambda_2}^{Y} =& -12 Y_b^4 -12 Y_t^4 +\left( 12 Y_b^2 +12 Y_t^2 \right) \lambda_2 \nonumber\\
\end{align*}

\begin{align}
16 \pi ^2 \beta _{\lambda_3}^{b} =& \frac{3}{4} g_1^4 -\frac{3}{2} g_1^2 g_2^2 +\frac{9}{4} g_2^4 -3g_1^2 \lambda_3 -9 g_2^2 \lambda_3 \nonumber \\
& +(\lambda_1 +\lambda_2 ) \left( 6  \lambda_3 +2\lambda_4\right) +4 \lambda_3^2 +2\lambda_4^2 +2\lambda_5^2 \nonumber\\
16 \pi ^2 \beta _{\lambda_3}^{Y} =& \left(6 Y_b^2 +6 Y_t^2 +2 Y_\tau^2\right) \lambda_3 \nonumber\\
16 \pi ^2 \beta _{\lambda_4}^{b} =& 3 g_1^2 g_2^2 -\left( 3 g_1^2 +9 g_2^2 \right) \lambda_4 +2 \lambda_1 \lambda_4+2 \lambda_2 \lambda_4+8 \lambda_3 \lambda_4+4 \lambda_4^2+8 \lambda_5^2 \nonumber\\
16 \pi ^2 \beta _{\lambda_4}^{Y} =& \left( 6 Y_b^2 +6 Y_t^2 +2 Y_\tau^2\right) \lambda_4 \nonumber\\
16 \pi ^2 \beta _{\lambda_5}^{b} =& \left(-3 g_1^2 -9 g_2^2 +2\lambda_1 +2\lambda_2 +8 \lambda_3 +12 \lambda_4\right) \lambda_5 \nonumber\\
16 \pi ^2 \beta _{\lambda_5}^{Y} = & \left(6 Y_b^2 +6 Y_t^2 +2Y_\tau^2\right) \lambda_5
\label{lambda_rge}
\end{align}

\noindent
Like before, the actual beta-function will be the sum of the bosonic and Yukawa components. 

\begin{equation}
\beta _{\lambda}= \beta _{\lambda}^{b} + \beta _{\lambda}^{Y}
\end{equation}

\noindent
One should note, since the Yukawa couplings depend on the specific kinds of 2HDM, it is obvious that their evolution as well as those of the quartic couplings are model-dependent. This is obvious from Equations.~\ref{yuk_rge} and \ref{lambda_rge}.

\subsection{Coupling trajectories and inference drawn from them}

In this subsection, the running of various couplings will be illustrated in terms of a few chosen benchmark points. A brief justification for choosing those will be given shortly. Based on the discussion in the preceding subsection, we will present here the full two-loop results for our benchmark points(BP). Our chosen benchmarks are consistent with theoretical as well as experimental constraints. 


We have seen that, in Scenario 1, the requirement of low branching fraction of SM-like Higgs to two pseudoscalars along with other constraints leads us to $m_A > \frac{m_h}{2}$ in the RS region. However, it is possible to get allowed points in the whole range of $m_A$ in the WS regime. Keeping this in mind, we choose three benchmarks BP1, BP2 and BP3 for scenario 1. BP1 corresponds to WS region with $m_A > \frac{m_h}{2}$. BP2 corresponds to WS region and $m_A < \frac{m_h}{2}$. For BP3, we have taken RS with $m_A > \frac{m_h}{2}$. We present the benchmark points chosen for Scenario 1 in Table~\ref{bp_gt125_ws_lt62}.

\begin{table}[!hptb]
\begin{center}
\begin{footnotesize}
\begin{tabular}{| c | c | c | c |}
\hline
& BP1 & BP2 & BP3 \\
\hline
$m_H$ in GeV & 449.734 & 324.237 & 153.865\\
\hline
$m_A$ in GeV & 80.0 & 24.6997 & 63.0 \\
\hline
$m_{H^{\pm}}$ in GeV  & 453.895 & 331.34 & 176.152 \\
\hline
$\lambda_1$  & 0.095392 & 1.4963 & 0.52616 \\
\hline
$\lambda_2$  & 0.25788 & 0.25792 & 0.25773\\
\hline
$\lambda_3$ & 6.9130 & 3.5968 & 0.52559\\
\hline
$\lambda_4$ & -3.3549 & -1.8783 & -0.56774\\
\hline
$\lambda_5$ & 3.23062 & 1.72343 & 0.324993\\
\hline
$m^2_{12}$ in $GeV^2$  & 2696.2389 & 1992.85 & 353.226215 \\
\hline
$\tan \beta$ & 75.0 & 52.7154 & 67.0\\
\hline
$\sin(\beta - \alpha)$ & 0.9996 & 0.999163 & 0.999996\\
\hline
$y_h^{\ell} \times \sin(\beta - \alpha)$ & -1.12095144 & -1.15624366 & 0.81048833\\
\hline
\end{tabular}
\end{footnotesize}
\caption{\it Benchmark points for Scenario 1.}
\label{bp_gt125_ws_lt62}
\end{center}
\end{table}

As long as we are in the alignment limit with large $\tan \beta$, $\lambda_2$ is precisely determined by SM-like Higgs with a very small value($\approx \frac{{m^2_h}}{v^2}\approx 0.258$), which is the case for all the benchmarks in Table~\ref{bp_gt125_ws_lt62}. On the other hand, $\lambda_1$ and $\lambda_3$ depend on the the mass splitting between two CP-even scalars. Furthermore, $\lambda_1$ can be controlled by ${m^2_{12}}$, which gets an enhancement factor in the large $\tan \beta$ region. As for this parameter space, we have ${m^2_{12}} \sim \frac{{m^2_H}}{\tan \beta}$ with large $\tan \beta$, $\lambda_4$ is proportional to ${m^2_A} - 2{m^2_{H^\pm}} + m^2_H$ and takes a negative value for our benchmarks. Similarly, $\lambda_5$ takes a value close to $\lambda_4$ with a opposite sign, being proportional to $-{m^2_A} + m^2_H$. It is clearly seen that for degenerate $m_H$ and $m_{H^\pm}$, $\lambda_5\approx -\lambda_4$. The equality in magnitude is prominent in case of large $m_H$. For BP3 this does not apply. However, the mutual opposite sign between $\lambda_4$ and $\lambda_5$ still holds. We would like to mention here that all the benchmarks satisfy the limit on $y^{\ell}_h$ as well as $y^V_h$ from the alignment condition~\cite{Sirunyan:2018koj,Aad:2019mbh}.

\begin{figure}
\floatsetup[subfigure]{captionskip=10pt}
    \begin{subfigure}{.44\linewidth}
    \centering
    \includegraphics[width=7.6cm, height=5.5cm]{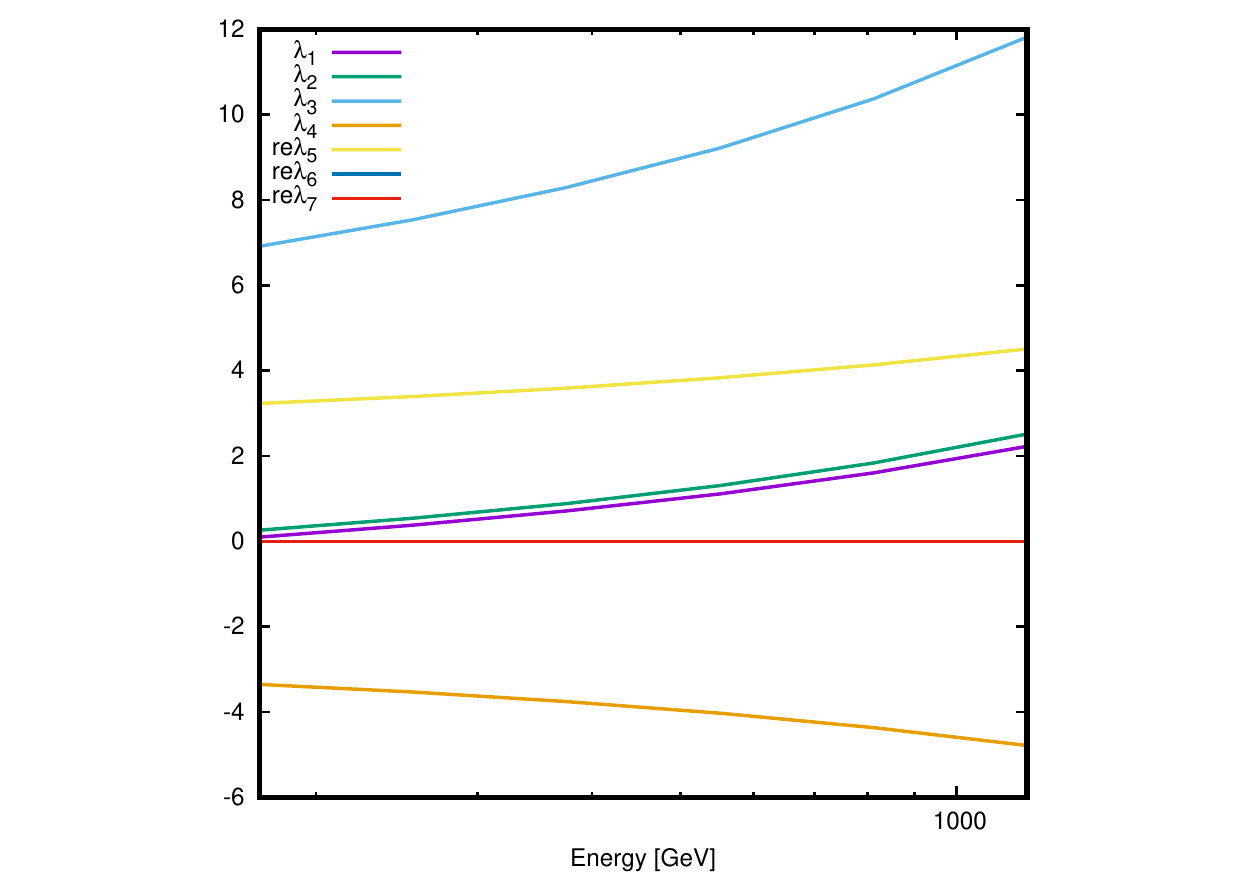}
    \caption{}\label{fig:image1}
    \end{subfigure} %
    \qquad
    \begin{subfigure}{.44\linewidth}
    \centering
    \includegraphics[width=7.6cm, height=5.5cm]{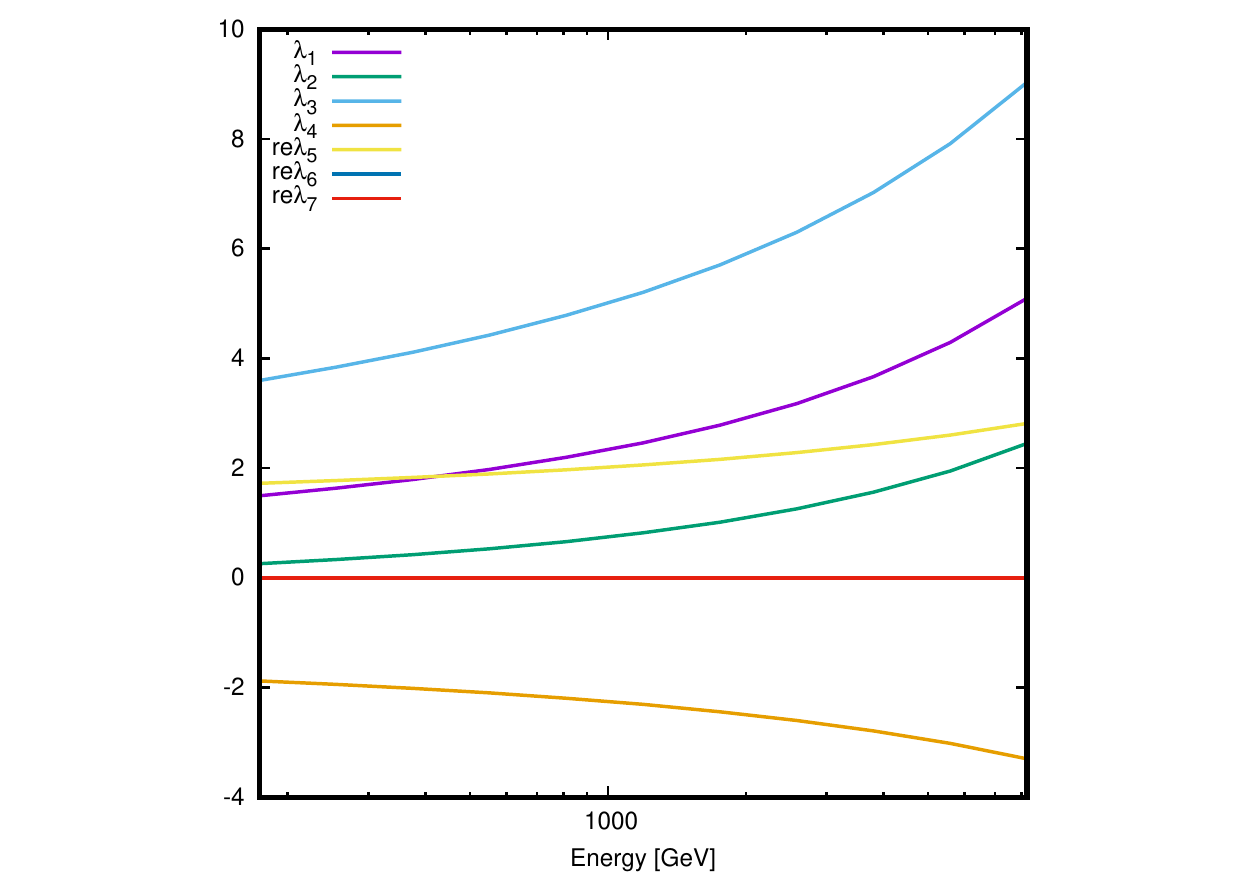}
    \caption{}\label{fig:image12}
   \end{subfigure}
\\[2ex]
  \begin{subfigure}{\linewidth}
  \centering
  \includegraphics[width=7.6cm, height=5.5cm]{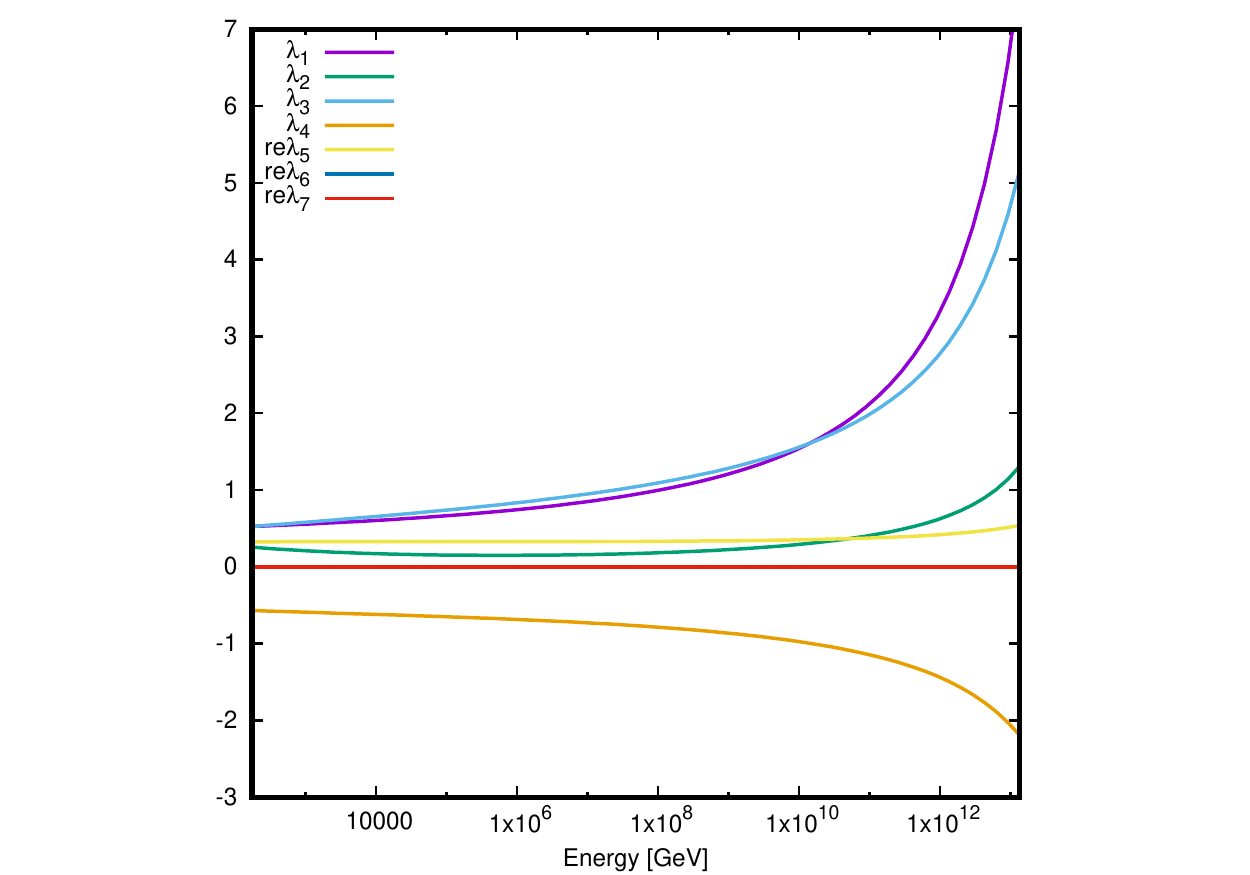}
  \caption{}\label{fig:image3}
  \end{subfigure} 
\RawCaption{\caption{\it Two-loop RG running of quartic couplings for the benchmarks (a) BP1, (b) BP2 and (c) BP3 from Scenario 1.}
\label{bp123}}
\end{figure}

In Figure~\ref{bp123}, we can see the two-loop RG running of quartic couplings for BP1, BP2 and BP3. For all these benchmarks tree level unitarity decides the value of $\Lambda^{cut-off}_{UV}$ which is denoted by the end scale in all figures, whereas stability and perturbativity can be satisfied even after that cut-off scale. It is clear from the running that the larger the value for any quatic coupling at the electroweak scale, the quicker it breaks the unitarity criteria. For both BP1 and BP2, $\lambda_3$ is becomes largest among the quartic couplings at the breakdown scale, whereas in BP3 $\lambda_1$ plays this role. Also from Figure~\ref{bp123}(c) it is clear that starting from nearly same value, $\lambda_1$ can increase faster than $\lambda_3$ as energy increases. On the other hand, the runnings of other $\lambda$'s show a flat nature compared to $\lambda_1$ and $\lambda_3$.
As we do not allow hard $Z_2$-breaking, $\lambda_6$ and $\lambda_7$ do not change with energy and are fixed at zero. In explicit terms, the RG equations for $\lambda_6$ and $\lambda_7$, always carry the terms proportional to these two $\lambda'$s and therefore the relation $\frac{d\lambda}{d\mu} = 0$ remains valid throughout the running .\\

A complementary picture is noticed in Scenario 2. Here the requirement of low branching fraction of SM-like Higgs to a pair of pseudoscalars along with other constraints pushes $m_A > \frac{m_H}{2}$ in the WS region. On the other hand, in the RS case, it is possible to get a low BR($h_{SM}\rightarrow AA)$ in the entire range of $m_A$. To examine Scenario 2 on a case by case basis, we choose three benchmarks BP4, BP5, BP6. BP4 corresponds to RS region with $m_A > \frac{m_H}{2}$, BP5 corresponds to RS region with $m_A < \frac{m_H}{2}$. We consider WS region with $m_A > \frac{m_H}{2}$ in BP6. The benchmarks for Scenario 2 are listed in Table~\ref{bp_lt125_rs_lt62}. We mention here that although it is possible to get a few points in the WS region, with $m_A \lsim \frac{m_H}{2}$, in the resonant region with severe fine-tuning, we do not consider this region further in our analysis.

\begin{table}[!hptb]
\begin{center}
\begin{footnotesize}
\begin{tabular}{| c | c | c | c |}
\hline
& BP4 & BP5 & BP6 \\
\hline
$m_H$ in GeV & 117.409 & 93.6073 & 121.448\\
\hline
$m_A$ in GeV & 70.0 & 15.7859 & 63.0 \\
\hline
$m_{H^{\pm}}$ in GeV  & 142.529 & 135.00 & 139.871 \\
\hline
$\lambda_1$  & 0.07121 & 1.0251 & 0.082024 \\
\hline
$\lambda_2$  & 0.25774 & 0.25767 & 0.25774\\
\hline
$\lambda_3$ & 0.46960 & 0.58636 & 0.38712\\
\hline
$\lambda_4$ & -0.3372 & -0.45412 & -0.33662\\
\hline
$\lambda_5$ & 0.121841 & 0.138905 & 0.177861\\
\hline
$m^2_{12}$ in $GeV^2$  & 168.10299 & 393.28757 & 204.844987\\
\hline
$\tan \beta$ & 82.0 & 22.0 & 72.00\\
\hline
$\sin(\beta - \alpha)$ & -0.00141421 & 0.00601127 & -0.02828\\
\hline
$y_h^{\ell} \times \cos(\beta - \alpha)$ & 0.88403289597 & 1.13220955 & -1.036145\\
\hline
\end{tabular}
\end{footnotesize}
\caption{\it Benchmark points for Scenario 2.}
\label{bp_lt125_rs_lt62}
\end{center}
\end{table}

Our BP4 and BP6 have negative $\sin(\beta-\alpha)$ and large $\tan \beta$, where BP5 has positive $\sin(\beta - \alpha)$ and comparatively small $\tan \beta$. 
Here too, in the alignment limit, $\lambda_2$ is governed by the 125-GeV Higgs mass and therefore for all the benchmarks it gets similar values as scenario 1. In this case, $\lambda_3$ is comparatively smaller than the previous case due to smaller mass gap between $m_H$ and $m_h$, whereas $\lambda_1$ can get somewhat enhanced contribution from ${m^2_{12}}$ term at large $\tan \beta$.
On the other hand, $\lambda_4$ and $\lambda_5$ are opposite in sign, similar to the previous scenario.

Having thus identified our benchmark points, we further note that all the six aforesaid benchmark points fall in the yellowish interior region of Figure~\ref{muon_anomaly}. Therefore our analyses based on them are legitimate, both with the older data and on taking the very recent results into account, as far as constraints from $g_{\mu} - 2$ is concerned. For our subsequent analysis on UV-completion, we shall use the combined data of $g_{\mu}-2$ as the constraining factor on the parameter space.

\begin{figure}
\floatsetup[subfigure]{captionskip=10pt}
    \begin{subfigure}{.44\linewidth}
    \centering
    \includegraphics[width=7.6cm, height=5.5cm]{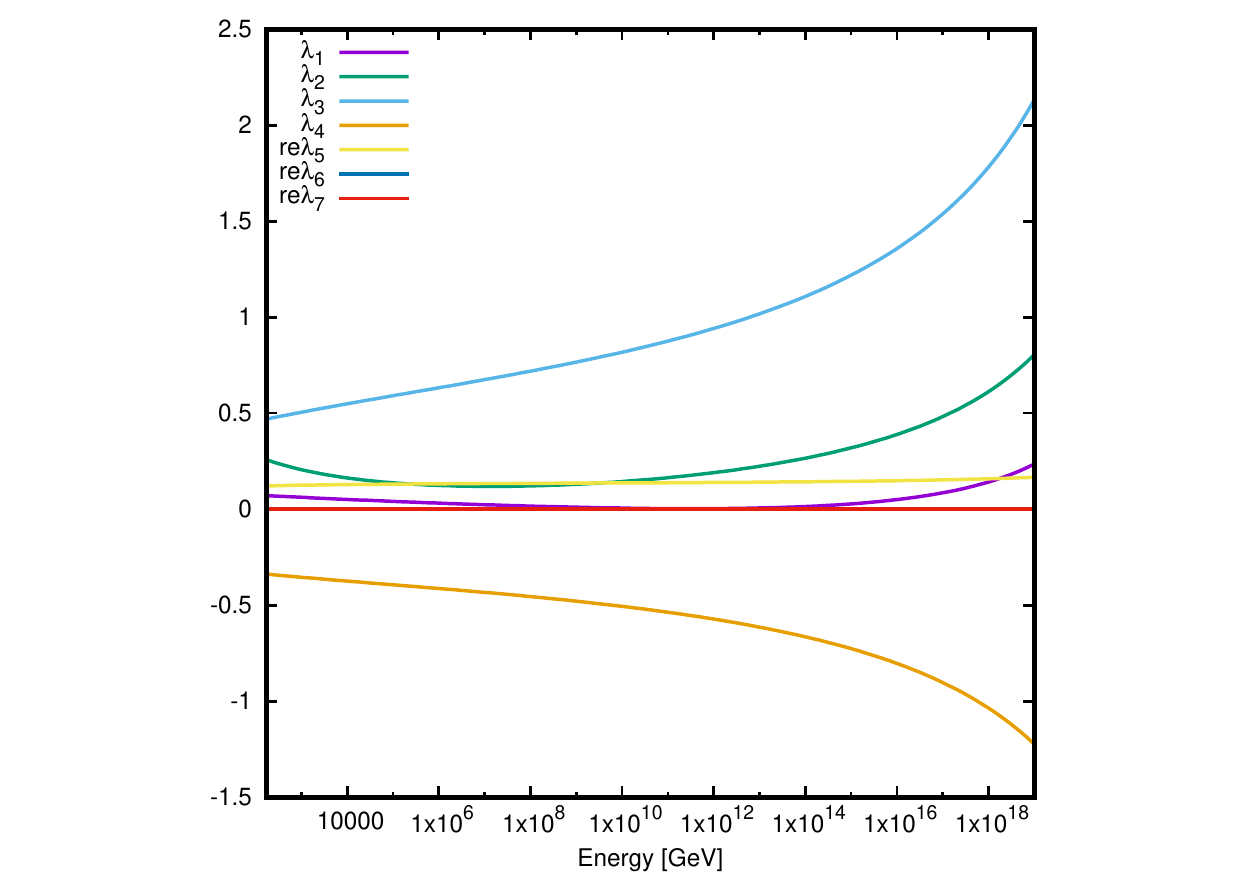}
    \caption{}\label{fig:image1}
    \end{subfigure} %
    \qquad
    \begin{subfigure}{.44\linewidth}
    \centering
    \includegraphics[width=7.6cm, height=5.5cm]{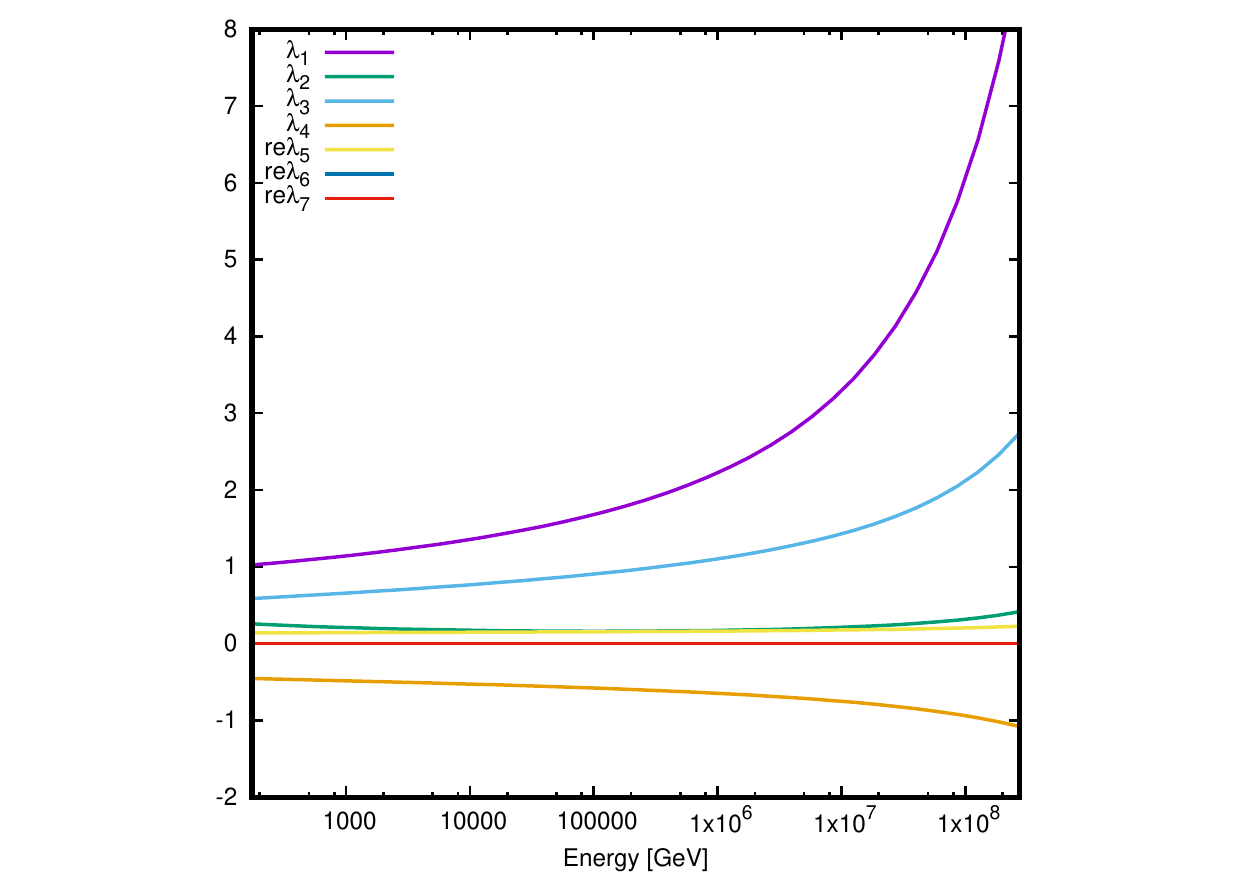}
    \caption{}\label{fig:image12}
   \end{subfigure}
\\[2ex]
  \begin{subfigure}{\linewidth}
  \centering
  \includegraphics[width=7.6cm, height=5.5cm]{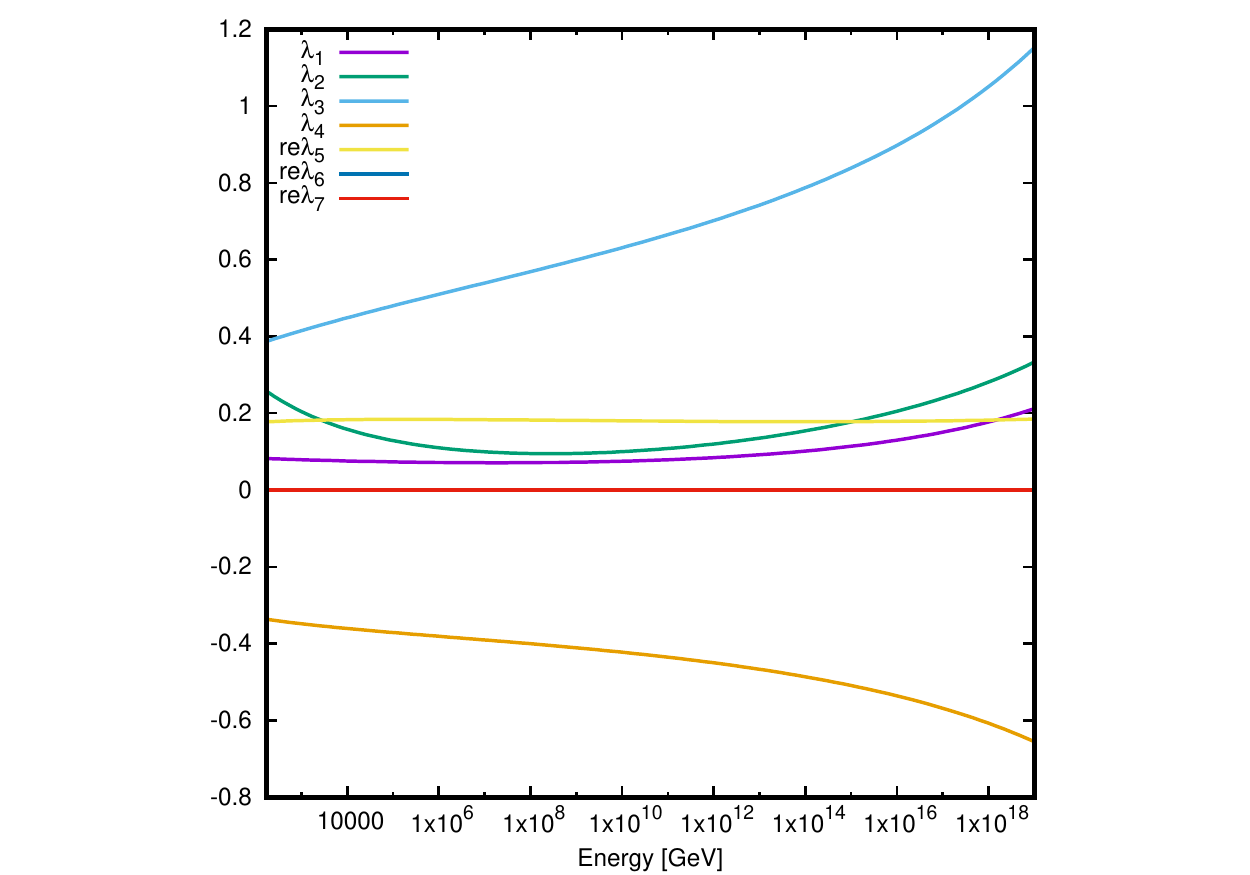}
  \caption{}\label{fig:image3}
  \end{subfigure} 
\RawCaption{\caption{\it Two-loop RG running of quartic couplings for the benchmarks (a) BP4, (b) BP5 and (c) BP6 from Scenario 2 .}
\label{bp456}}
\end{figure}

In Figure~\ref{bp456}, we display two-loop RG running of quatic couplings for BP4, BP5 and BP6. We can see that for all the three benchmarks $\Lambda^{cut-off}_{UV}$ is again decided by tree level unitarity. For BP4 and BP6, comparatively smaller values of $\lambda'$s at the electroweak scale ensure tree-level unitarity as well as perturbativity and stability upto  very high scale$(\sim 10^{19}$ GeV). We can see that in general it is easier to achieve UV-completion for scenario 2 than 1.\\

We have noticed in Figures~\ref{bp123} and \ref{bp456} that for all the $\lambda'$s, the negative contribution to the running of $\lambda'$s comes from the combination $(3g_1^2 + 9g_2^2)\lambda$ and terms involving Yukawa couplings. If to this we couple the information that $g_2$ falls at higher energies, while $g_1$ has at best marginal rise and the Yukawa couplings remain more or less constant, one finally has all quartic couplings rising with energy in this scenario. This feature which is generic to 2HDMs is due to the proliferation of bosonic degrees of freedom in the RG equations. Thus the stronger constraint almost invariably comes from perturbative unitarity.

The quartic coupling $\lambda_2$ shows a unique behavior. For some benchmarks (namely BP1 and BP2) it shows the usual monotonically increasing trend. But for the other BPs (BP3 to BP6) it decreases initially and then increases. The reason behind this behavior is the following: in case of BP1 and BP2 the magnitude of $\lambda_3, \lambda_4$ and $\lambda_5$ are much larger compared to the rest of the benchmarks. The terms proportional to $\lambda_3, \lambda_4$ and $\lambda_5$ control the positive contribution to the beta function for $\lambda_2$. Therefore depending on their values the cancellation between the positive and negative terms can sometimes, take place. However, here the dominant negative contribution comes from terms involving Yukawa couplings and their strengths drop at higher energies. Thus $\lambda_2$ starts to increase at high energies for all the benchmarks. One can also note that this behavior is correlated to the mass difference between the two neutral scalars $h$ and $H$ as the coupling $\lambda_3$ is proportional to this mass difference. It is evident from Table~\ref{bp_gt125_ws_lt62} that in case of BP1 and BP2, this mass difference is much larger. Consequently, the beta function for $\lambda_2$ takes a positive value in these cases at all energies.

On the other hand, in case of BP3, we see that $\lambda_1$ and $\lambda_3$ start from similar low energy values, but $\lambda_1$ tends to grow faster. As in this case $\lambda_1, \lambda_3 > \lambda_2, \lambda_4$, the bosonic contribution is larger in case of $\lambda_1$ compared to $\lambda_3$. The Yukawa contributions drop with energy whereas the bosonic contribution keeps growing. Therefore, beyond a certain energy ($\sim 10^{10}$ GeV), $\lambda_1$ becomes larger than $\lambda_3$.

Another interesting feature can be observed if we compare the running of $\lambda_4$ and $\lambda_5$ in case of BP1 and BP2. In these two cases, at the EW scale, $|\lambda_4| \approx |\lambda_5|$ as can be seen from Table~\ref{bp_gt125_ws_lt62}. It can be checked from Equation~\ref{lambda_rge} that in this limit, the beta functions for $\lambda_4$ and $\lambda_5$ become almost equal to each other in magnitude and opposite in sign. This behavior is clearly seen in Figure~\ref{bp123}.

\begin{figure}
\floatsetup[subfigure]{captionskip=10pt}
    \begin{subfigure}{.44\linewidth}
    \centering
    \includegraphics[width=7.6cm, height=5.5cm]{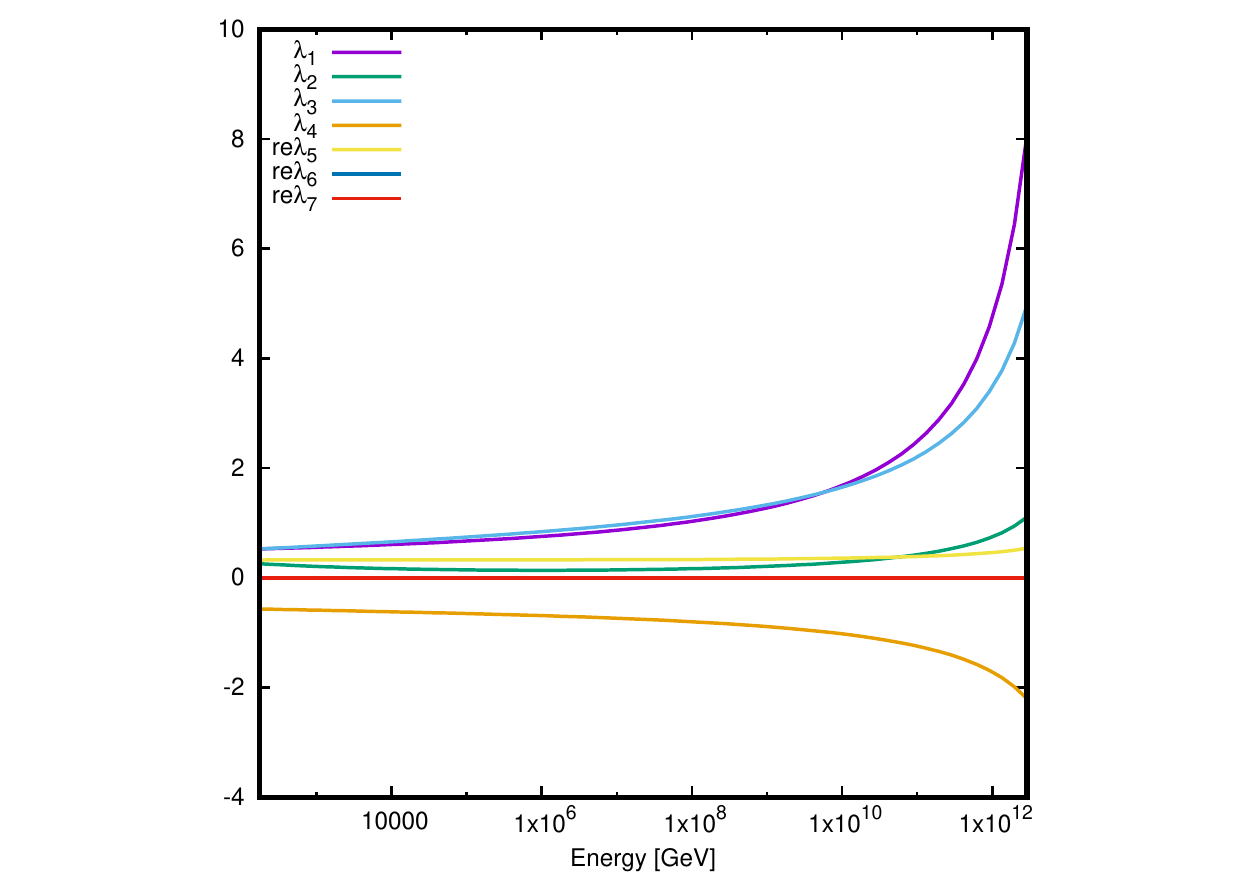}
    \caption{}\label{fig:image1}
    \end{subfigure} %
    \qquad
    \begin{subfigure}{.44\linewidth}
    \centering
    \includegraphics[width=7.6cm, height=5.5cm]{rg_bp3_rs_g125_g62_new_2.pdf}
    \caption{}\label{fig:image12}
   \end{subfigure}
\\[2ex]
\RawCaption{\caption{\it (a) One-loop and (b) two-loop RG running of quartic couplings for BP3.}
\label{RG_1loop_2loop}}
\end{figure}

A comparison has been made between one-loop and two-loop RG running of quartic couplings in Figure~\ref{RG_1loop_2loop}, for a representative benchmark (BP3). We have seen that in case of one-loop RG evolution unitarity breaks down faster than the two-loop case. However, the breaking scale is of the order of $10^{13}$ GeV in both cases. The values of quartic couplings, too, are very similar at high scales. Nonetheless, it is seen that the two-loop contribution helps us achieve somewhat higher UV cut-off scales.

\begin{figure}
\floatsetup[subfigure]{captionskip=10pt}
    \begin{subfigure}{.44\linewidth}
    \centering
    \includegraphics[width=5.6cm, height=5.5cm]{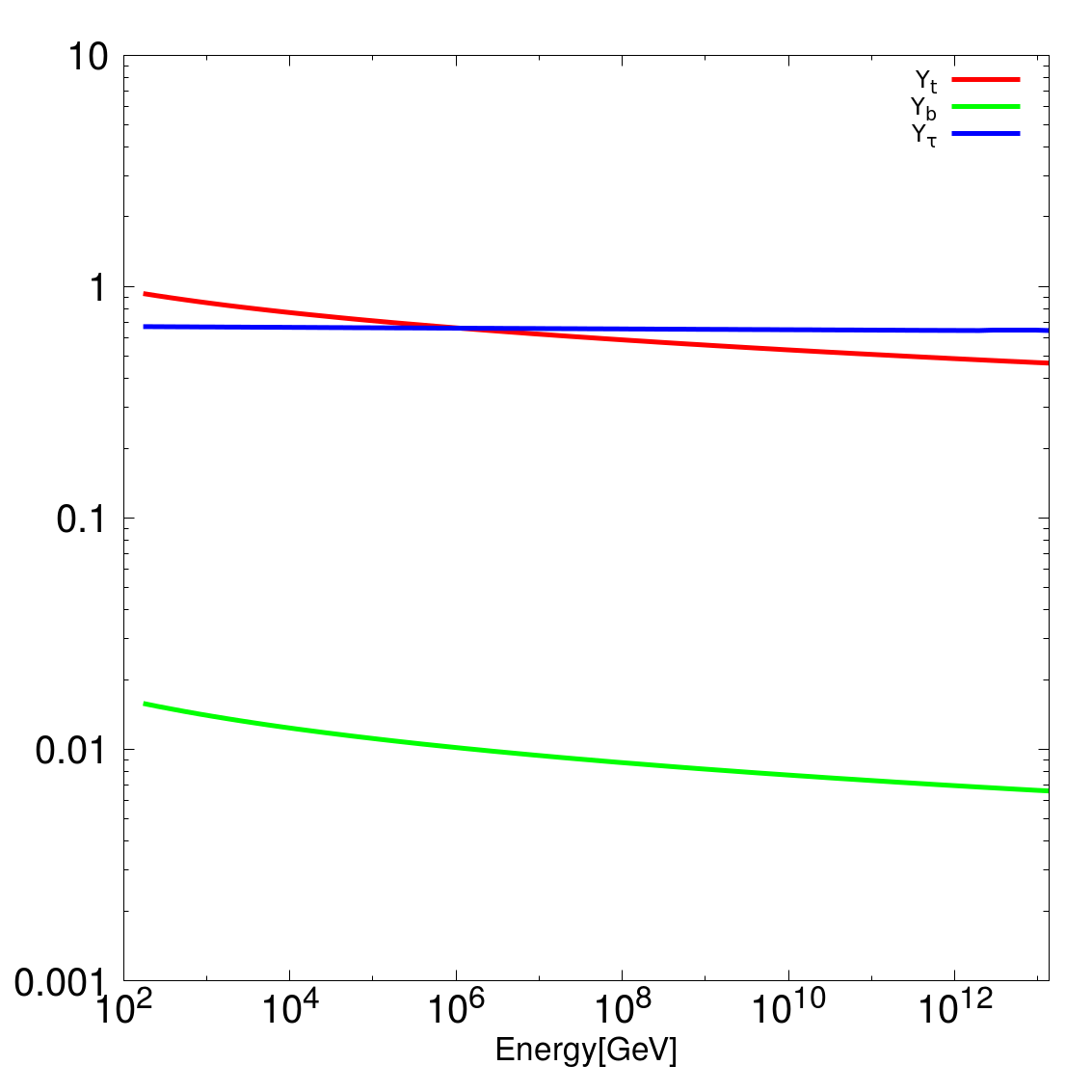}
    \caption{}\label{fig:image1}
    \end{subfigure} %
    \qquad
    \begin{subfigure}{.44\linewidth}
    \centering
    \includegraphics[width=5.6cm, height=5.5cm]{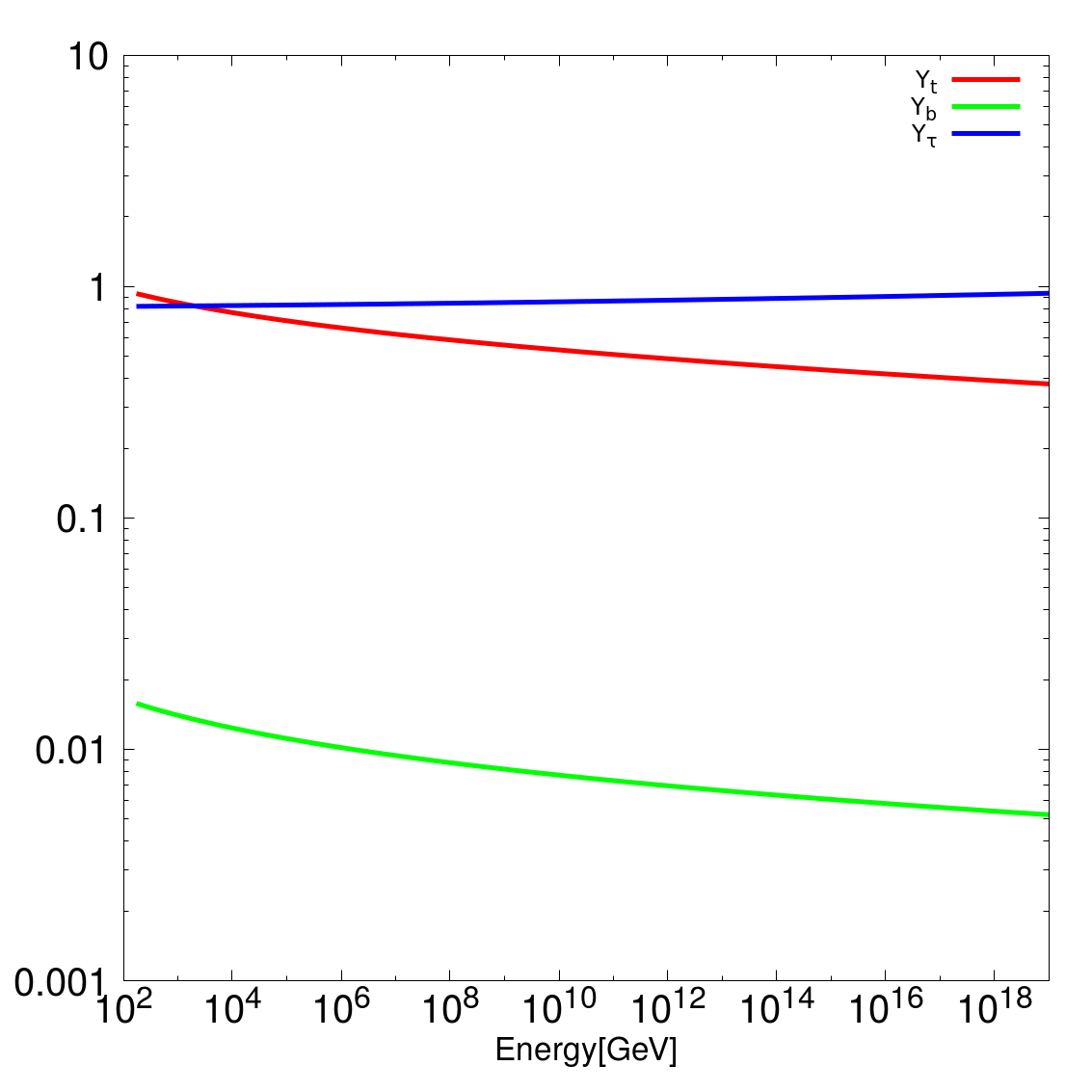}
    \caption{}\label{fig:image12}
   \end{subfigure}
\\[2ex]
    \begin{subfigure}{.44\linewidth}
    \centering
    \includegraphics[width=7.6cm, height=5.5cm]{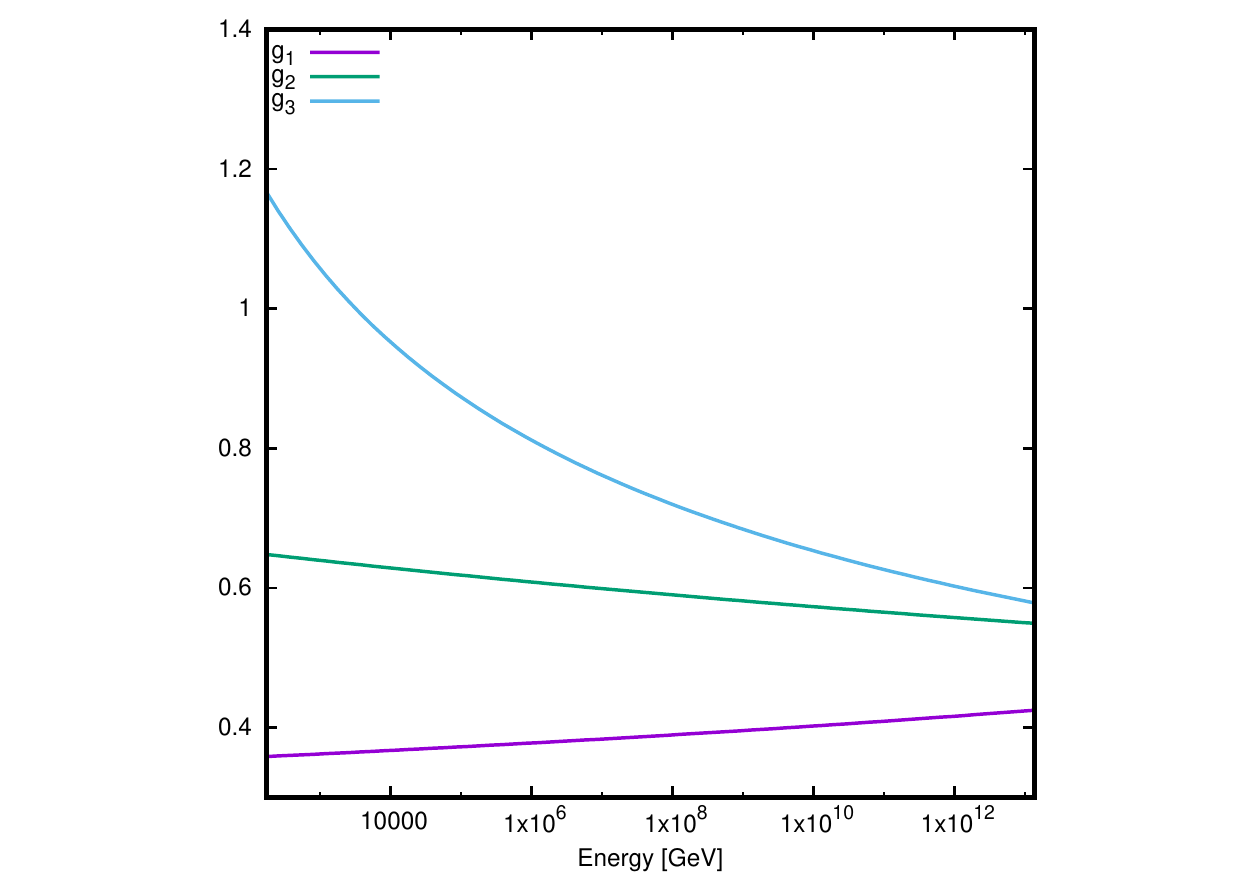}
    \caption{}\label{fig:image1}
    \end{subfigure} %
    \qquad
    \begin{subfigure}{.44\linewidth}
    \centering
    \includegraphics[width=7.6cm, height=5.5cm]{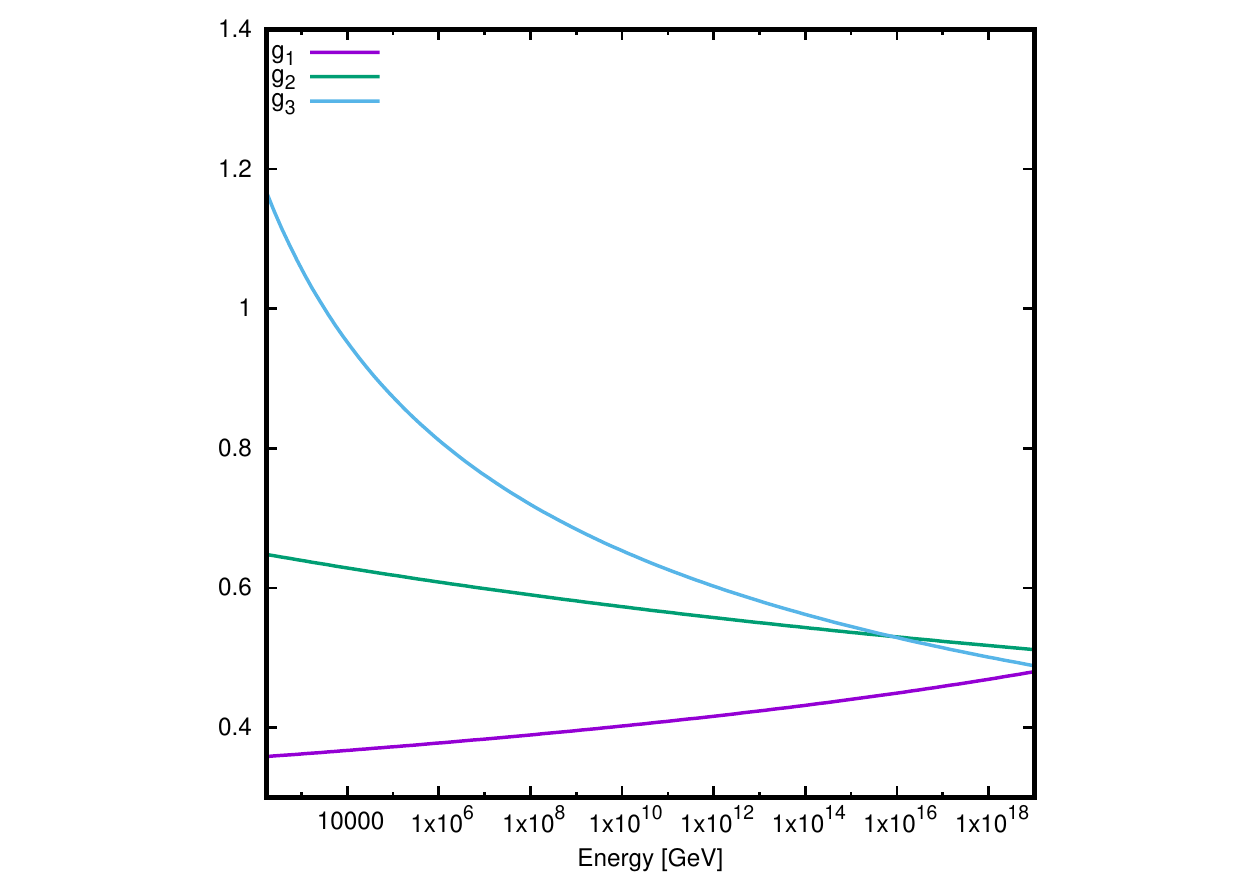}
    \caption{}\label{fig:image12}
   \end{subfigure}
\RawCaption{\caption{\it Two-loop RG running of third generation Yukawa couplings for (a) BP3 and (b) BP4 and gauge couplings for (c) BP3 and (d) BP4 respectively.}
\label{RG_gauge_yukawa_2loop}}
\end{figure}


The running of the gauge and Yukawa couplings are shown in case of BP3 and BP4 in Figure~\ref{RG_gauge_yukawa_2loop}. The qualitative nature of the running will be same for all the benchmarks. The variation in the top- and bottom-Yukawa couplings are significant, as can be seen through the logarithmic plots, since they are affected by 
 strong  interaction, unlike what happens to the $\tau$-Yukawa. The
 $\tau$-Yukawa interaction overrides even the top-Yukawa coupling at high scales, by virtue of the fact that we are considering benchmark points with large $\tan\beta$.

Some remarks are in order on the evolution of the gauge couplings,
especially in the context of possible embedding of the Type-X 2HDM in
a Grand Unified Theory (GUT).
As far as the gauge interactions are concerned, the evolution patterns are largely similar to the SM trajectories, if one remembers that  GUT normalization has not been used for the $U(1)$ gauge coupling. It should also be noted that
one loses perturbative unitarity of quartic couplings at around $10^{13}$ GeV for BP3, and even the 
two-loop RGEs cease to be trustworthy beyond that. So long as perturbativity is held to be sacrosanct, one therefore needs the intervention of new physics within approximately $10^{13}$ GeV in this case, and that intervening physics should have a role in ensuring grand unification, if at all. For BP4, on the other hand, no such requirement arises since the interactions are perturbative all the way to the Planck scale. However, the merger of the three kinds of gauge interaction at the GUT scale still requires some additional threshold effects, as much as they do in the standard model, a requirement eminently fulfilled, for example, by supersymmetry broken at the TeV-scale.

\section{Allowed regions with various cut-off scales}
\label{sec6}

After discussing the RG-evolutions of all the relevant couplings in the model, we proceed to scan
the model parameter space and look for points which satisfy all the theoretical constraints, namely perturbativity, unitarity and vacuum stability upto cut-off scale
$\Lambda^{cut-off}_{UV}$($\sim 10^4, 10^8, 10^{16}, 10^{19}$ GeV). To maintain consistency in the discussion we divide our analysis in four previously considered scenarios namely, 

\medskip
\noindent
1) Case 1: Scenario 1 with WS Yukawa,\\
2) Case 2: Scenario 1 with RS Yukawa,\\
3) Case 3: Scenario 2 with WS Yukawa,\\
4) Case 4: Scenario 2 with RS Yukawa.\\

\begin{figure}[!hptb]
\floatsetup[subfigure]{captionskip=10pt}
    \begin{subfigure}{.44\linewidth}
    \centering
    \includegraphics[width=7.0cm, height=5.5cm]{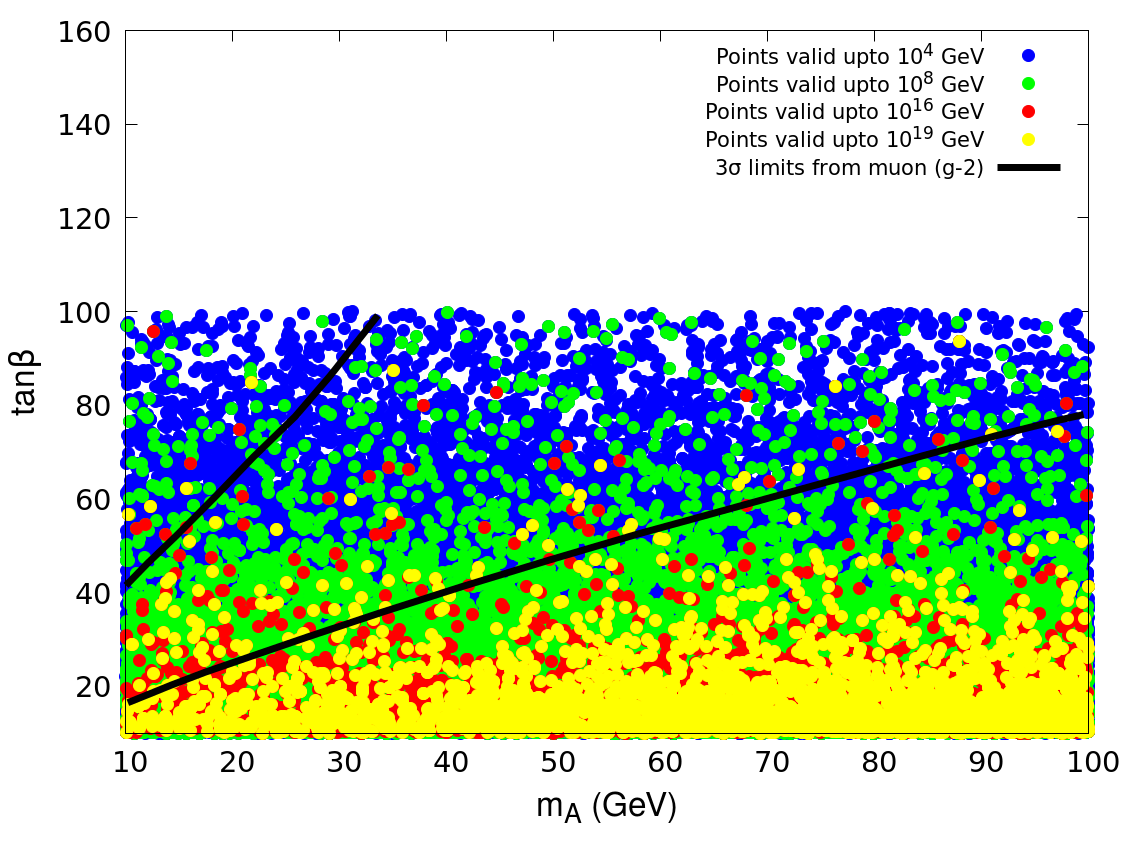}
    \caption{}\label{fig:image1}
    \end{subfigure} %
    \qquad
    \begin{subfigure}{.44\linewidth}
    \centering
    \includegraphics[width=7.0cm, height=5.5cm]{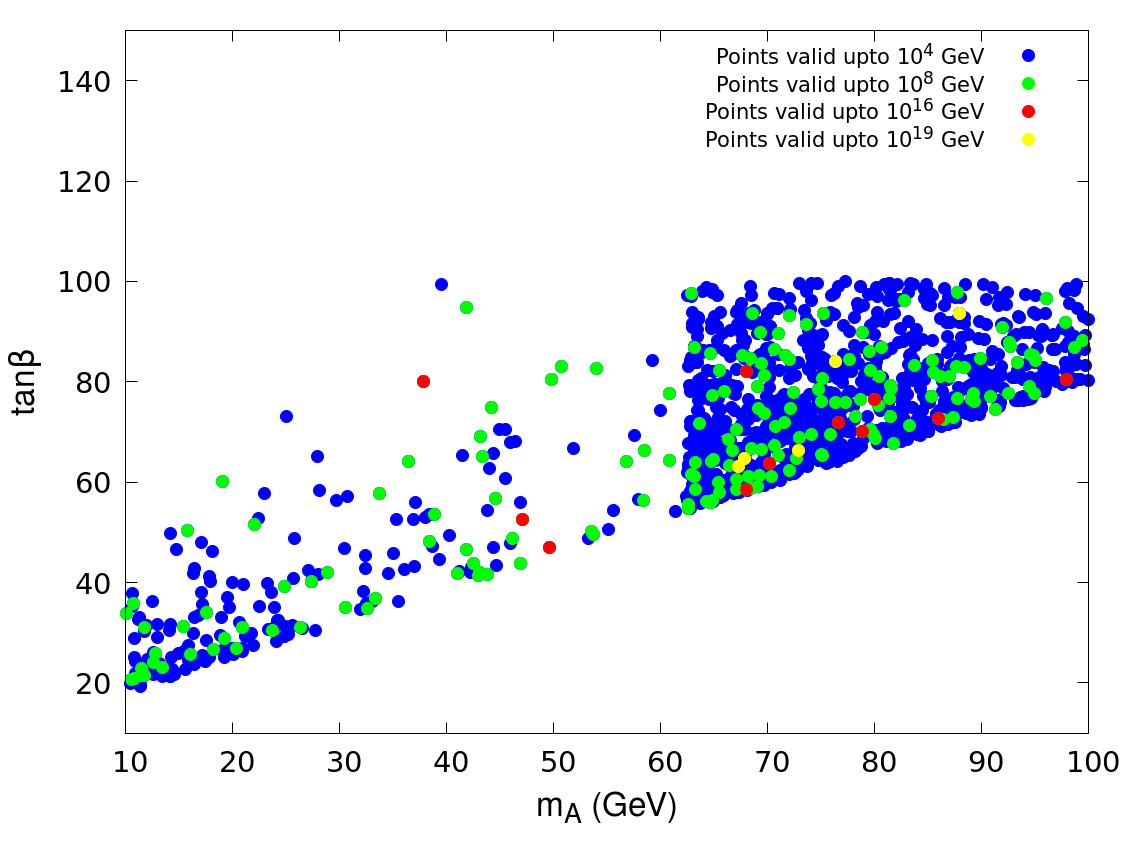}
    \caption{}\label{fig:image12}
   \end{subfigure}
\\[2ex]
\RawCaption{\caption{\it $m_A-\tan \beta$ plane, valid upto different energy scales after applying (a) theoretical constraints (b) theoretical constraints + $(g_{\mu}-2)$ at 3$\sigma$ + $BR(h_{SM} \rightarrow A A)$ bounds for Case 1.}
\label{hsvalidity_mA_tb_ws_gt125}}
\end{figure}

\begin{figure}[!hptb]
\floatsetup[subfigure]{captionskip=10pt}
    \begin{subfigure}{.44\linewidth}
    \centering
    \includegraphics[width=7.0cm, height=5.5cm]{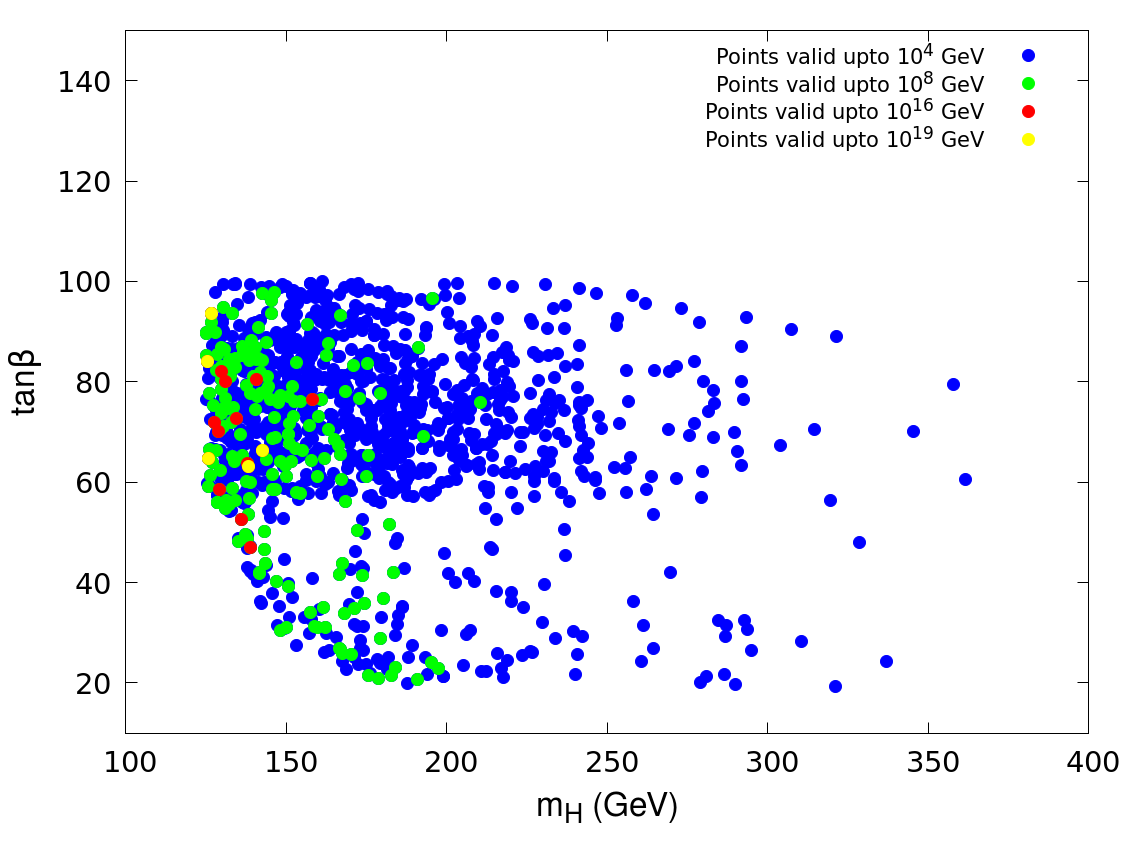}
    \caption{}\label{fig:image1}
    \end{subfigure} %
    \qquad
    \begin{subfigure}{.44\linewidth}
    \centering
    \includegraphics[width=7.0cm, height=5.5cm]{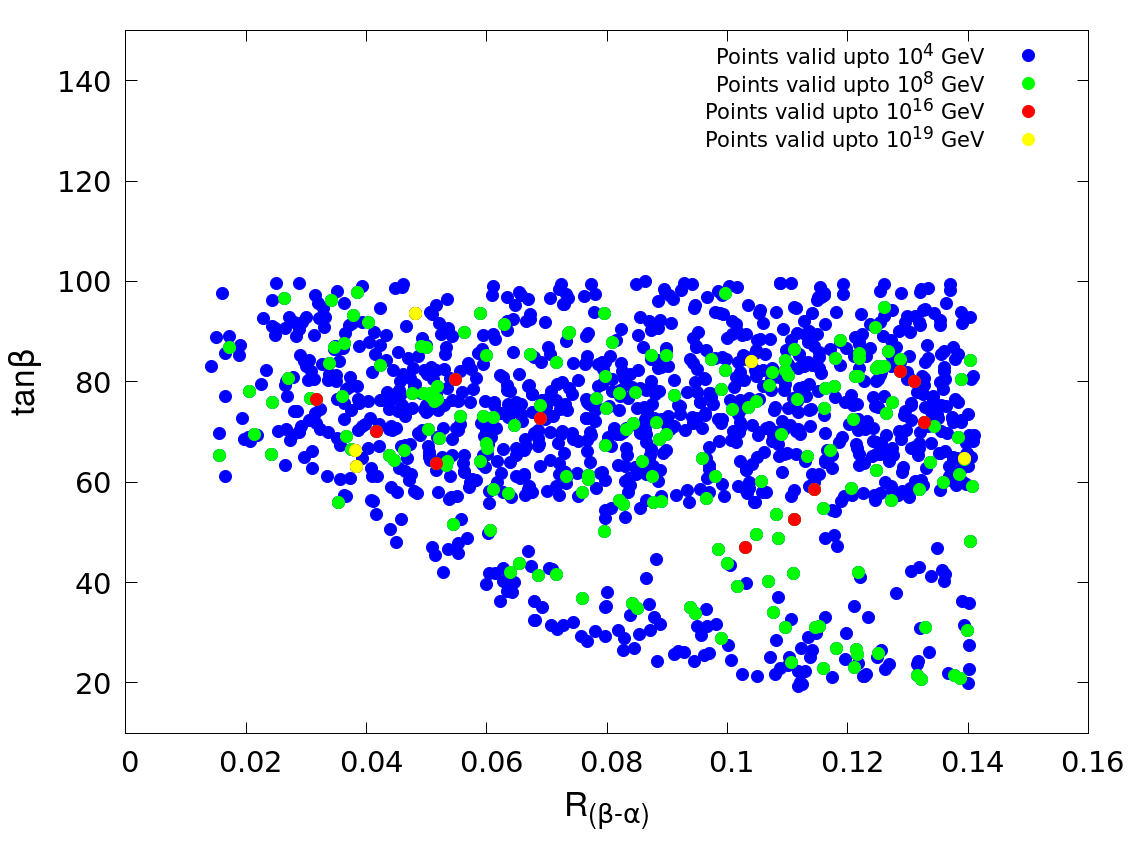}
    \caption{}\label{fig:image12}
   \end{subfigure}
\\[2ex]
  \begin{subfigure}{\linewidth}
  \centering
  \includegraphics[width=7.0cm, height=5.5cm]{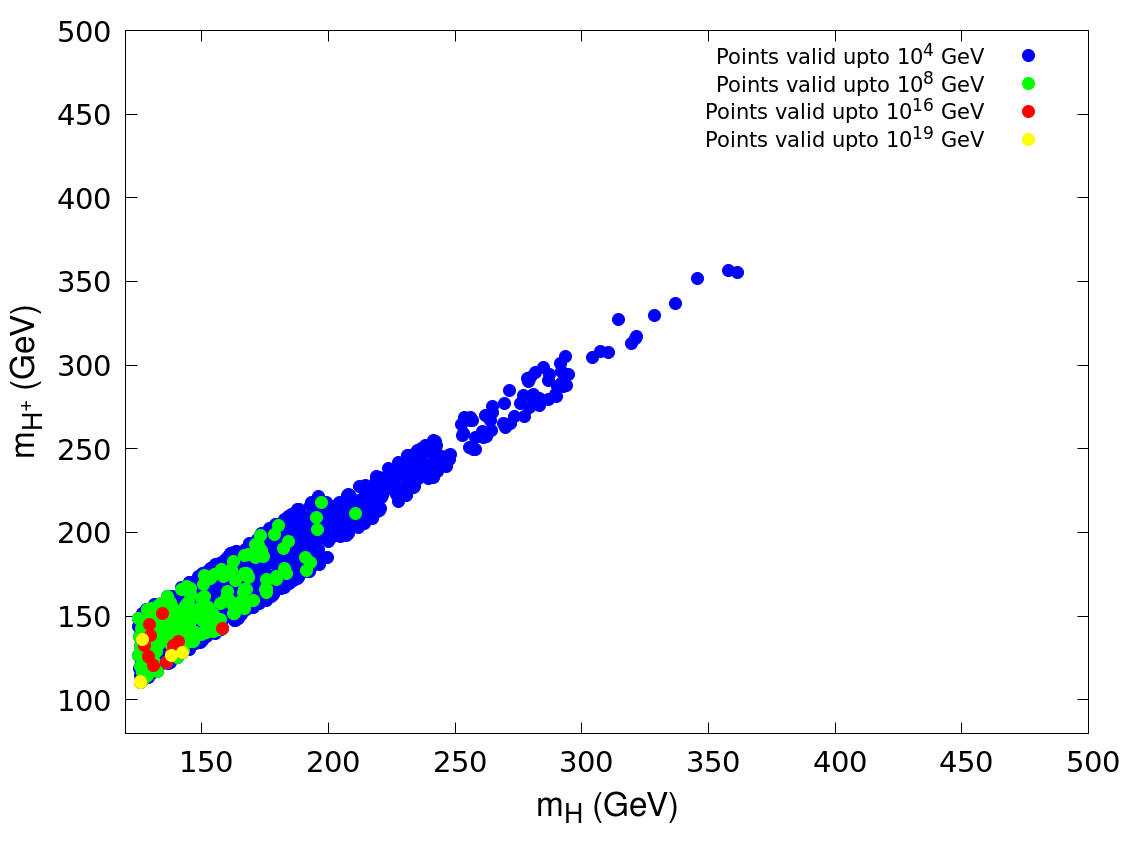}
  \caption{}\label{fig:image3}
  \end{subfigure} 
\RawCaption{\caption{\it (a) $m_H-\tan \beta$, (b) $R_{(\beta-\alpha)}-\tan \beta$ and (c) $m_H - m_{H^{\pm}}$ plane, valid upto different energy scales after applying theoretical constraints + $(g_{\mu}-2)$ at 3$\sigma$ + $BR(h_{SM} \rightarrow A A)$ bounds for Case 1.}
\label{hsvalidity_ws_gt125_others}}
\end{figure}

\begin{figure}[!hptb]
\floatsetup[subfigure]{captionskip=10pt}
    \begin{subfigure}{.44\linewidth}
    \centering
    \includegraphics[width=7.0cm, height=5.5cm]{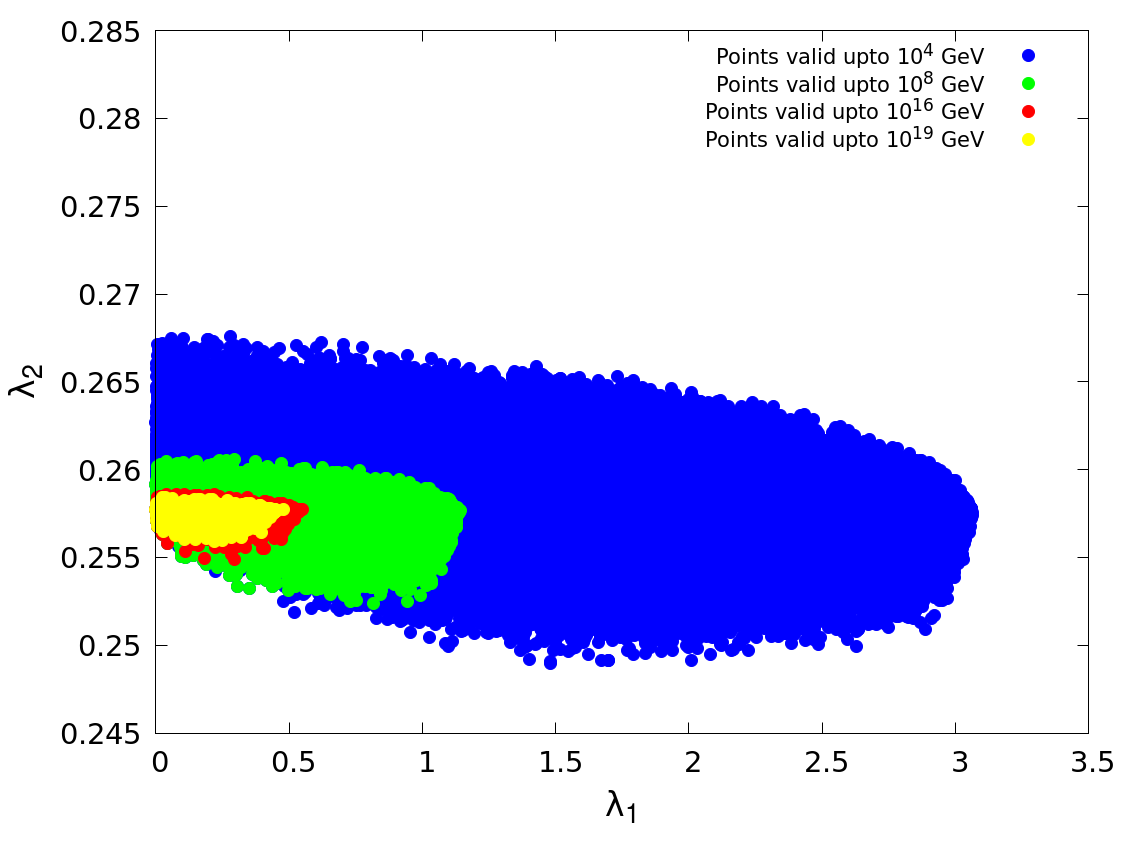}
    \caption{}\label{fig:image1}
    \end{subfigure} %
    \qquad
    \begin{subfigure}{.44\linewidth}
    \centering
    \includegraphics[width=7.0cm, height=5.5cm]{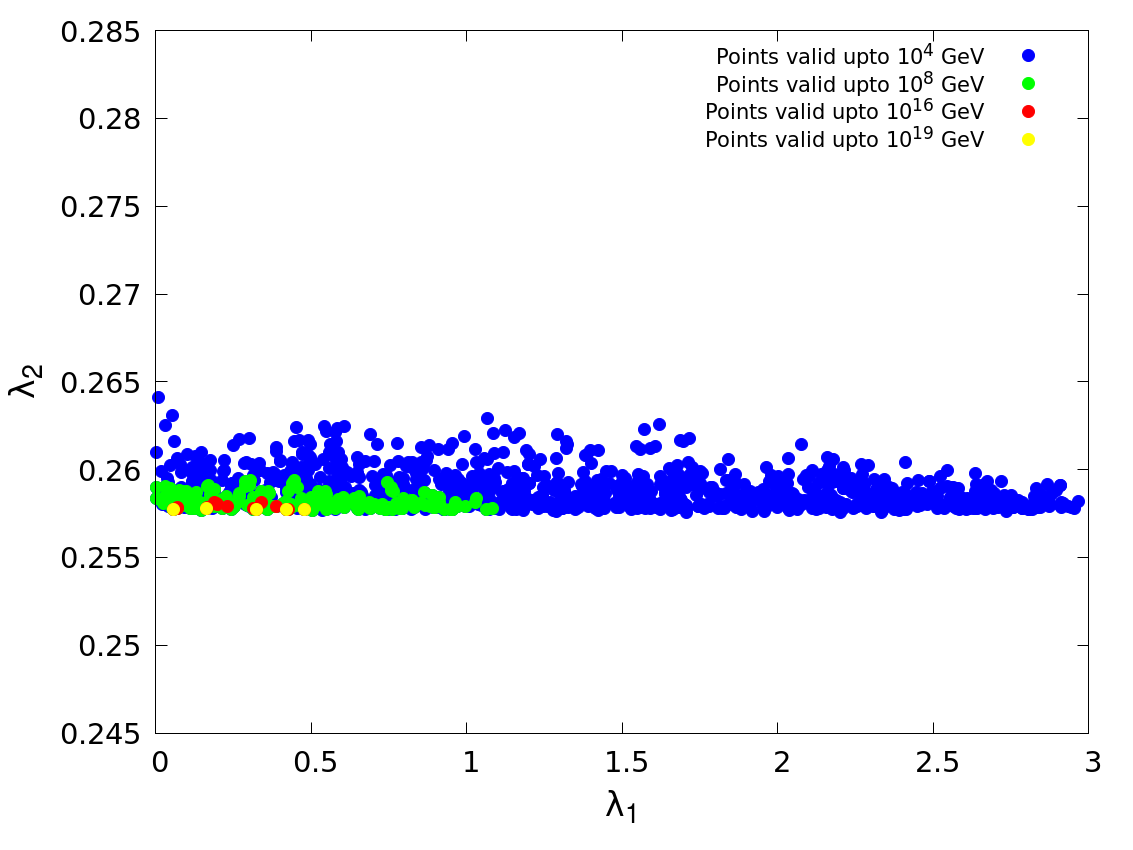}
    \caption{}\label{fig:image12}
   \end{subfigure}
\\[2ex]
    \begin{subfigure}{.44\linewidth}
    \centering
    \includegraphics[width=7.0cm, height=5.5cm]{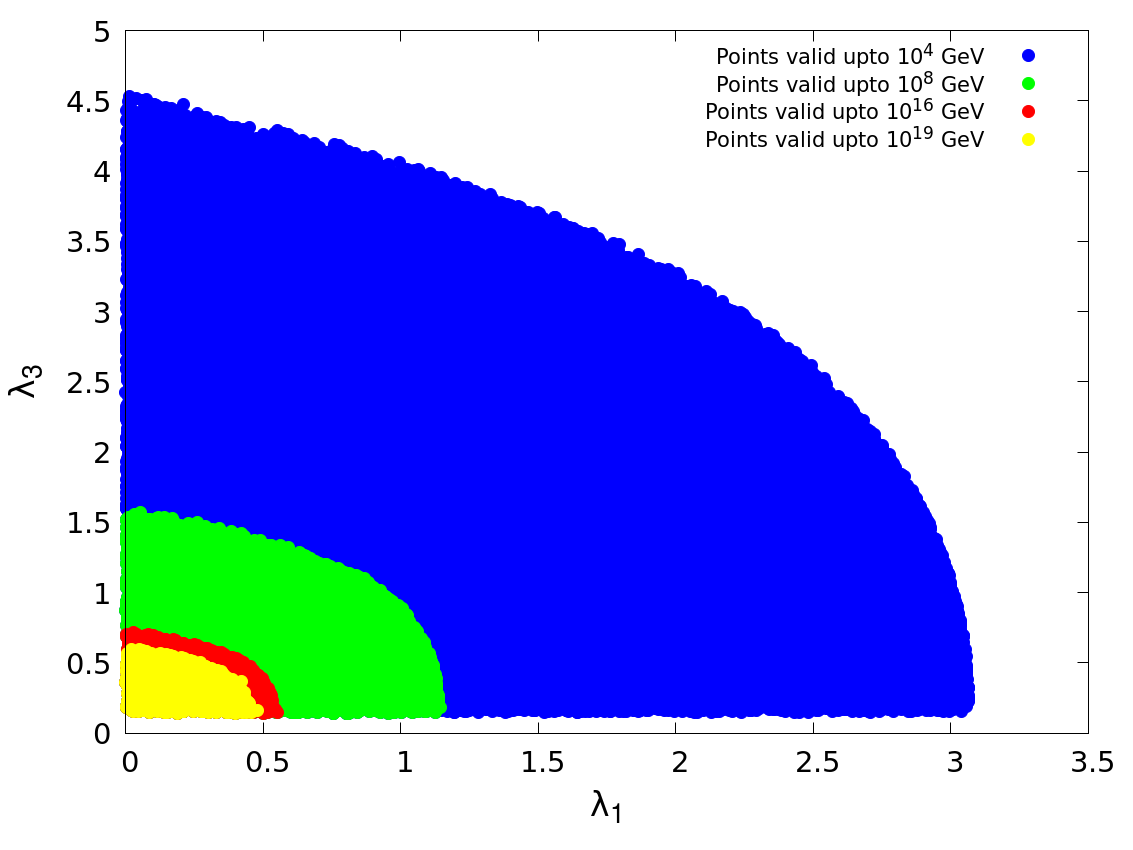}
    \caption{}\label{fig:image1}
    \end{subfigure} %
    \qquad
    \begin{subfigure}{.44\linewidth}
    \centering
    \includegraphics[width=7.0cm, height=5.5cm]{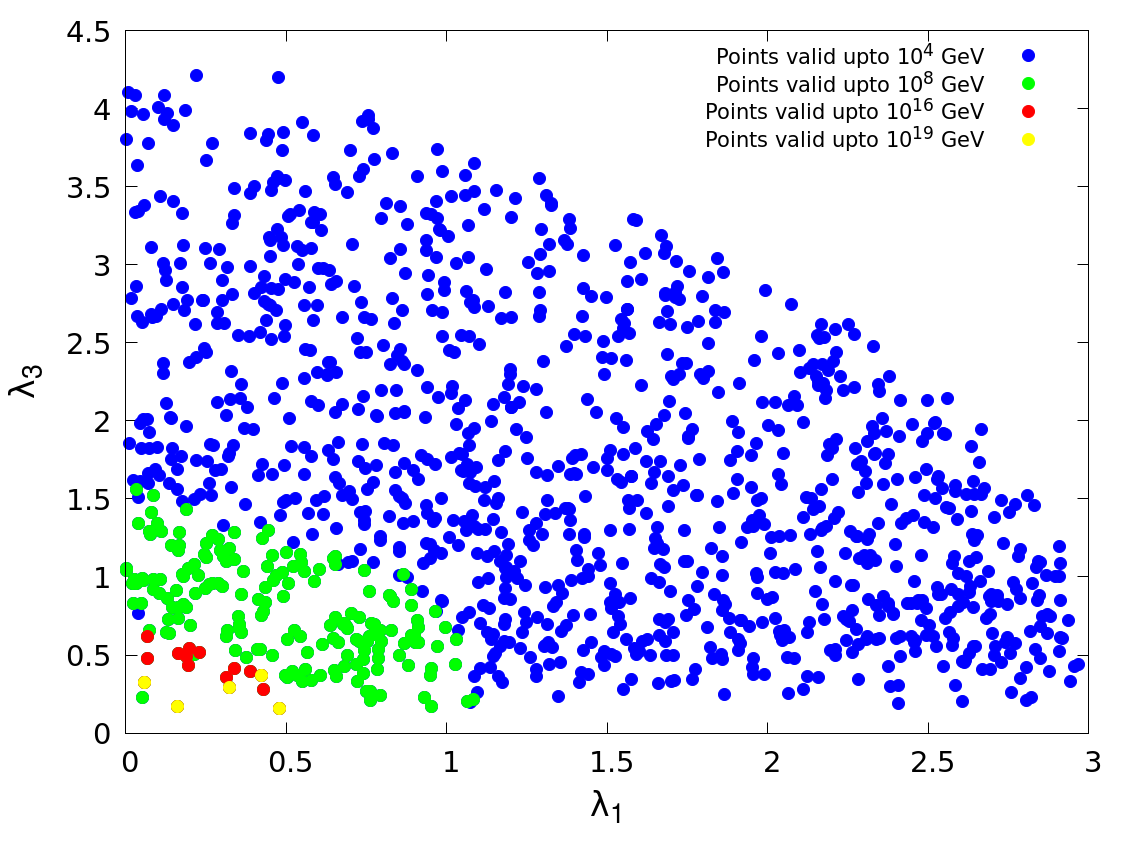}
    \caption{}\label{fig:image12}
   \end{subfigure}
\\[2ex]
    \begin{subfigure}{.44\linewidth}
    \centering
    \includegraphics[width=7.0cm, height=5.5cm]{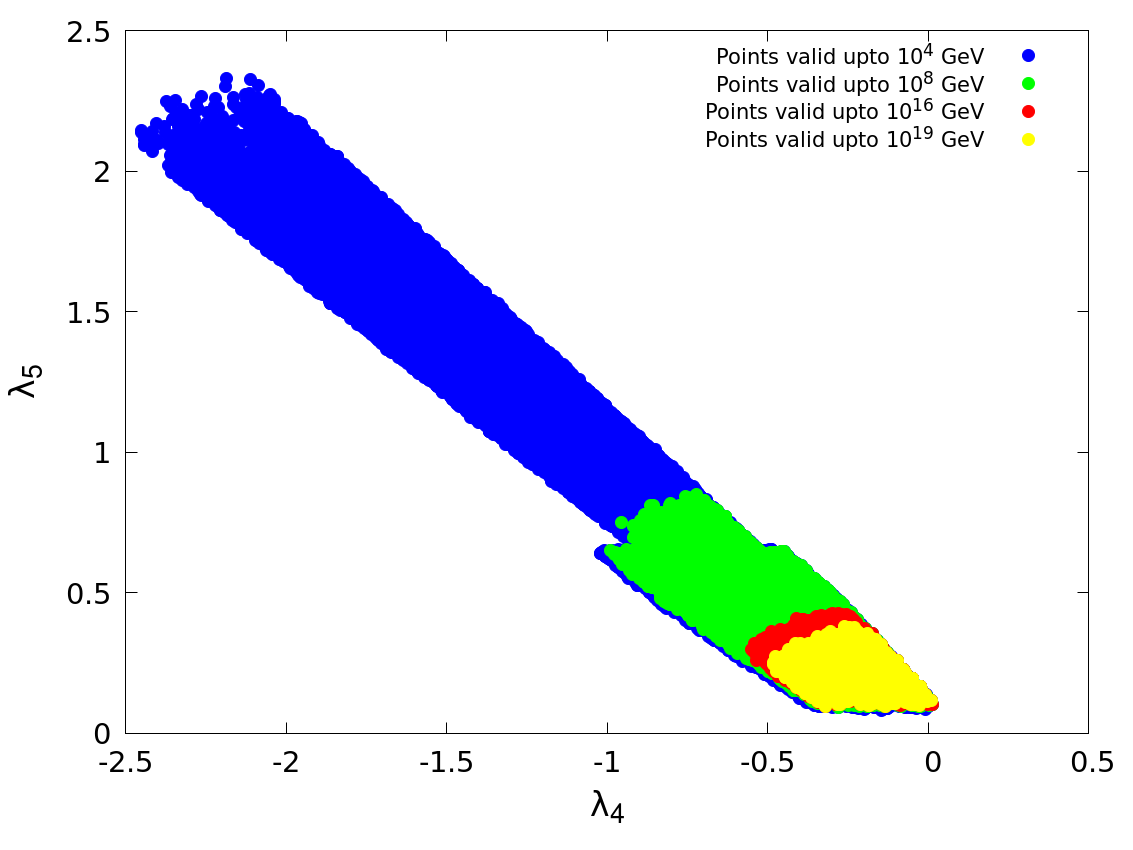}
    \caption{}\label{fig:image1}
    \end{subfigure} %
    \qquad
    \begin{subfigure}{.44\linewidth}
    \centering
    \includegraphics[width=7.0cm, height=5.5cm]{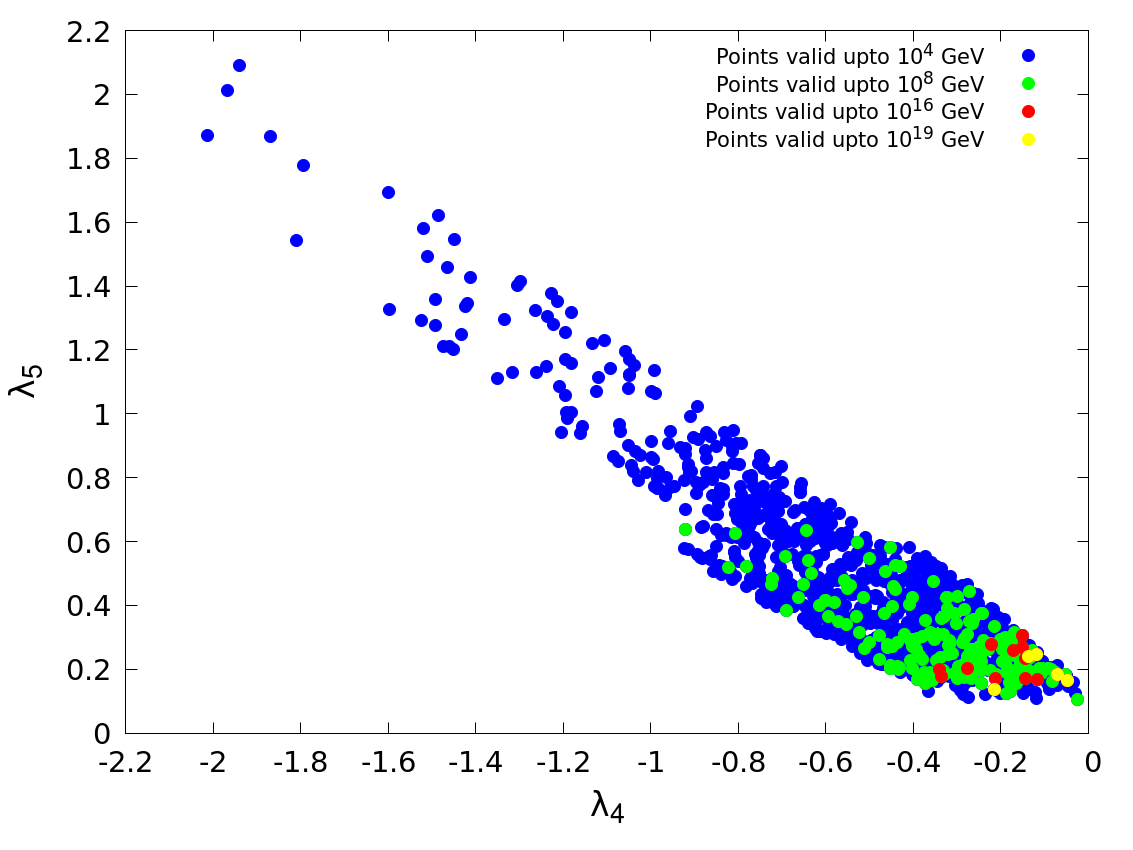}
    \caption{}\label{fig:image12}
   \end{subfigure}
\\[2ex]
\RawCaption{\caption{\it Quartic couplings valid upto different energy scales after applying (a),(c) and (e) theoretical constraints and (b),(d) and (f) theoretical constraints+$(g_{\mu}-2)$ at 3$\sigma$ + $BR(h_{SM} \rightarrow A A)$ bounds for Case 1.}
\label{hsvalidity_l1_l2_ws_gt125}}
\end{figure}

\begin{figure}[!hptb]
\floatsetup[subfigure]{captionskip=10pt}
    \begin{subfigure}{.44\linewidth}
    \centering
    \includegraphics[width=7.0cm, height=5.1cm]{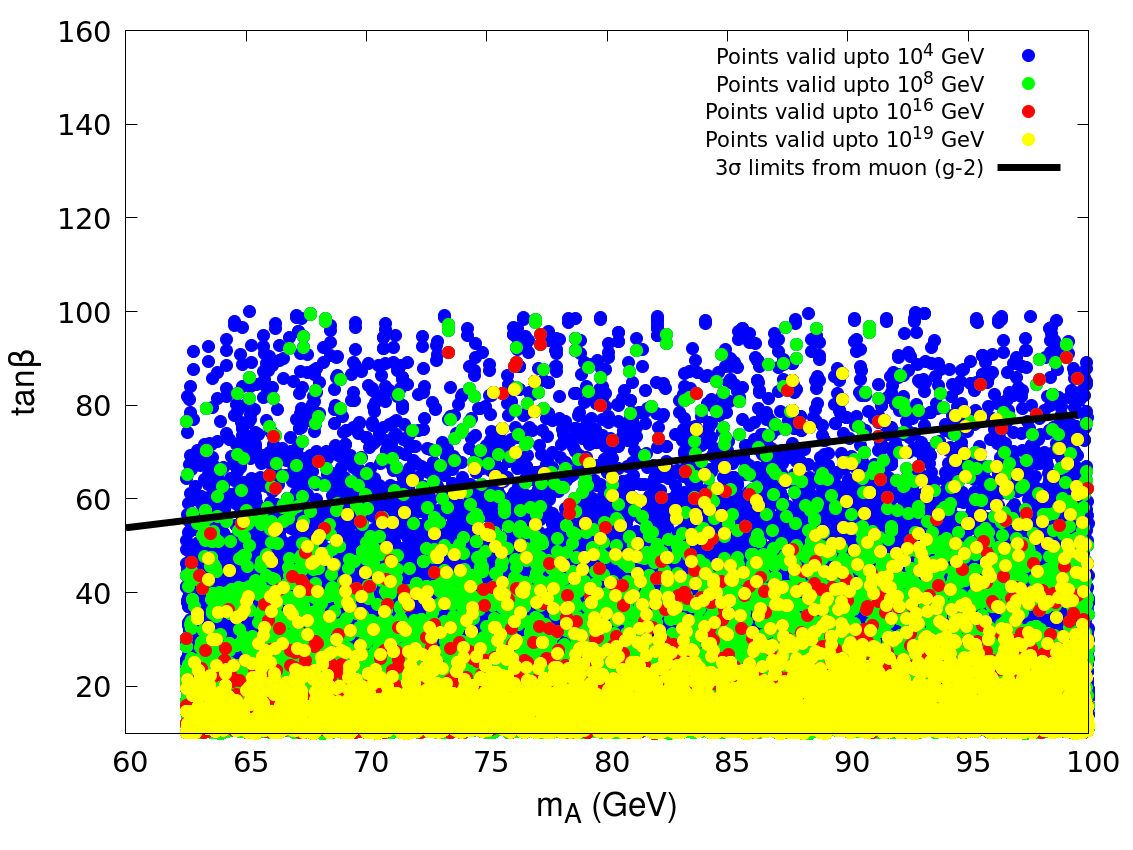}
    \caption{}\label{fig:image1}
    \end{subfigure} %
    \qquad
    \begin{subfigure}{.44\linewidth}
    \centering
    \includegraphics[width=7.0cm, height=5.1cm]{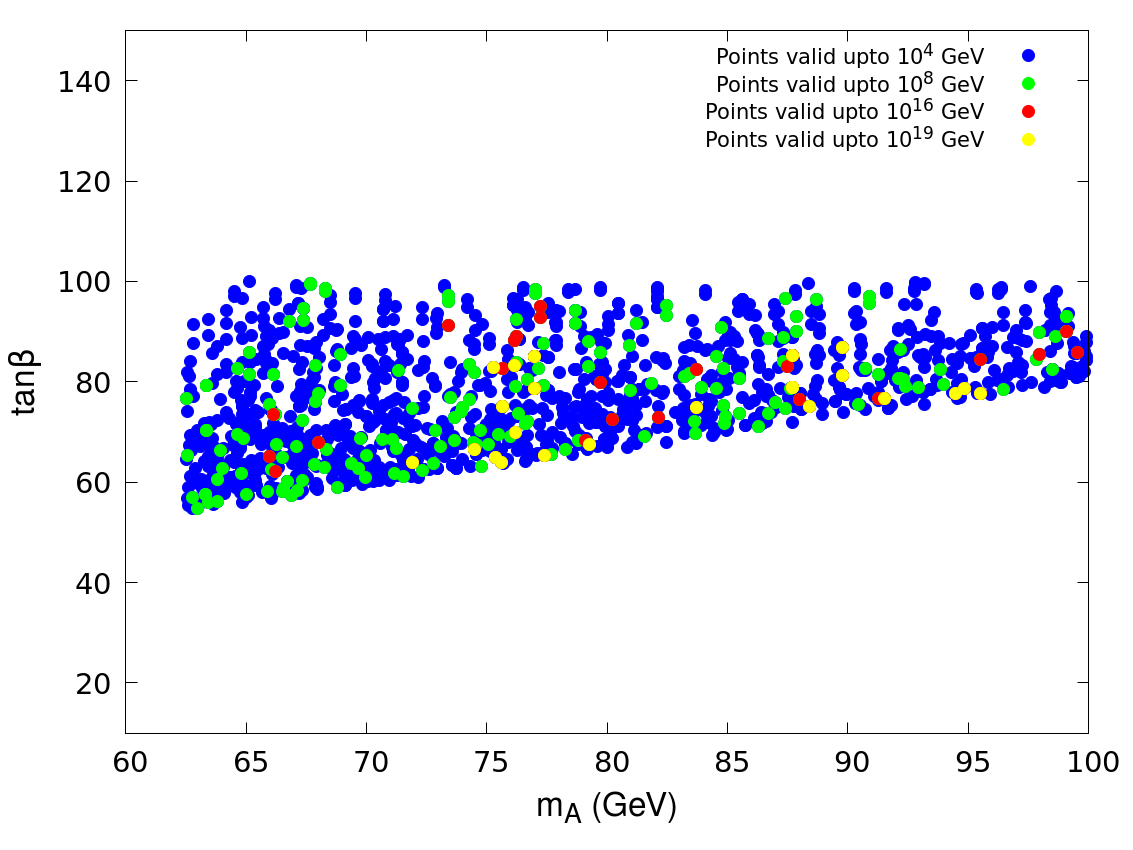}
    \caption{}\label{fig:image12}
   \end{subfigure}
\\[2ex]
\RawCaption{\caption{\it $m_A-\tan \beta$ plane, valid upto different energy scales after applying (a) theoretical constraints (b) theoretical constraints + $(g_{\mu}-2)$ at 3$\sigma$ + $BR(h_{SM} \rightarrow A A)$ bounds for Case 2.}
\label{hsvalidity_mA_tb_rs_gt125}}
\end{figure}

\begin{figure}[!hptb]
\floatsetup[subfigure]{captionskip=10pt}
    \begin{subfigure}{.44\linewidth}
    \centering
    \includegraphics[width=7.0cm, height=5.1cm]{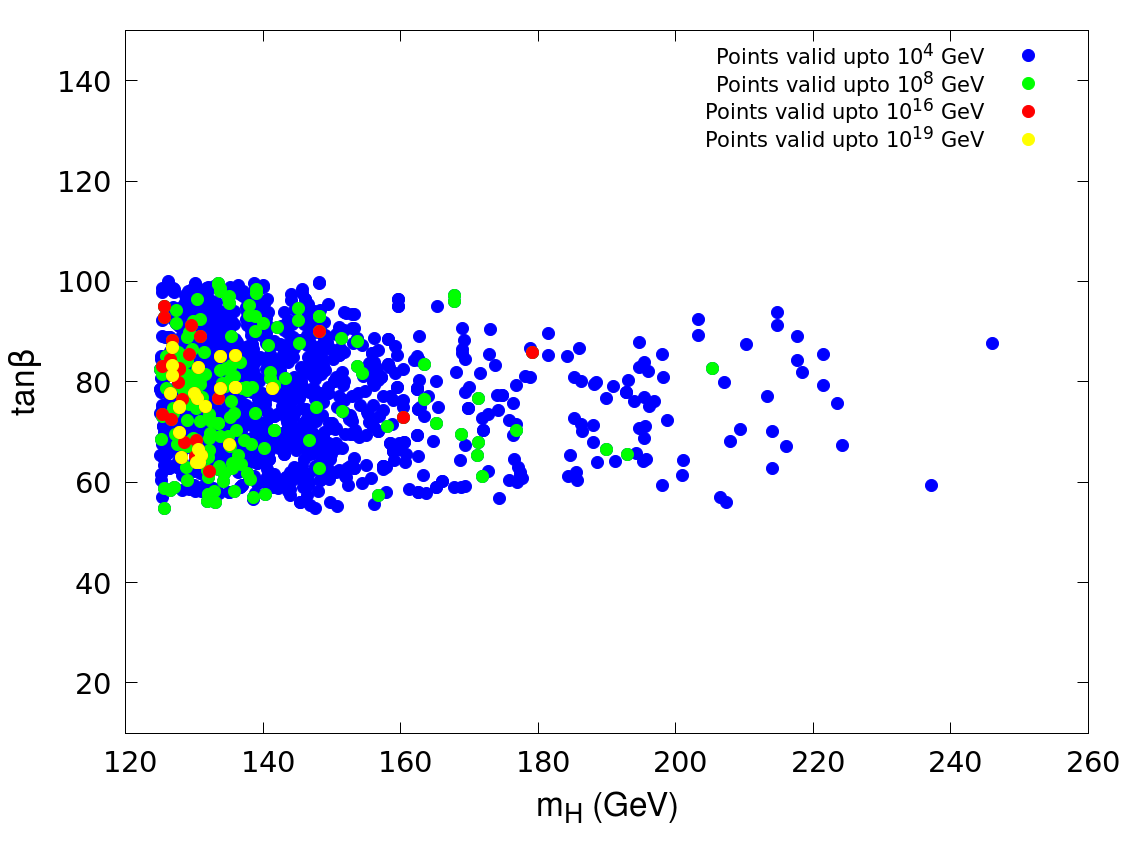}
    \caption{}\label{fig:image1}
    \end{subfigure} %
    \qquad
    \begin{subfigure}{.44\linewidth}
    \centering
    \includegraphics[width=7.0cm, height=5.1cm]{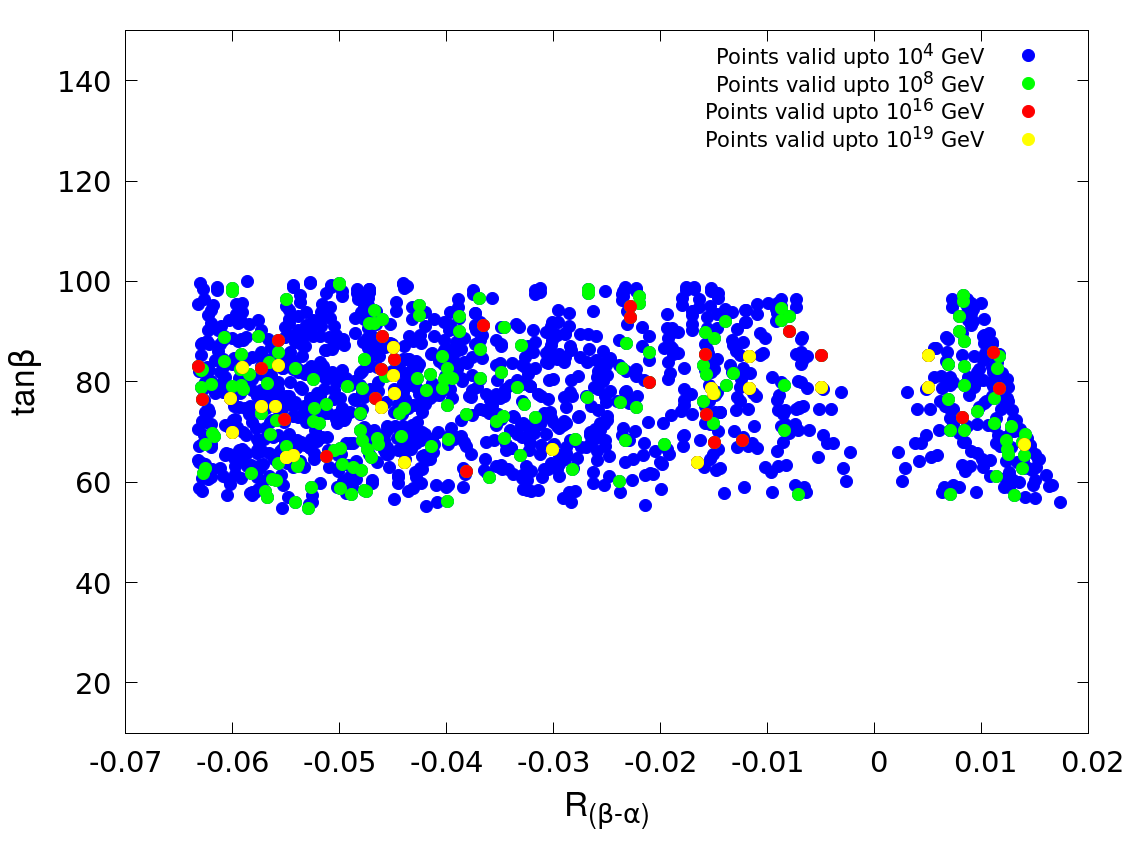}
    \caption{}\label{fig:image12}
   \end{subfigure}
\\[2ex]
  \begin{subfigure}{\linewidth}
  \centering
  \includegraphics[width=7.0cm, height=5.1cm]{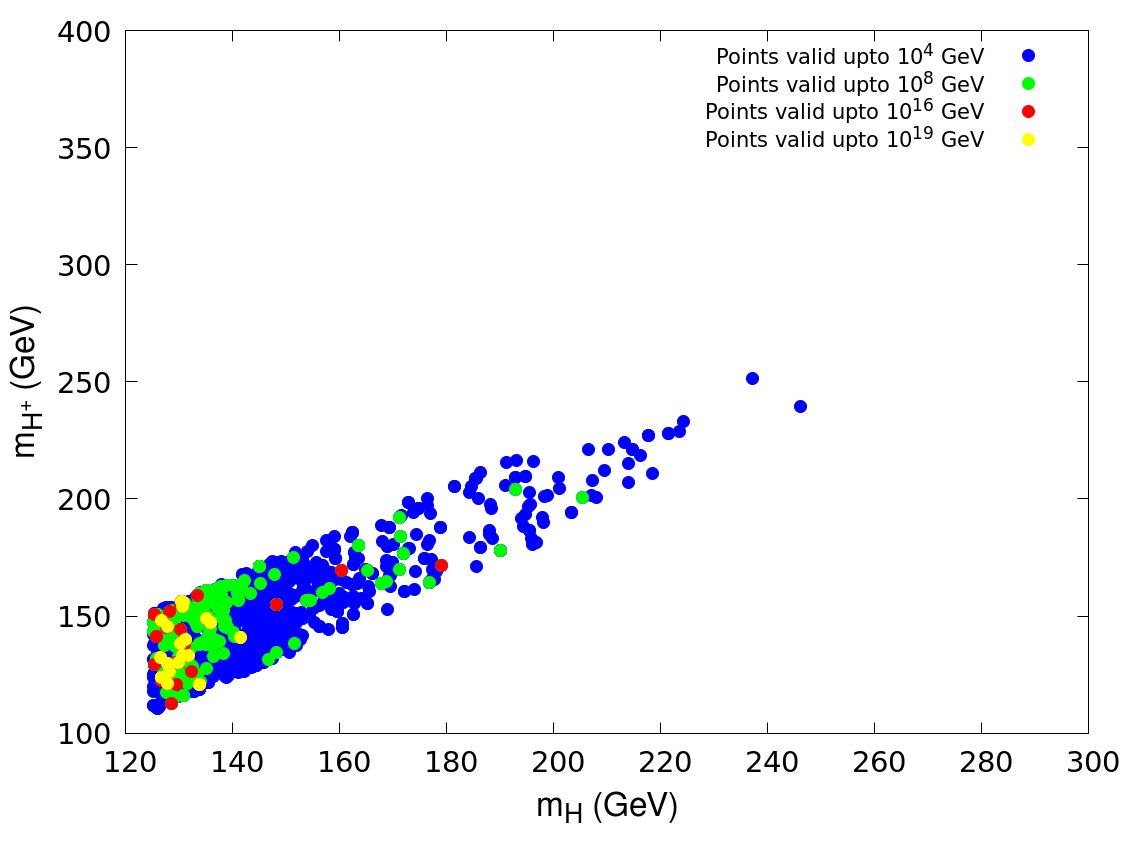}
  \caption{}\label{fig:image3}
  \end{subfigure} 
\RawCaption{\caption{\it (a) $m_H-\tan \beta$, (b) $R_{(\beta-\alpha)}-\tan \beta$ and (c) $m_H - m_{H^{\pm}}$ plane, valid upto different energy scales after applying theoretical constraints + $(g_{\mu}-2)$ at 3$\sigma$ + $BR(h_{SM} \rightarrow A A)$ bounds for Case 2.}
\label{hsvalidity_rs_gt125_others}}
\end{figure}

\begin{figure}[!hptb]
\floatsetup[subfigure]{captionskip=10pt}
    \begin{subfigure}{.44\linewidth}
    \centering
    \includegraphics[width=7.0cm, height=5.5cm]{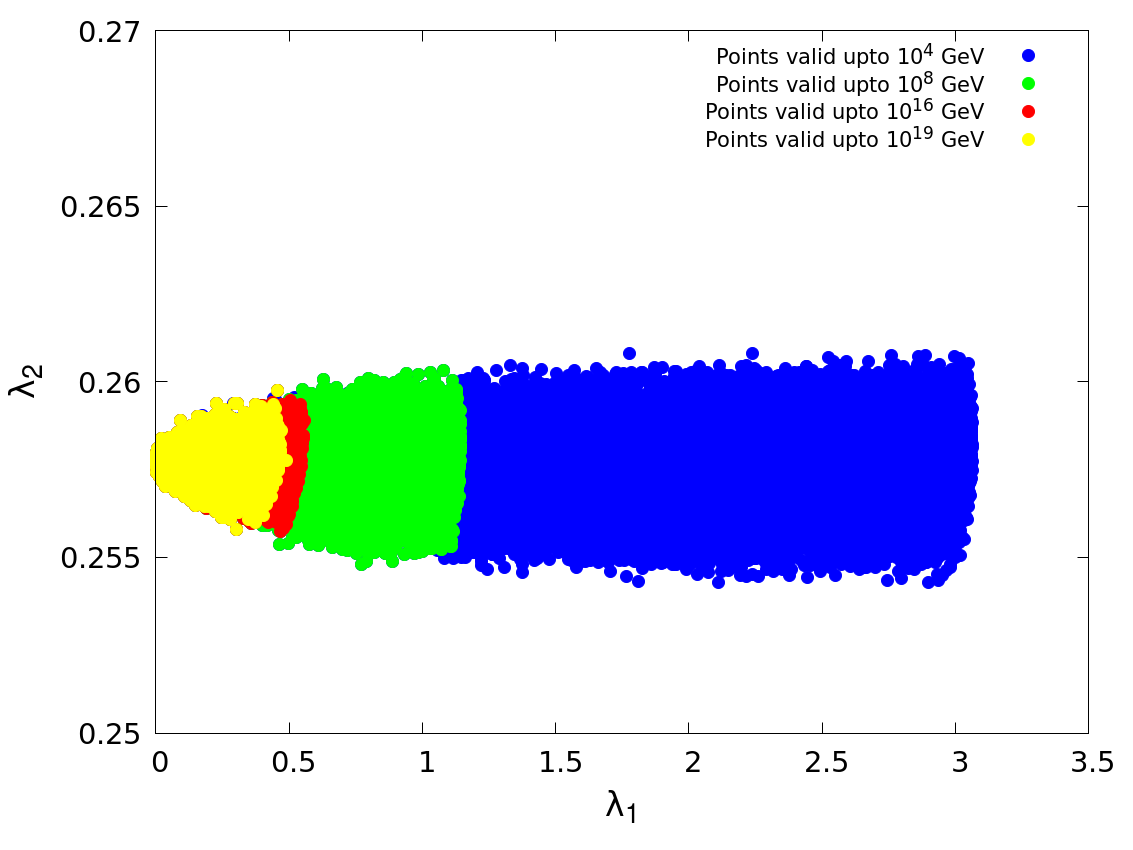}
    \caption{}\label{fig:image1}
    \end{subfigure} %
    \qquad
    \begin{subfigure}{.44\linewidth}
    \centering
    \includegraphics[width=7.0cm, height=5.5cm]{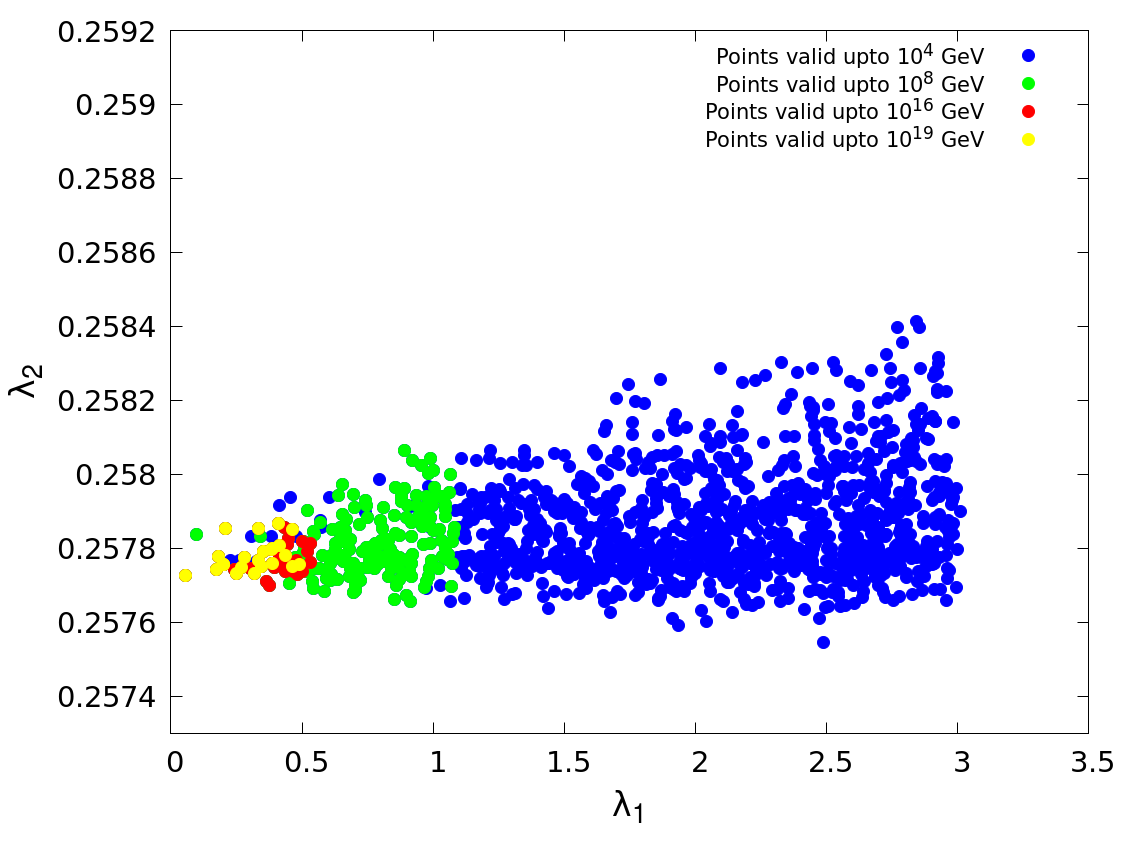}
    \caption{}\label{fig:image12}
   \end{subfigure}
\\[2ex]
    \begin{subfigure}{.44\linewidth}
    \centering
    \includegraphics[width=7.0cm, height=5.5cm]{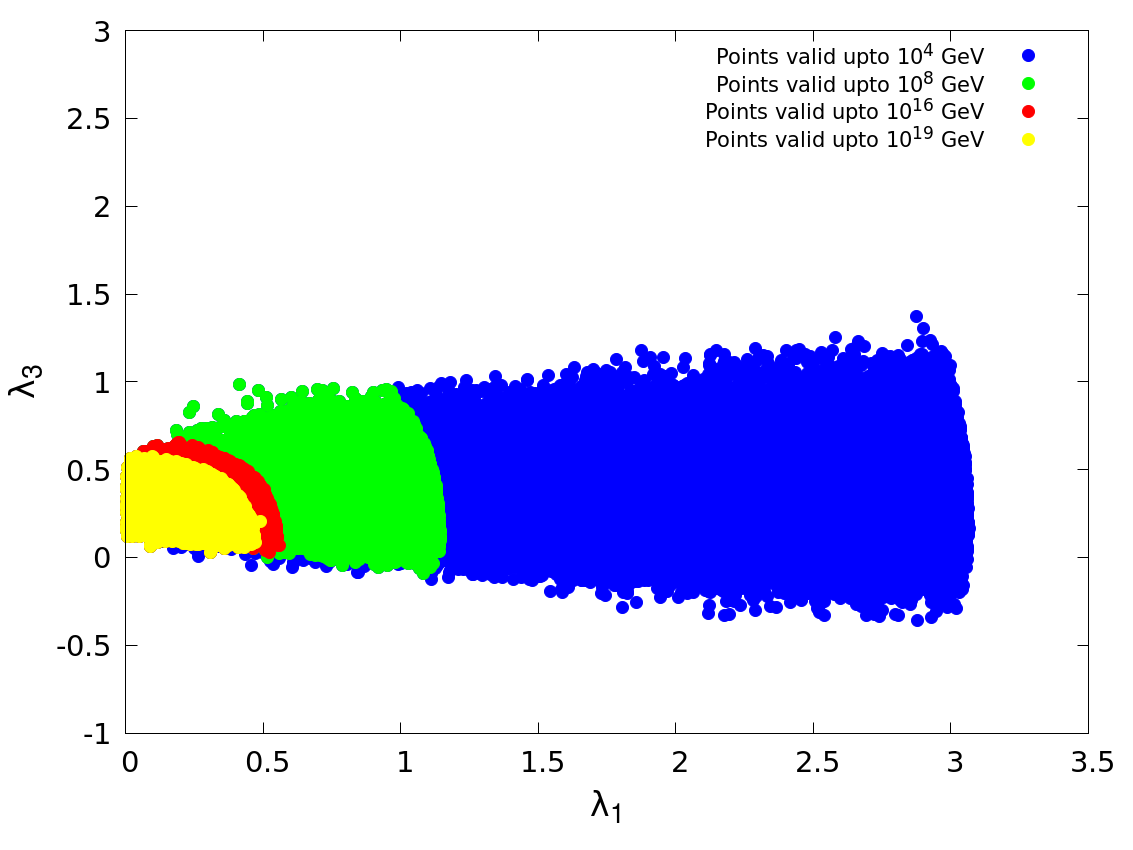}
    \caption{}\label{fig:image1}
    \end{subfigure} %
    \qquad
    \begin{subfigure}{.44\linewidth}
    \centering
    \includegraphics[width=7.0cm, height=5.5cm]{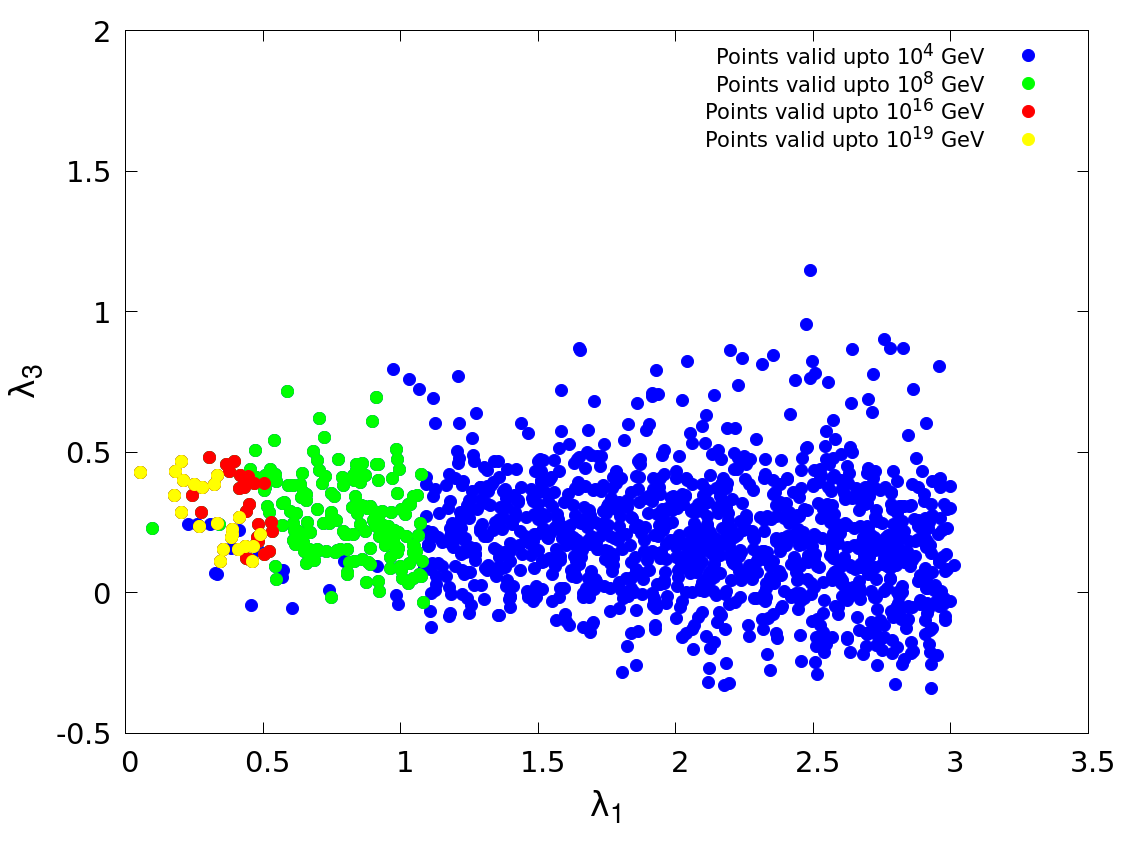}
    \caption{}\label{fig:image12}
   \end{subfigure}
\\[2ex]
    \begin{subfigure}{.44\linewidth}
    \centering
    \includegraphics[width=7.0cm, height=5.5cm]{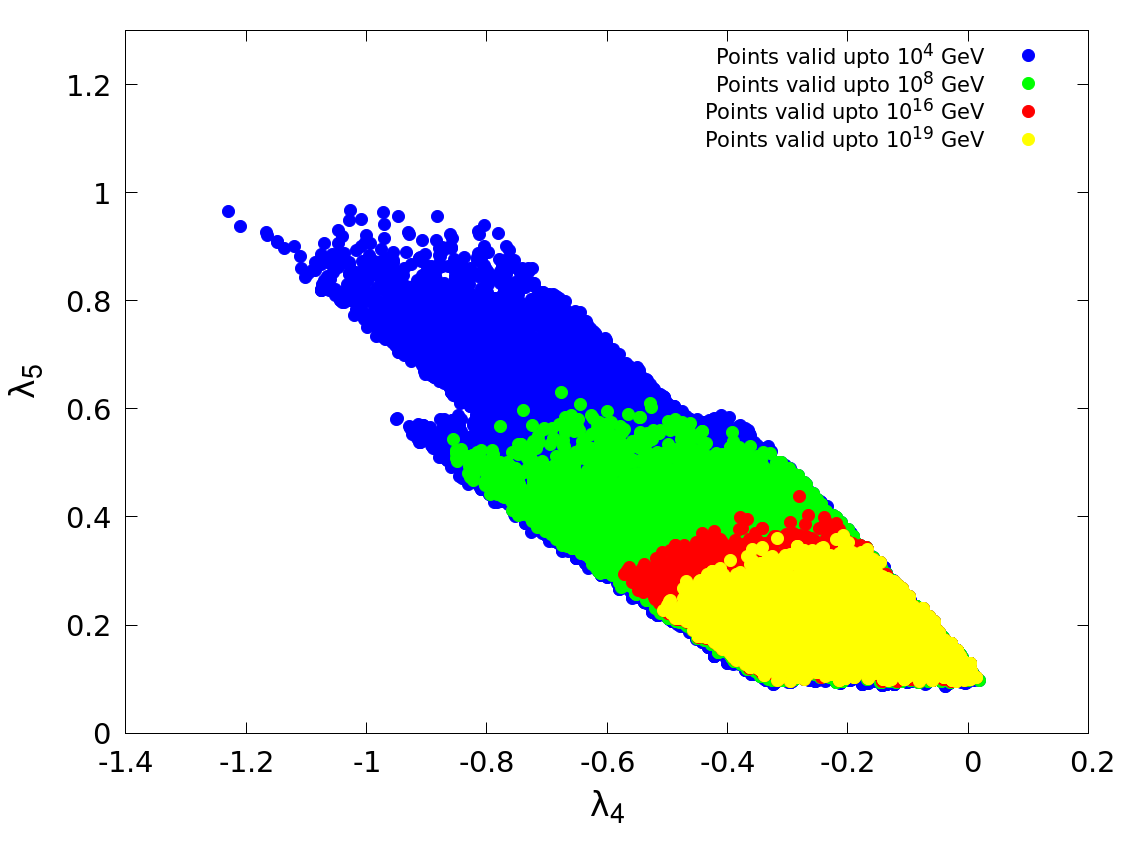}
    \caption{}\label{fig:image1}
    \end{subfigure} %
    \qquad
    \begin{subfigure}{.44\linewidth}
    \centering
    \includegraphics[width=7.0cm, height=5.5cm]{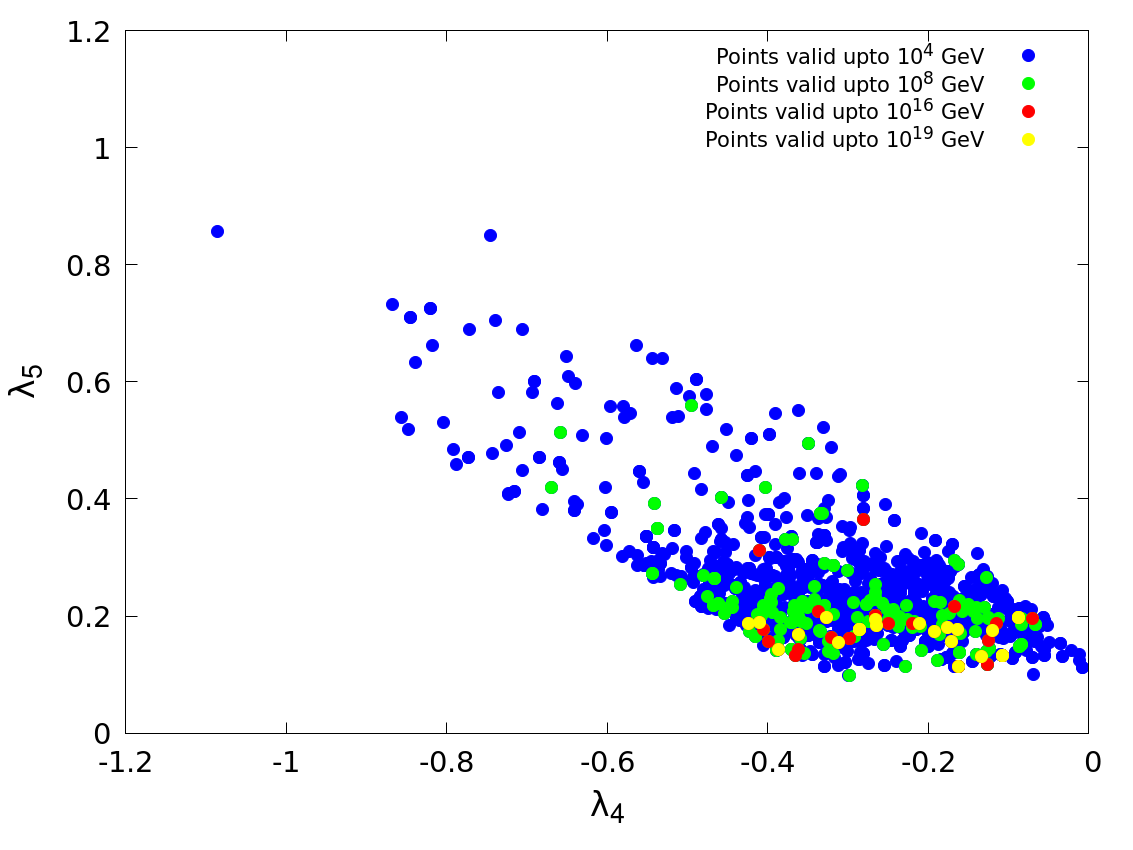}
    \caption{}\label{fig:image12}
   \end{subfigure}
\\[2ex]
\RawCaption{\caption{\it Quartic couplings valid upto different energy scales after applying (a),(c) and (e) theoretical constraints and (b),(d) and (f) theoretical constraints + $(g_{\mu}-2)$ at 3$\sigma$ + $BR(h_{SM} \rightarrow A A)$ bounds for Case 2.}
\label{hsvalidity_l1_l2_rs_gt125}}
\end{figure}

\begin{figure}[!hptb]
\floatsetup[subfigure]{captionskip=10pt}
    \begin{subfigure}{.44\linewidth}
    \centering
    \includegraphics[width=7.0cm, height=5.1cm]{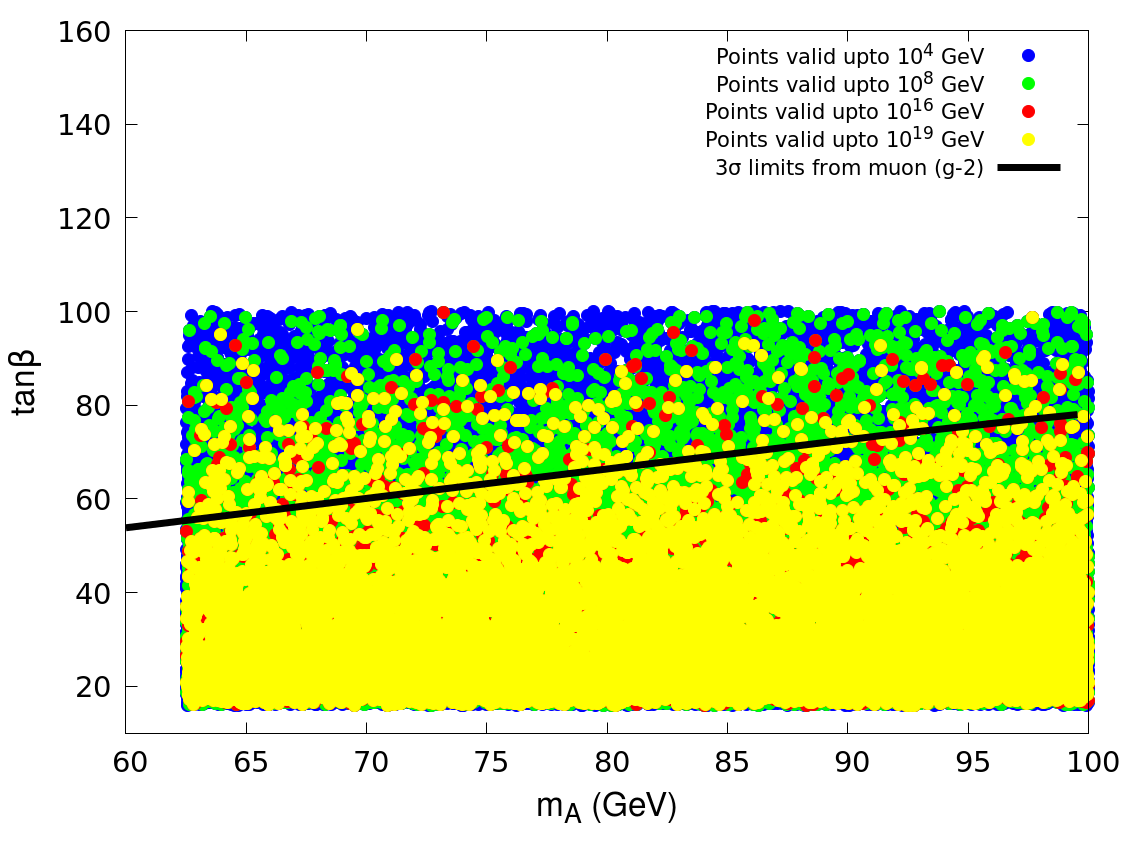}
    \caption{}\label{fig:image1}
    \end{subfigure} %
    \qquad
    \begin{subfigure}{.44\linewidth}
    \centering
    \includegraphics[width=7.0cm, height=5.0cm]{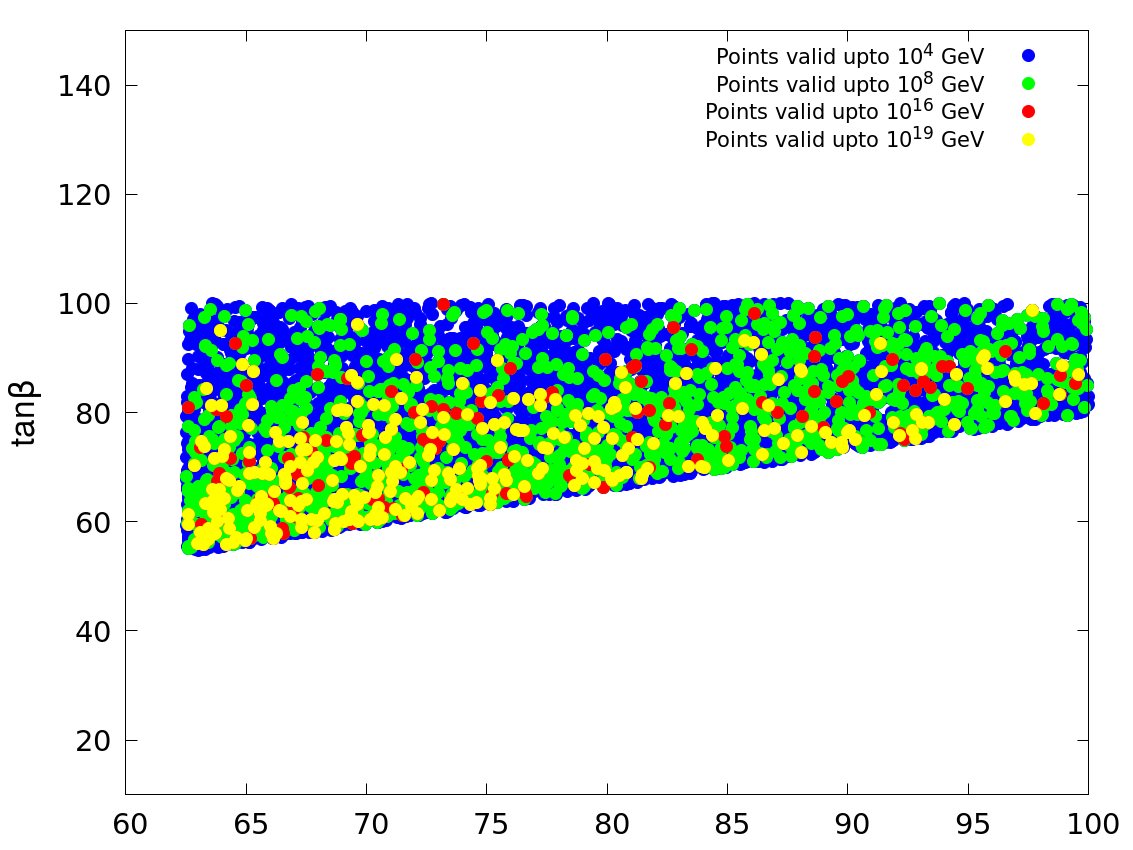}
    \caption{}\label{fig:image12}
   \end{subfigure}
\\[2ex]
\RawCaption{\caption{\it $m_A-\tan \beta$ plane, valid upto different energy scales after applying (a) theoretical constraints (b) theoretical constraints + $(g_{\mu}-2)$ at 3$\sigma$ + $BR(h_{SM} \rightarrow A A)$ bounds for Case 3.}
\label{hsvalidity_mA_tb_ws_lt125}}
\end{figure}

\begin{figure}[!hptb]
\floatsetup[subfigure]{captionskip=10pt}
    \begin{subfigure}{.44\linewidth}
    \centering
    \includegraphics[width=7.0cm, height=5.1cm]{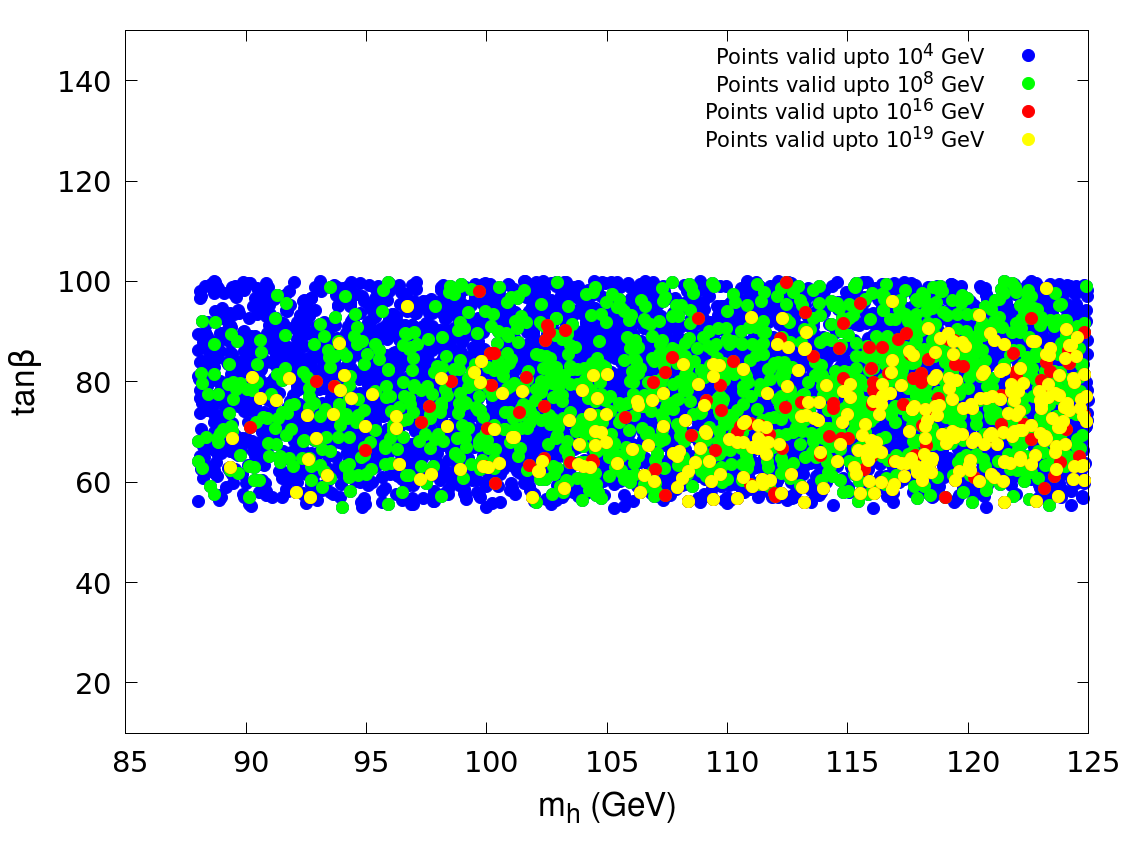}
    \caption{}\label{fig:image1}
    \end{subfigure} %
    \qquad
    \begin{subfigure}{.44\linewidth}
    \centering
    \includegraphics[width=7.0cm, height=5.1cm]{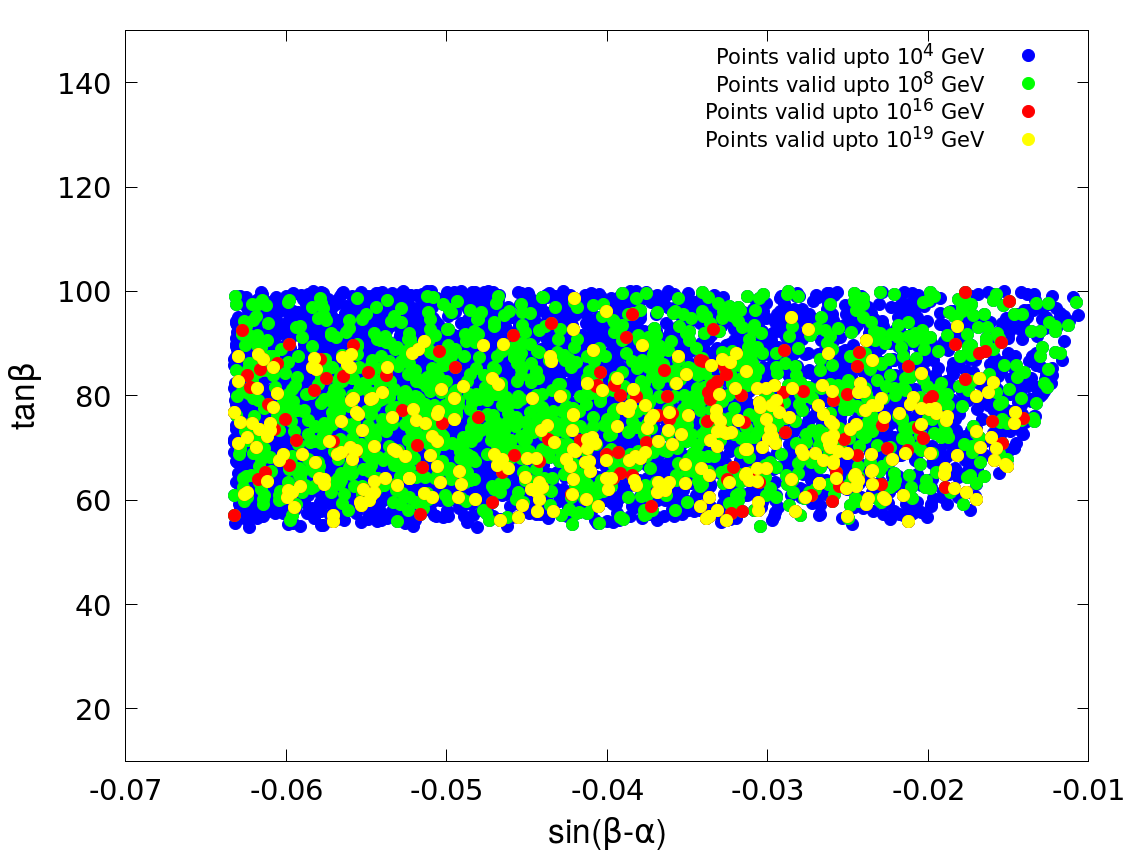}
    \caption{}\label{fig:image12}
   \end{subfigure}
\\[2ex]
  \begin{subfigure}{\linewidth}
  \centering
  \includegraphics[width=7.0cm, height=5.1cm]{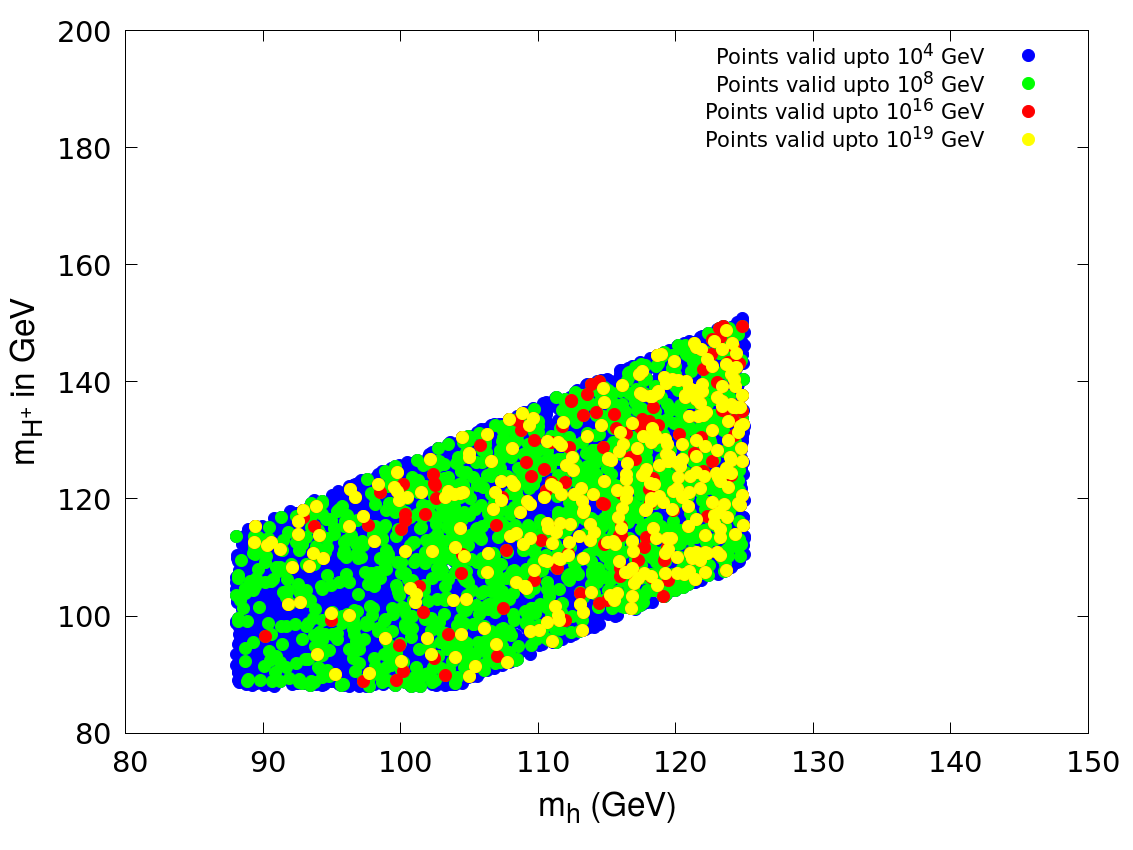}
  \caption{}\label{fig:image3}
  \end{subfigure} 
\RawCaption{\caption{\it (a) $m_h-\tan \beta$, (b) $\sin (\beta-\alpha)-\tan \beta$ and (c) $m_h - m_{H^{\pm}}$ plane, valid upto different energy scales after applying theoretical constraints+ $(g_{\mu}-2)$ at 3$\sigma$ + $BR(h_{SM} \rightarrow A A)$ bounds for Case 3.}
\label{hsvalidity_ws_lt125_others}}
\end{figure}

\begin{figure}[!hptb]
\floatsetup[subfigure]{captionskip=10pt}
    \begin{subfigure}{.44\linewidth}
    \centering
    \includegraphics[width=7.0cm, height=5.5cm]{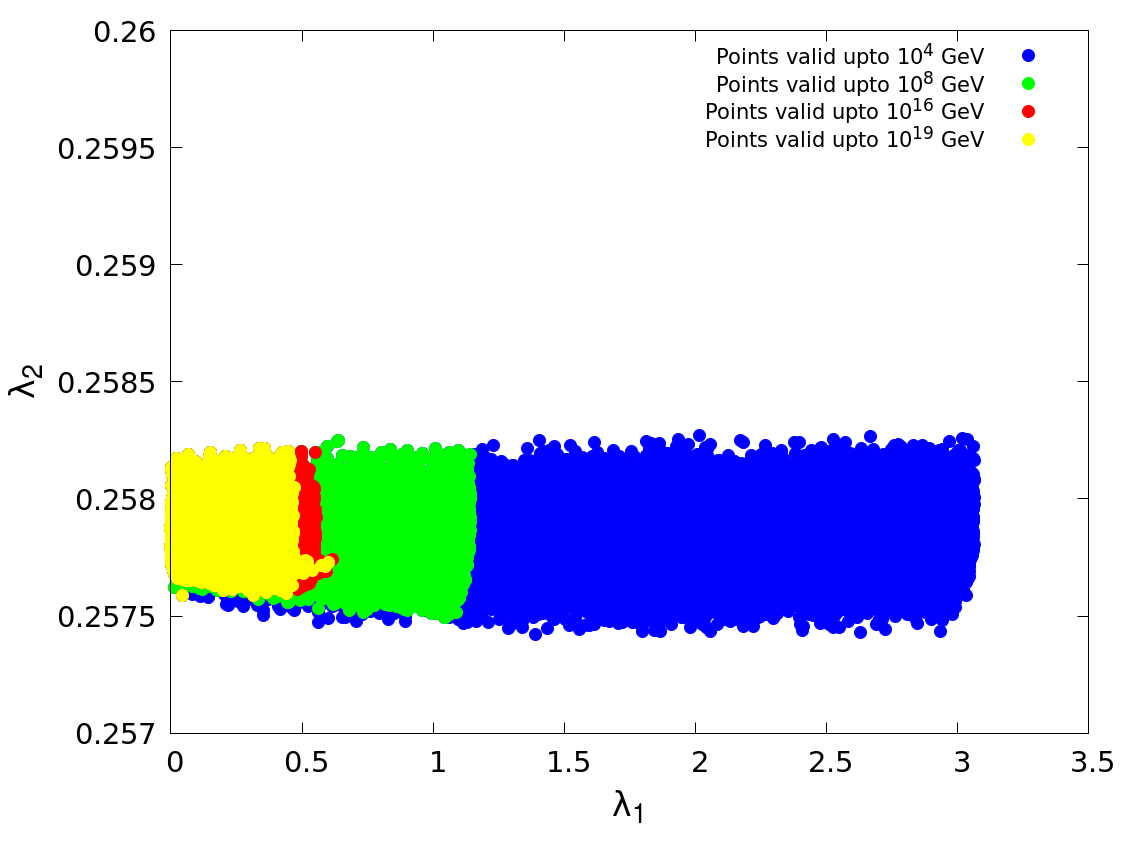}
    \caption{}\label{fig:image1}
    \end{subfigure} %
    \qquad
    \begin{subfigure}{.44\linewidth}
    \centering
    \includegraphics[width=7.0cm, height=5.5cm]{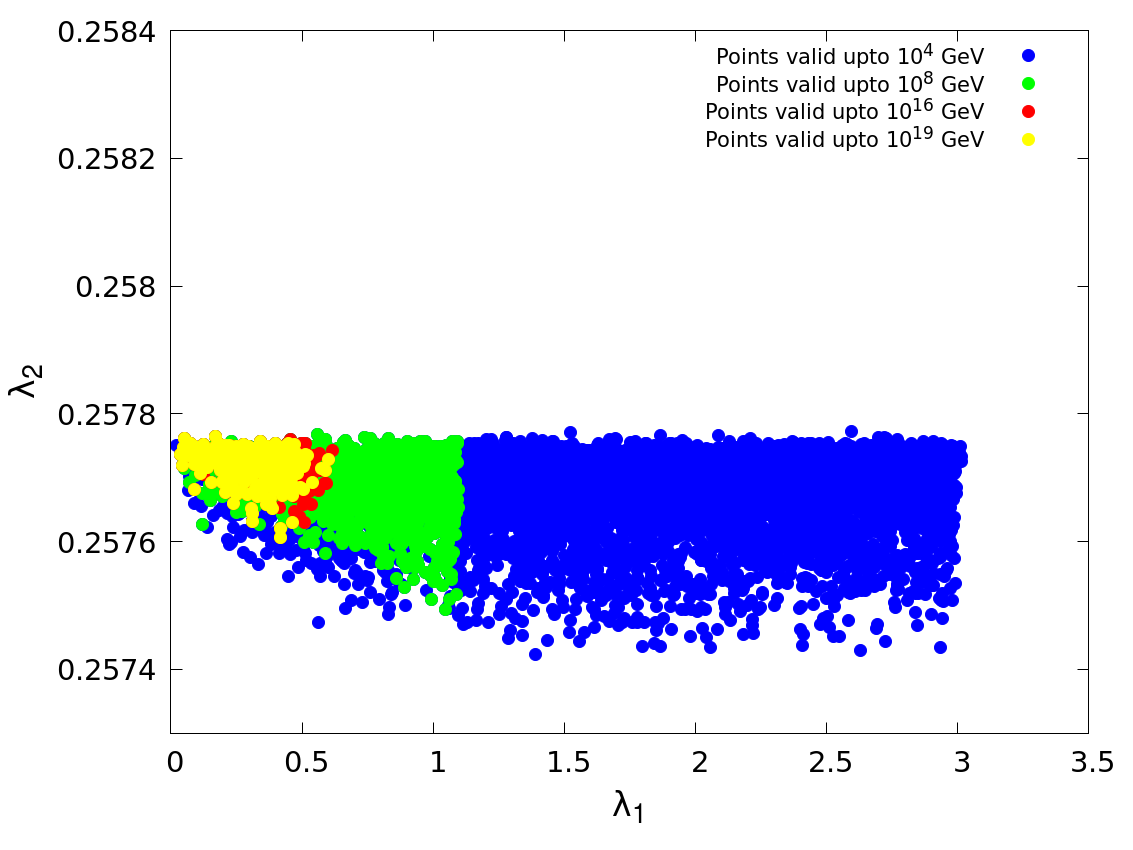}
    \caption{}\label{fig:image12}
   \end{subfigure}
\\[2ex]
    \begin{subfigure}{.44\linewidth}
    \centering
    \includegraphics[width=7.0cm, height=5.5cm]{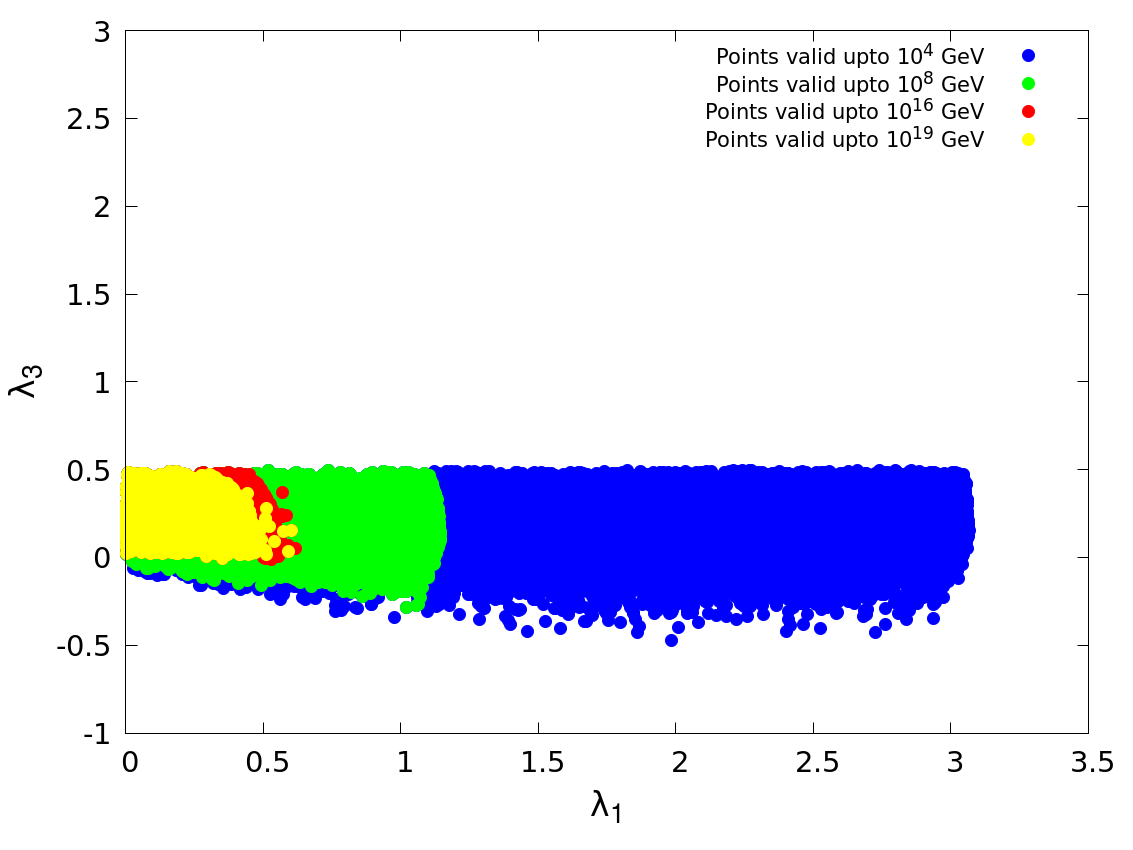}
    \caption{}\label{fig:image1}
    \end{subfigure} %
    \qquad
    \begin{subfigure}{.44\linewidth}
    \centering
    \includegraphics[width=7.0cm, height=5.5cm]{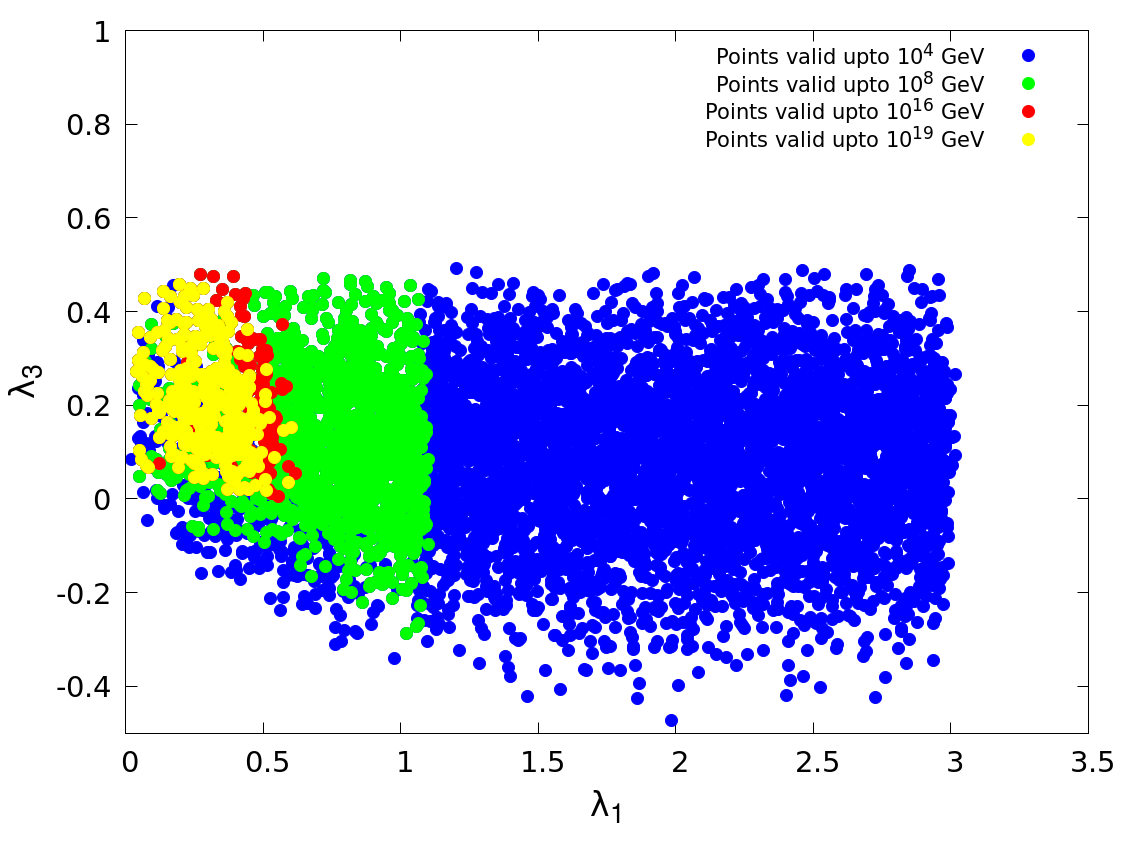}
    \caption{}\label{fig:image12}
   \end{subfigure}
\\[2ex]
    \begin{subfigure}{.44\linewidth}
    \centering
    \includegraphics[width=7.0cm, height=5.5cm]{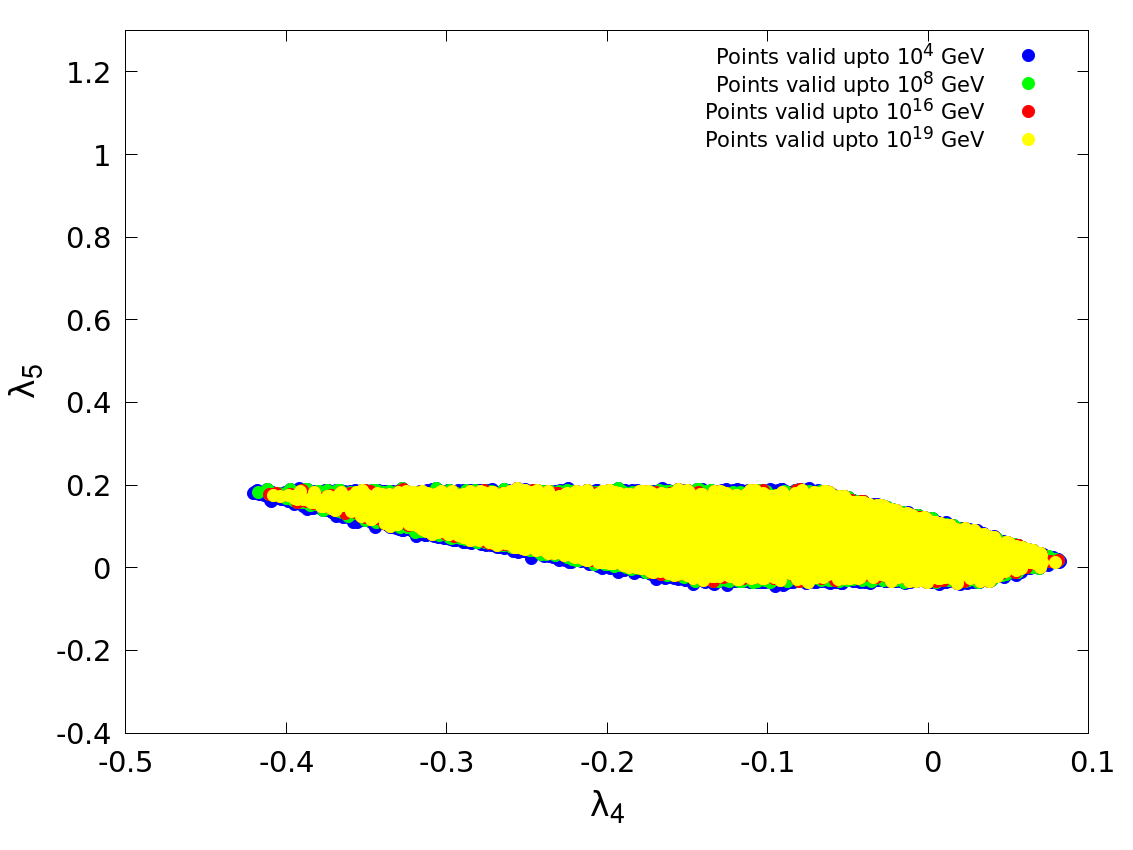}
    \caption{}\label{fig:image1}
    \end{subfigure} %
    \qquad
    \begin{subfigure}{.44\linewidth}
    \centering
    \includegraphics[width=7.0cm, height=5.5cm]{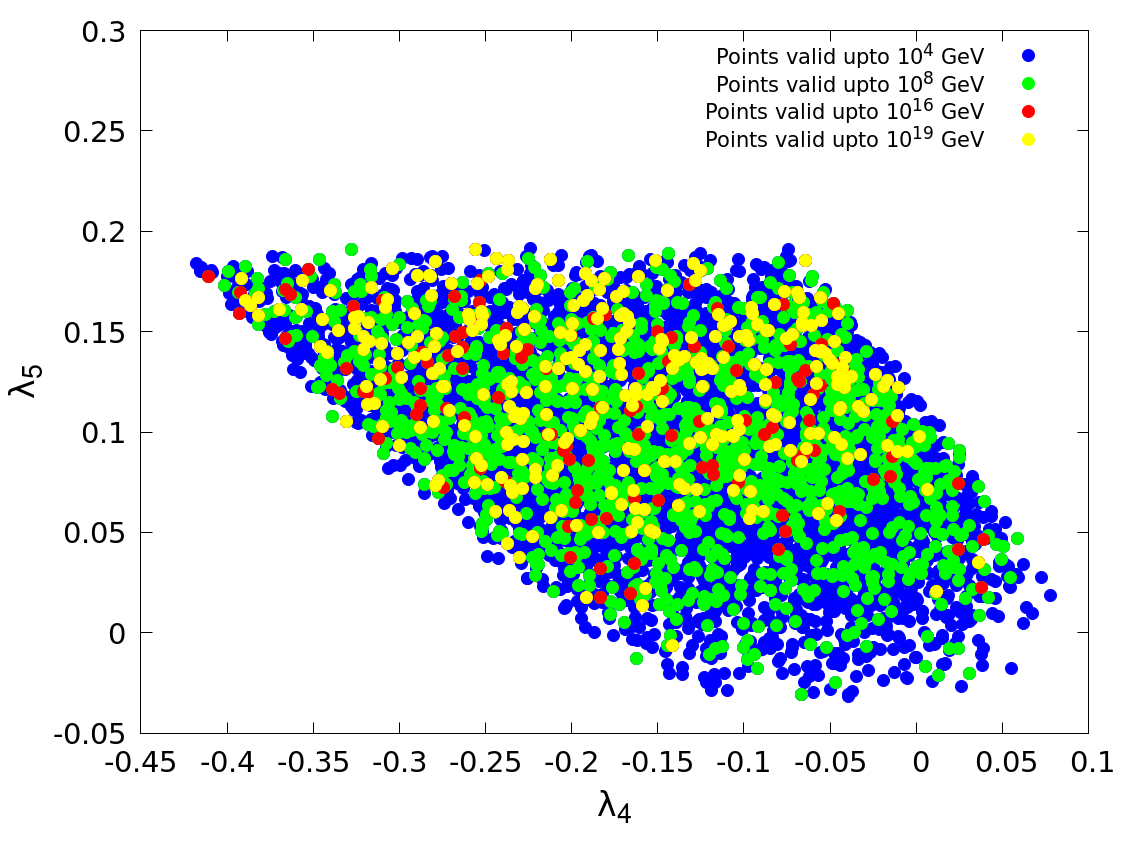}
    \caption{}\label{fig:image12}
   \end{subfigure}
\\[2ex]
\RawCaption{\caption{\it Quartic couplings valid upto different energy scales after applying (a),(c) and (e) theoretical constraints and (b),(d) and (f) theoretical constraints + $(g_{\mu}-2)$ at 3$\sigma$ + $BR(h_{SM} \rightarrow A A)$ bounds for Case 3.}
\label{hsvalidity_l1_l2_ws_lt125}}
\end{figure}

\begin{figure}[!hptb]
\floatsetup[subfigure]{captionskip=10pt}
    \begin{subfigure}{.44\linewidth}
    \centering
    \includegraphics[width=7.0cm, height=5.1cm]{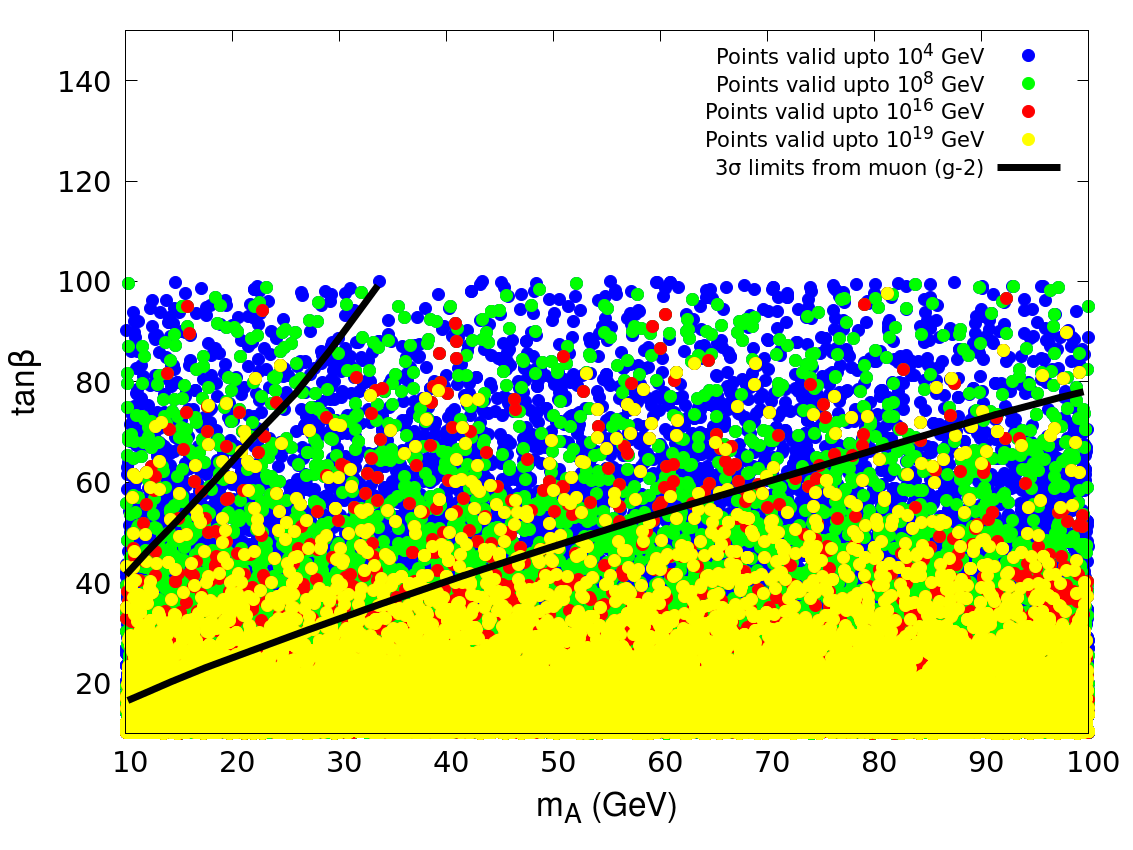}
    \caption{}\label{fig:image1}
    \end{subfigure} %
    \qquad
    \begin{subfigure}{.44\linewidth}
    \centering
    \includegraphics[width=7.0cm, height=5.1cm]{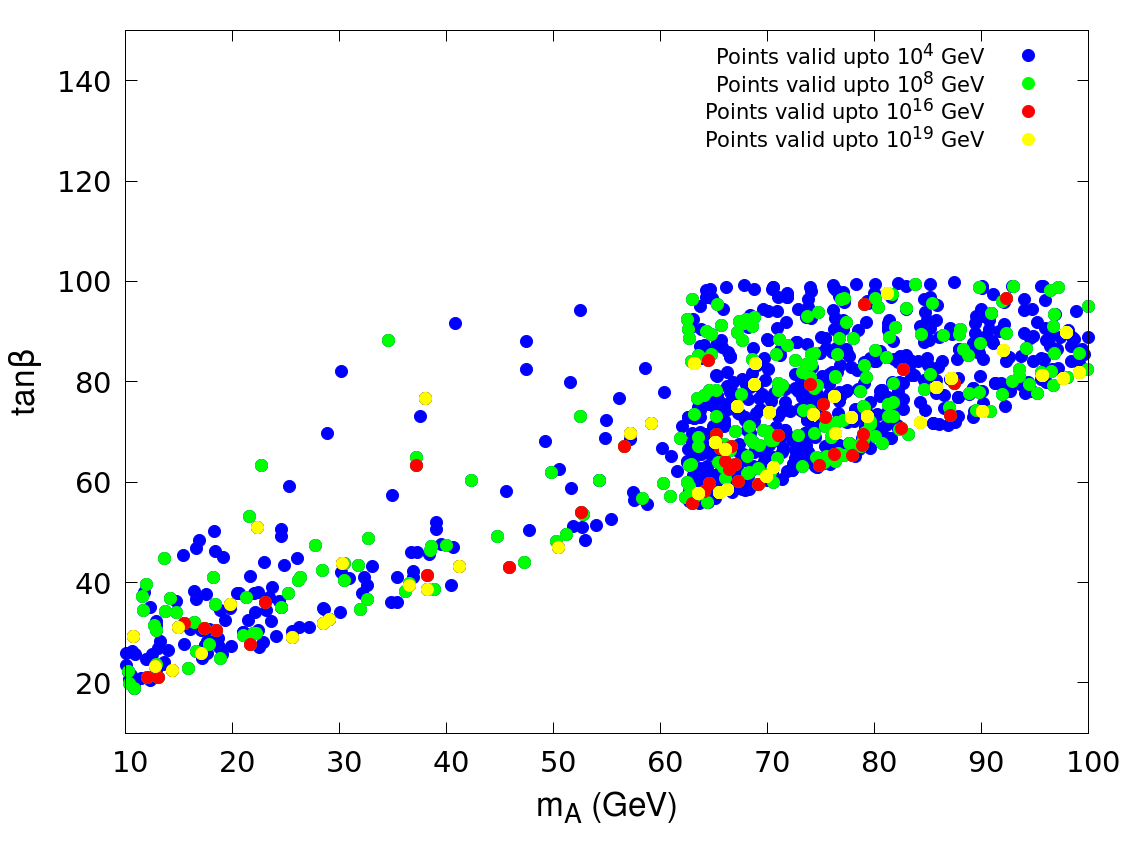}
    \caption{}\label{fig:image12}
   \end{subfigure}
\\[2ex]
\RawCaption{\caption{\it $m_A-\tan \beta$ plane, valid upto different energy scales after applying (a) theoretical constraints (b) theoretical constraints + $(g_{\mu}-2)$ at 3$\sigma$ + $BR(h_{SM} \rightarrow A A)$ bounds for Case 4.}
\label{hsvalidity_mA_tb_rs_lt125}}
\end{figure}

\begin{figure}[!hptb]
\floatsetup[subfigure]{captionskip=10pt}
    \begin{subfigure}{.44\linewidth}
    \centering
    \includegraphics[width=7.0cm, height=5.1cm]{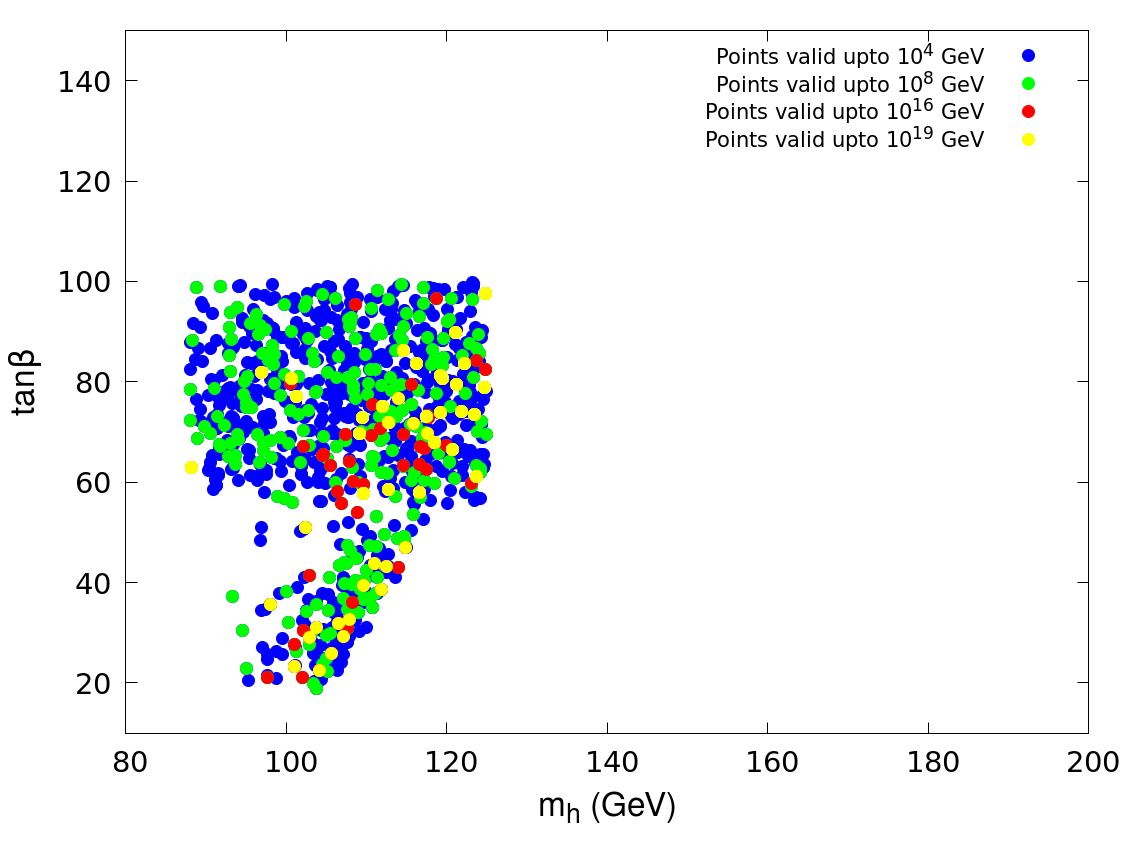}
    \caption{}\label{fig:image1}
    \end{subfigure} %
    \qquad
    \begin{subfigure}{.44\linewidth}
    \centering
    \includegraphics[width=7.0cm, height=5.1cm]{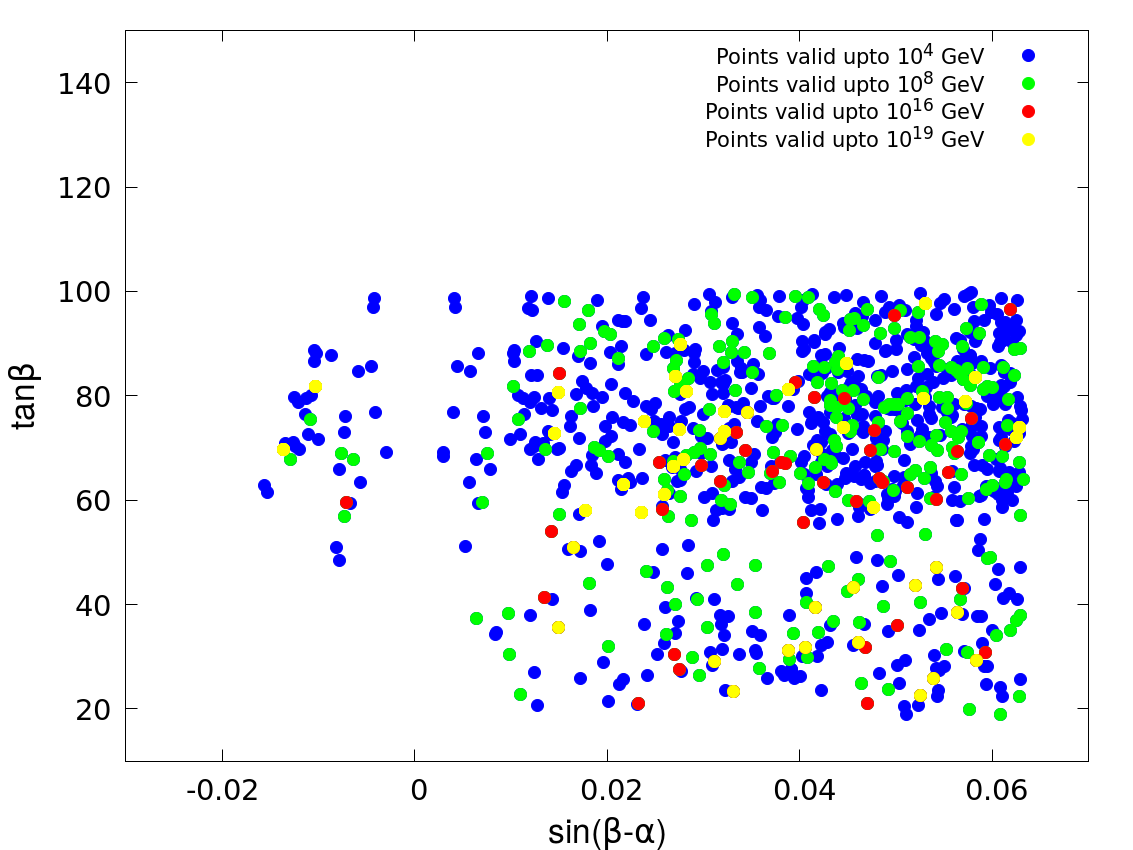}
    \caption{}\label{fig:image12}
   \end{subfigure}
\\[2ex]
  \begin{subfigure}{\linewidth}
  \centering
  \includegraphics[width=7.0cm, height=5.1cm]{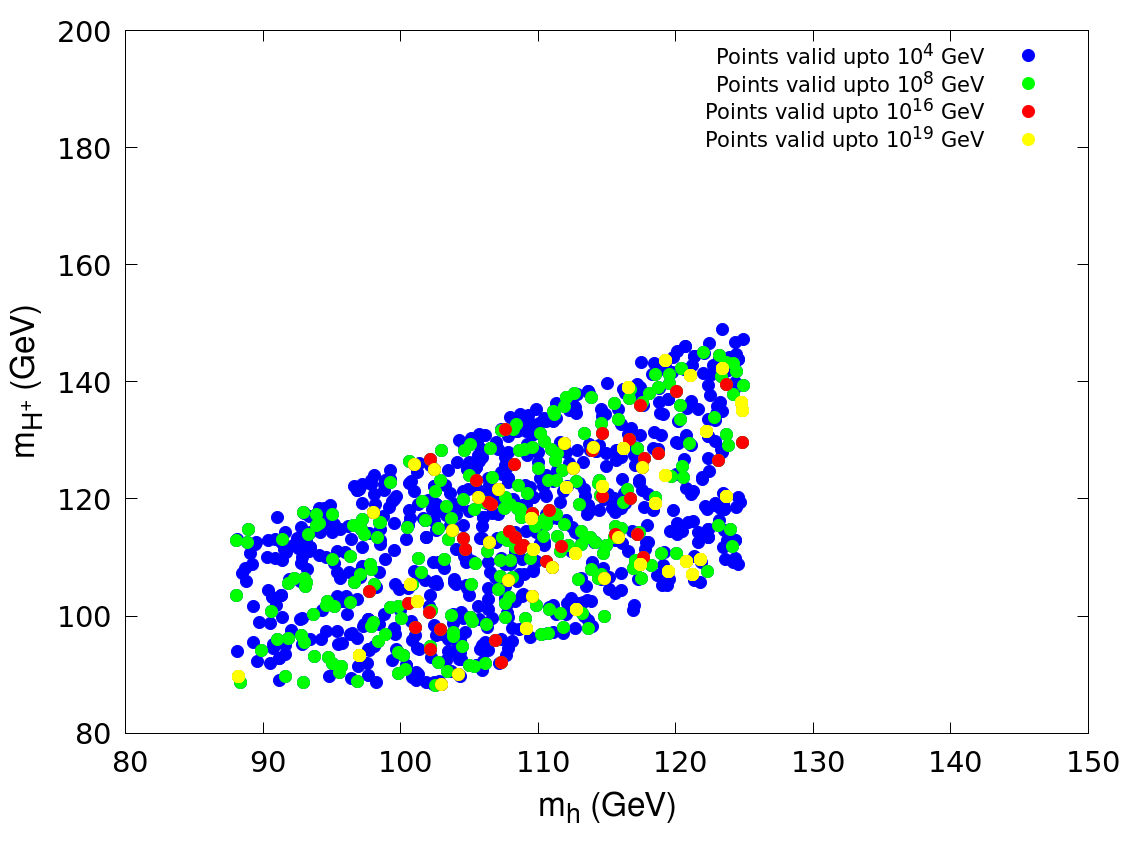}
  \caption{}\label{fig:image3}
  \end{subfigure} 
\RawCaption{\caption{\it (a) $m_h -\tan \beta$, (b) $\sin (\beta-\alpha)-\tan \beta$ and (c) $m_h - m_{H^{\pm}}$ plane, valid upto different energy scales after applying theoretical constraints + $(g_{\mu}-2)$ at 3$\sigma$ + $BR(h_{SM} \rightarrow A A)$ bounds for Case 4.}
\label{hsvalidity_rs_lt125_others}}
\end{figure}

\begin{figure}[!hptb]
\floatsetup[subfigure]{captionskip=10pt}
    \begin{subfigure}{.44\linewidth}
    \centering
    \includegraphics[width=7.0cm, height=5.5cm]{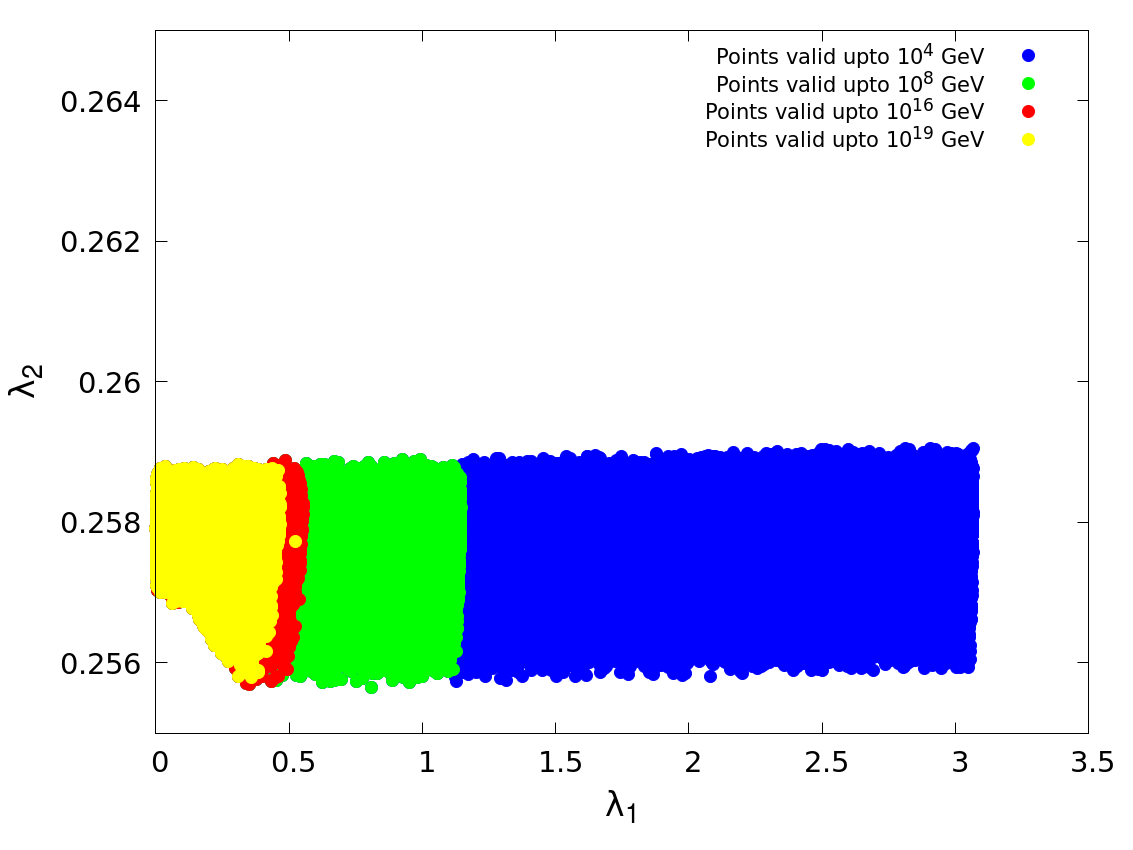}
    \caption{}\label{fig:image1}
    \end{subfigure} %
    \qquad
    \begin{subfigure}{.44\linewidth}
    \centering
    \includegraphics[width=7.0cm, height=5.5cm]{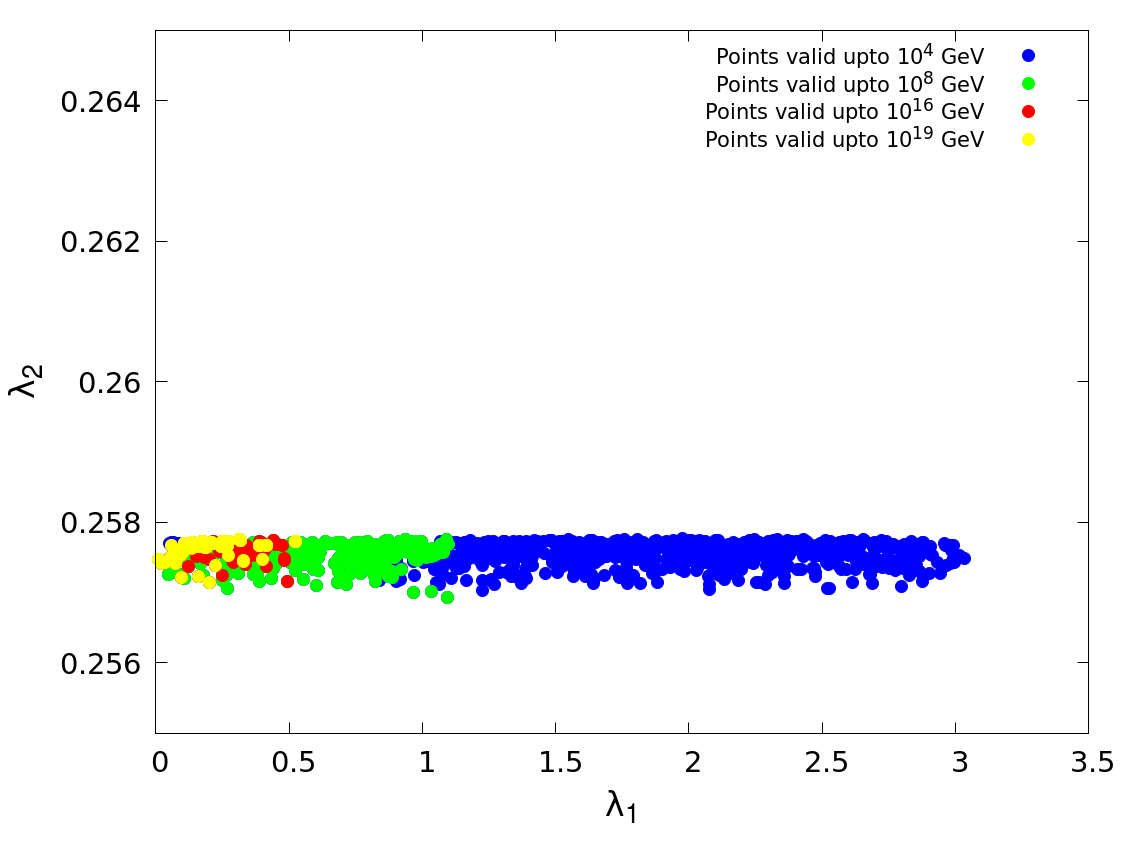}
    \caption{}\label{fig:image12}
   \end{subfigure}
\\[2ex]
    \begin{subfigure}{.44\linewidth}
    \centering
    \includegraphics[width=7.0cm, height=5.5cm]{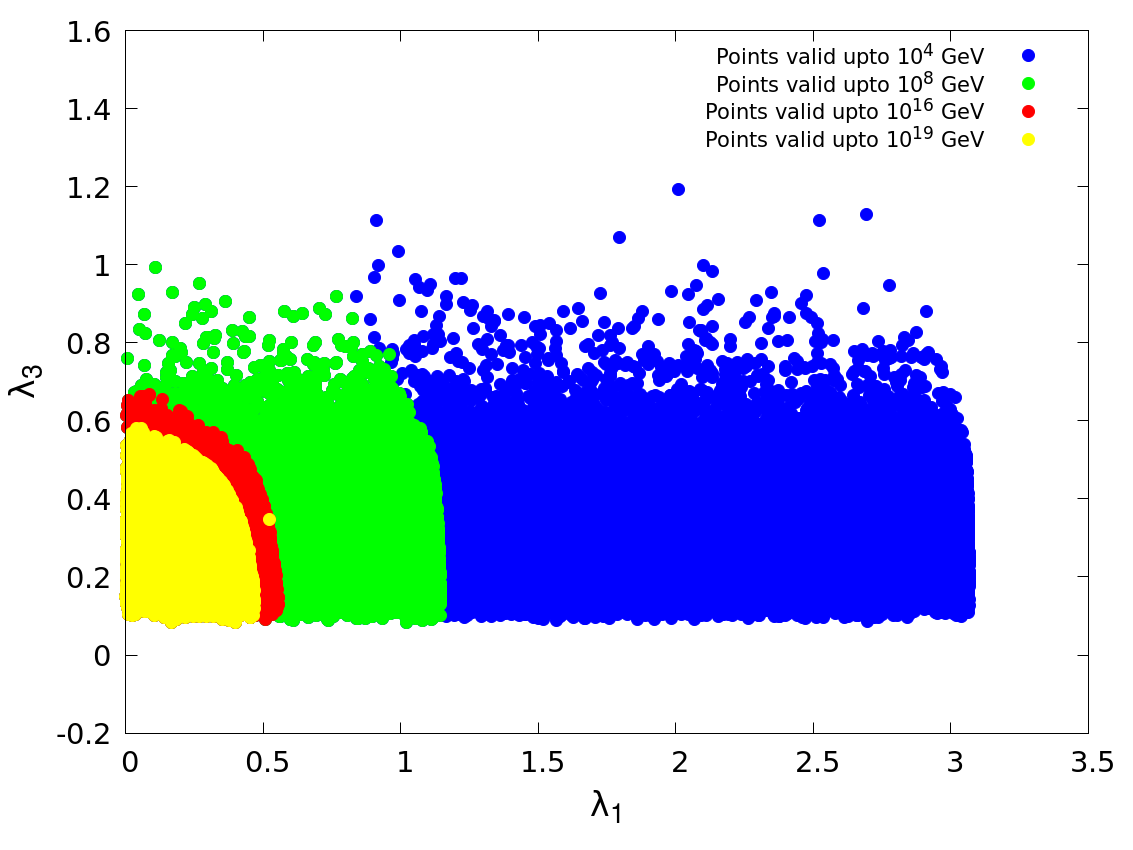}
    \caption{}\label{fig:image1}
    \end{subfigure} %
    \qquad
    \begin{subfigure}{.44\linewidth}
    \centering
    \includegraphics[width=7.0cm, height=5.5cm]{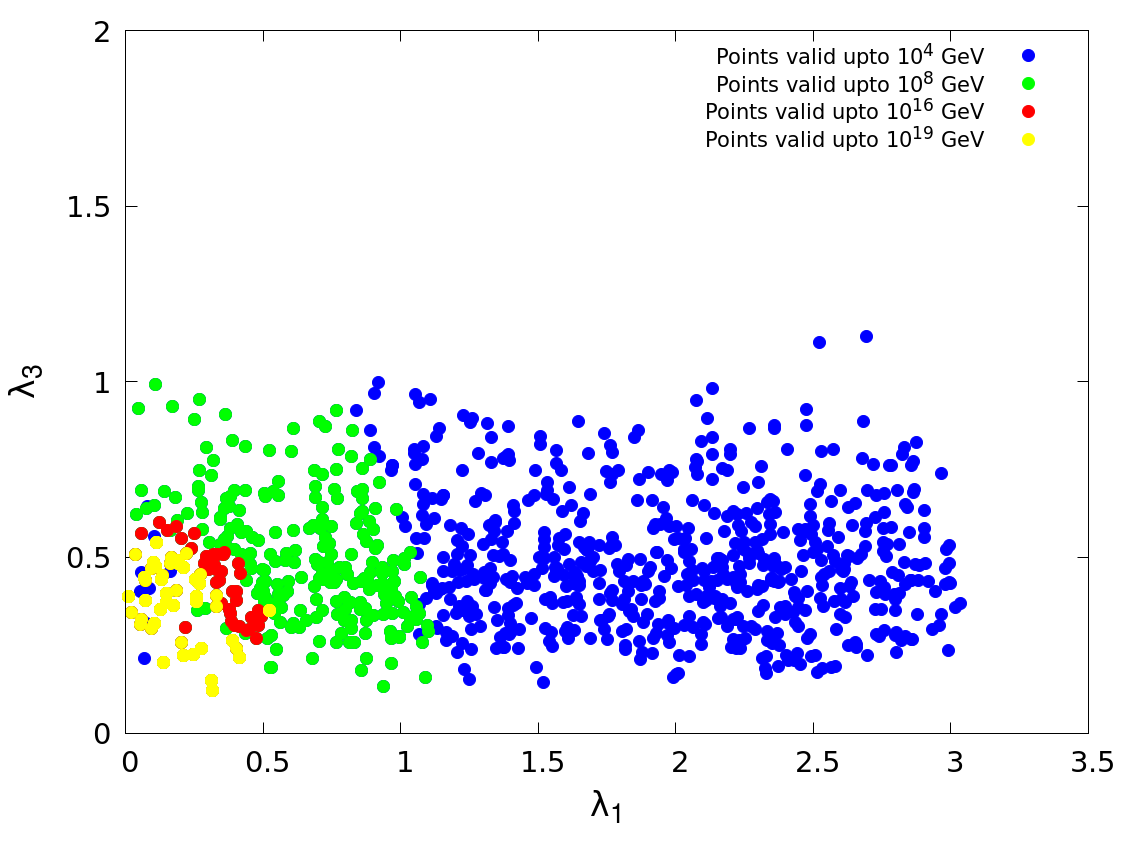}
    \caption{}\label{fig:image12}
   \end{subfigure}
\\[2ex]
    \begin{subfigure}{.44\linewidth}
    \centering
    \includegraphics[width=7.0cm, height=5.5cm]{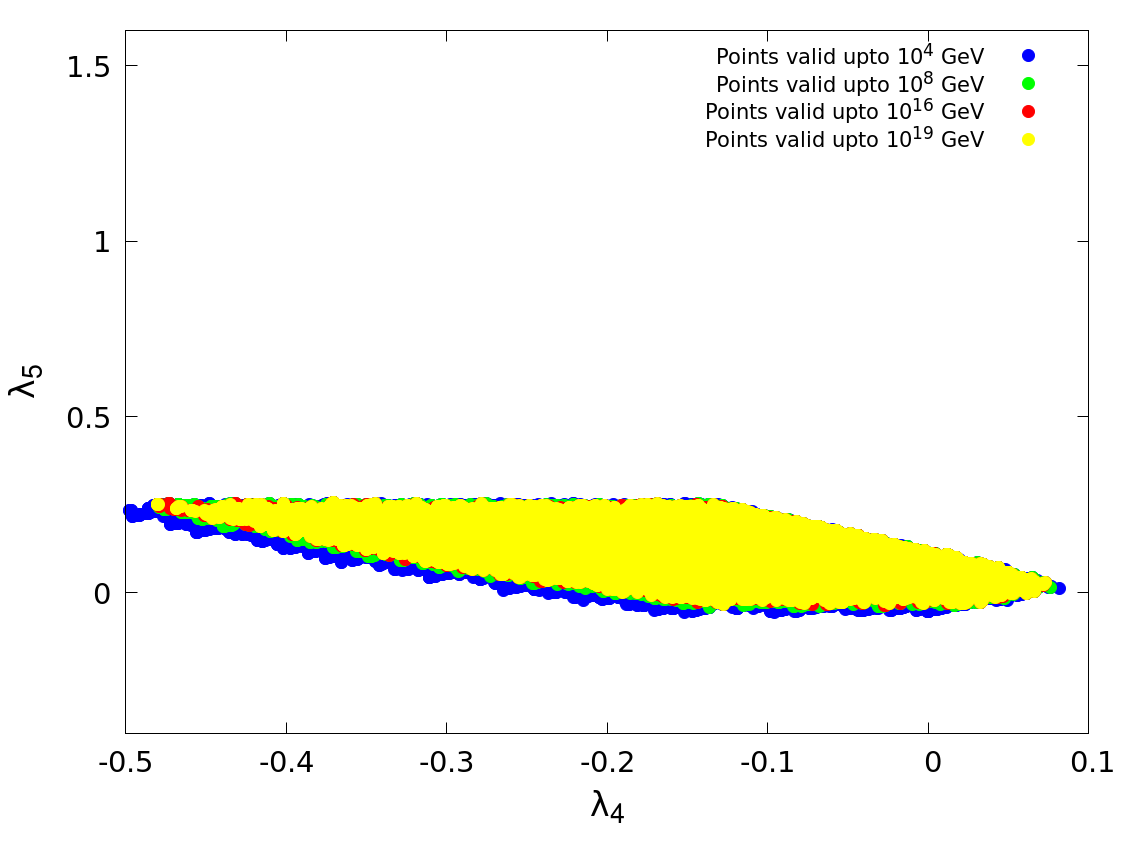}
    \caption{}\label{fig:image1}
    \end{subfigure} %
    \qquad
    \begin{subfigure}{.44\linewidth}
    \centering
    \includegraphics[width=7.0cm, height=5.5cm]{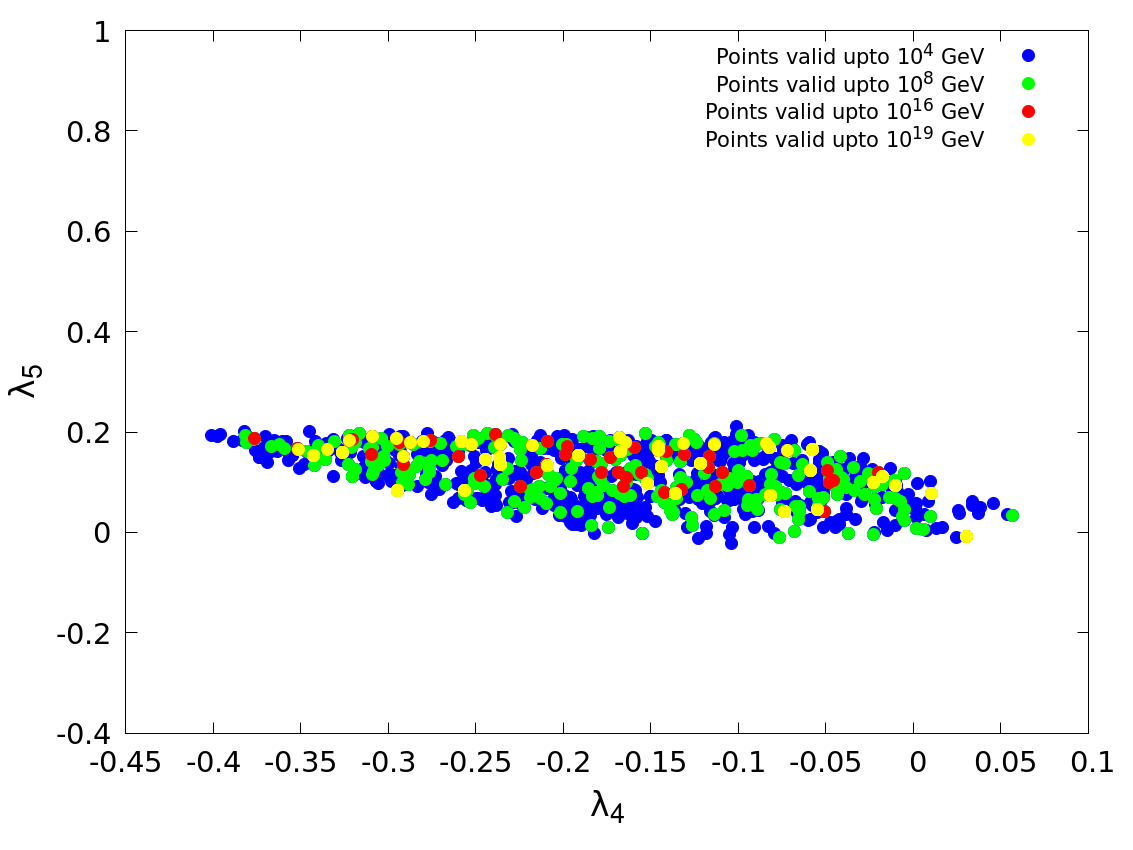}
    \caption{}\label{fig:image12}
   \end{subfigure}
\\[2ex]
\RawCaption{\caption{\it Quartic couplings valid upto different energy scales after applying (a),(c) and (e) theoretical constraints and (b),(d) and (f) theoretical constraints + $(g_{\mu}-2)$ at 3$\sigma$ + $BR(h_{SM} \rightarrow A A)$ bounds for Case 4.}
\label{hsvalidity_l1_l2_rs_lt125}}
\end{figure}

\noindent
We will identify the allowed parameter spaces for each of these cases in two-dimensional planes of relevant physical model parameters as well as the quartic couplings $\lambda'$s. In all the plots in Figure~\ref{hsvalidity_mA_tb_ws_gt125}-\ref{hsvalidity_l1_l2_rs_lt125} the blue, green, red and yellow points represent the regions valid upto $ 10^4, 10^8, 10^{16}, 10^{19}$ GeV respectively.

\bigskip

\noindent
$\bullet$ {\bf Case 1:}\\

\noindent
In Figure~\ref{hsvalidity_mA_tb_ws_gt125}(a), the two black lines represent the upper and lower bounds from $g_{\mu}-2$ anomaly at $3\sigma$. Figure~\ref{hsvalidity_mA_tb_ws_gt125}(b) shows the allowed parameter space which satisfy the observed $g_{\mu}-2$ as well as the strong upper limit from Br($h_{SM} \rightarrow AA)$.
 It is clear from this plot that high-scale validity upto the Planck scale demands $\tan \beta \lsim 30$. At the electroweak scale we have seen that large $\tan \beta$ regions are disfavored from the requirement of perturbative unitarity, since large $\tan \beta$ eventually results in large $\lambda'$s. For high-scale validity, $\lambda'$s need to be small at electroweak scale. Naturally, relatively small values of $\tan \beta$ are favored from the stand-point of high-scale validity. On the other hand, the observed $g_{\mu}-2$ prefers $\tan \beta$ on the higher side, rendering a very tiny region valid upto the Planck scale, which is clear from Figure~\ref{hsvalidity_mA_tb_ws_gt125}(b). In the same figure, we see a sharp discontinuity in the allowed parameter space around $m_A \approx m_h/2$. This is because of the fact that when $m_A \lsim \frac{m_h}{2}$, one is strongly restricted by the limit BR$(h_{SM} \rightarrow A A) \lsim 4\%$. This constraint is particularly severe for large $\tan \beta$, a feature we have already seen in Section~\ref{sec4}.

In Figures ~\ref{hsvalidity_ws_gt125_others}(a), (b) and (c), we show the high-scale validity in the $(m_H-\tan \beta)$, $(\sin(\beta-\alpha)-\tan \beta)$ and $(m_H-m_{H^{\pm}})$ planes respectively. We can see that the high-scale validity demands smaller $m_H$. The major reason behind this is the following. As $\lambda_3$ increases with $m_H$ in this region (see Equation~\ref{eq:paratran}), the requirement of small $\lambda_3$ at the electroweak scale (which is necessary for high-scale validity) pushes us towards small $m_H$ values. One more feature from the figure is that, when $\tan \beta \lsim 50$ there is a discontinuity in the allowed points. The reason behind this is the following. The parameter space with $\tan \beta \lsim 50$ and $m_A > \frac{m_h}{2}$, albeit allowed by the BR($h_{SM} \rightarrow AA$) constraints, faces severe constraint from the lower limit on $(g_{\mu} - 2)$ (see Figure~\ref{hsvalidity_mA_tb_ws_gt125}(a)). On the other hand, the small strip below $\tan \beta \lsim 50$ corresponds to the points where $m_A \lsim \frac{m_h}{2}$ and BR($h_{SM} \rightarrow AA$) upper limit is satisfied. Similar feature is observed in Figure ~\ref{hsvalidity_ws_gt125_others}(b) where the small strip below $\tan \beta \lsim 50$ corresponds to $m_A \lsim \frac{m_h}{2}$. From Figure~\ref{hsvalidity_ws_gt125_others}(c), we can see that the high-scale validity puts a strong upper bound on $m_H$, which also follows from our understanding of the perturbativity and unitarity condition at the electroweak scale. The degeneracy between $m_H$ and $m_{H^{\pm}}$ mass naturally pushes the charged scalar mass to smaller values, at the high scales which is evident from the Figure~\ref{hsvalidity_ws_gt125_others}(c).


Let us now discuss the high-scale validity in the planes spanned by the quartic couplings, as they play the key role in this regard. In Figure~\ref{hsvalidity_l1_l2_ws_gt125}(a), we can see that $\lambda_1$ controls the high-scale behavior much more than $\lambda_2$. This happens because $\lambda_2$ at the electroweak scale is solely determined by the 125-GeV Higgs mass and varies only slightly with energy, a behavior we have already seen. With the variation in scale from $10^4$ GeV to $10^{19}$ GeV, the allowed range of $\lambda_2$ varies only slightly around its electroweak value. On the contrary, allowed range for $\lambda_1$ varies from 3 to 0.5 with the same variation in scale. In Figure~\ref{hsvalidity_l1_l2_ws_gt125}(b), we have shown the region allowed after the constraints from $g_{\mu}-2$ and BR($h_{SM} \rightarrow AA)$ are applied. We have seen from our earlier discussions that $g_{\mu} - 2$ favors large $\tan \beta$ while the upper limit on BR($h_{SM} \rightarrow AA)$ favors low $\tan \beta$. $\lambda_2$ is inversely proportional to $\tan \beta$ in the alignment region. Therefore higher values of $\lambda_2$ are disfavored by the observed $g_{\mu}-2$ data, while the lower $\lambda_2$ gets 
constrained from the BR($h_{SM} \rightarrow AA)$.

In Figure~\ref{hsvalidity_l1_l2_ws_gt125}(c), we demonstrate regions with different levels of high-scale validity in the parameter space spanned by $\lambda_1$ and $\lambda_3$. Their high-scale behavior appears to be strongly correlated with each other and the allowed range in the $\lambda_1-\lambda_3$ plane shows elliptic contours. Figure~\ref{hsvalidity_l1_l2_ws_gt125}(d) shows the allowed region after the imposition of $g_{\mu} - 2$ and BR($h_{SM} \rightarrow AA$) constraints. We can see that these two constraints do not affect these couplings directly, but only reduces the density of points uniformly, depending on the other quartic couplings.

 In Figure~\ref{hsvalidity_l1_l2_ws_gt125}(e), we plot the high-scale validity in the $\lambda_4-\lambda_5$ plane. We have seen in our earlier discussion that the mass degeneracy between $m_H$ and $m_{H^{\pm}}$ implies $\lambda_4 \approx -\lambda_5$. As the perturbative unitarity condition favors this mass degeneracy, this correlation between $\lambda_4$ and $\lambda_5$ is also favored for high-scale validity. The $g_{\mu}-2$ and BR($h_{SM} \rightarrow AA$) constraints result in only uniform reduction of allowed points, the nature of the allowed region remaining unaltered (see Figure~\ref{hsvalidity_l1_l2_ws_gt125}(f)).

It is clear from the discussion in the plane of quatic couplings that, the requirement of validity of the theory upto higher scales, pushes the quartic couplings to smaller values.  

\bigskip

\noindent
$\bullet$ {\bf Case 2:}\\

In Figure~\ref{hsvalidity_mA_tb_rs_gt125}, we show the high-scale validity in $m_A - \tan \beta$ plane in the right-sign region of Scenario 1. Here the nature of high-scale validity is same as case 1 and for the same reason. The black line in the Figure~\ref{hsvalidity_mA_tb_rs_gt125}(a) denotes the lower limit coming from the $g_{\mu} - 2$ data. We have shown only the region $m_A > \frac{m_h}{2}$ here, because from the upper limit on BR($h_{SM} \rightarrow AA$), this is the only allowed region in this case, as discussed in Section~\ref{sec5}.

In Figure~\ref{hsvalidity_rs_gt125_others}(a), (b) and (c) we show the high-scale validity in the $(m_H-\tan \beta)$, $(R_{(\beta - \alpha)}-\tan \beta)$ and $(m_H-m_{H^{\pm}})$ planes respectively, after imposing the $g_{\mu}-2$ constraints and the upper limit from BR($h_{SM} \rightarrow AA$). Here too, we observe similar behavior as case 1 and the same discussion follows. We note here that, $\tan \beta \lsim 50$ is completely disfavored in this case unlike case 1. The reason behind this is in case 2, we do not have a region with $m_A \lsim \frac{m_h}{2}$ that satisfies the upper limit on BR($h_{SM} \rightarrow AA$) and $m_A > \frac{m_h}{2}$ region gets severely constrained by the lower limit from $g_{\mu}-2$ when $\tan \beta \lsim 50$.

We report next on the high-scale validity in the RS region of Scenario 1 in the parameter space spanned by the quartic couplings. In Figure~\ref{hsvalidity_l1_l2_rs_gt125}(a)-(f), we see similar features as in case 1. However, in this case, large values of $\lambda_3$ become disfavored even at the electroweak scale as can be seen from Figure~\ref{hsvalidity_l1_l2_rs_gt125}(c) and (d), since in the RS case a stronger upper bound is imposed on the $m_{H^{\pm}}$ and $m_H$, compared to WS case, in the pseudoscalar mass range of our interest. In Figure~\ref{hsvalidity_l1_l2_rs_gt125}(e) and (f), we see, in this case, the lower masses of $m_H$ and $m_{H^{\pm}}$ restrict the upper limits on $\lambda_4$ and $\lambda_5$ to lower values compared to the WS case.

\bigskip

\noindent
$\bullet$ {\bf Case 3:}\\

We now proceed to scenario 2 (ie. $m_H = 125$ GeV), in the WS region. In Scenario 2, the charged scalar and the non-standard CP-even scalar masses are kept at a lower range compared to Scenario 1. In Figure~\ref{hsvalidity_mA_tb_ws_lt125}, we show the the high-scale validity in the $m_A-\tan \beta$ plane. Like the previous cases, here too, the tension between the high-scale validity and the observed $g_{\mu}-2$ continues. We have not shown the region $m_A \lsim \frac{m_H}{2}$ in Figure~\ref{hsvalidity_mA_tb_ws_lt125}, because in Scenario 2, WS region, this region does not satisfy BR$(h_{SM} \rightarrow AA)$ upper limit. On the other hand, $m_A > \frac{m_H}{2}$ trivially satisfies this bound.

In Figure~\ref{hsvalidity_ws_lt125_others}(a), (b) and (c), we show the parameter space allowed by all the aforementioned constraints in the $(m_h-\tan \beta)$, $(\sin(\beta-\alpha)-\tan \beta)$ and $(m_h-m_H{^{\pm}})$ plane respectively. As a low mass range for the non-standard CP-even scalar ($h$) is considered in this case, the entire mass range is valid upto very high scales. However, the $g_{\mu}-2$ data disfavors the region below $\tan \beta \lsim 50$ for $m_A > \frac{m_h}{2}$ whereas $m_A \lsim \frac{m_h}{2}$ is disfavored from the upper limit on BR($h_{SM} \rightarrow AA$), therefore we see no point in the range $\tan \beta \lsim 50$ in Figure~\ref{hsvalidity_ws_lt125_others}(a) and (b).

The behavior of the quartic couplings in the context of high-scale validity is similar to the previous cases considered, as we can see from Figure~\ref{hsvalidity_l1_l2_ws_lt125}(a)-(f). The apparently stronger upper limit on $\lambda_3$ in Figure~\ref{hsvalidity_l1_l2_ws_lt125}(c) and (d) follows from the fact that the CP-even non-standard scalar mass ($m_h$) is much lower in Scenario 2, compared to Scenario 1, irrespective of WS and RS. In Figure~\ref{hsvalidity_l1_l2_ws_lt125}(e) and (f), we see a different behavior compared to scenario 1. The correlation between $\lambda_4$ and $\lambda_5$ here is not very clear. The reason again being, in Scenario 2, we are confined within small range for $m_h$ and therefore the degeneracy, which is responsible for the correlation between $\lambda_4$ and $\lambda_5$, is not very apparent in this case. As the non-standard scalar masses are already small, almost the entire region considered is allowed upto a very high scale($10^{19}$ GeV).

\bigskip

\noindent
$\bullet$ {\bf Case 4:}\\

\noindent
Now we will focus on the RS region of scenario 2 ie. $m_H = 125$ GeV. Here too, the charged Higgs and the non-standard CP-even scalar masses are kept on the lower side.   
In Figure~\ref{hsvalidity_mA_tb_rs_lt125}, similar behavior as the previous cases is observed. One may note, similar to case 1, here we get a small region in the range $m_A \lsim \frac{m_H}{2}$, mostly in the low $\tan \beta$, which satisfies the constraint from BR($h_{SM} \rightarrow AA$).

In Figure~\ref{hsvalidity_rs_lt125_others}(a), (b) and (c), we show the allowed parameter space in the plane spanned by $(m_h-\tan \beta)$, $(\sin(\beta - \alpha)-\tan \beta)$ and $(m_h-m_{H^{\pm}})$ plane respectively. Due to low mass range of the non-standard scalars, the entire mass range considered is valid upto very high scales just like in case 3. The strip below $\tan \beta \lsim 50$ in Figure~\ref{hsvalidity_rs_lt125_others}(a) and (b) corresponds to the points with $m_A \lsim \frac{m_H}{2}$, that satisfy the limit from BR($h_{SM} \rightarrow AA$), as we have argued in case 1.

Next we show the region of parameter space in the plane of the quartic couplings, in Figure~\ref{hsvalidity_l1_l2_rs_lt125}(a)-(f). The qualitative nature of the allowed regions are very similar to case 3, precisely because of low non-standard scalar masses in both cases.

The most salient points of the discussion in the current section can be summarized as follows:

\begin{itemize}
\item Irrespective of the specific case at hand, the smaller the quartic couplings are at the electroweak scale, the higher is the scale of validity of a theory.
\item The requirement of small quartic couplings naturally implies moderate $\tan \beta$ and/or non-standard scalar masses on the lower side.
\item The observed $g_{\mu}-2$ data favor large $\tan \beta$, creating a tension with high-scale validity.
\item $\lambda_1$(also $\lambda_3$, although in a correlated manner with $\lambda_1$) and $\lambda_5$($\lambda_4$ shows a strong correlation with it), mainly control the high-scale behavior and remain practically unaffected by the constraints such as $g_{\mu}-2$ or the BR($h_{SM} \rightarrow AA$). 
\item $\lambda_2$ on the other hand, does not play a significant role in the high-scale validity, but remains heavily constrained from $g_{\mu}-2$ and BR($h_{SM} \rightarrow AA$).  
\item High-scale validity in general demands degeneracy between the non-standard scalar masses as well as their closeness to the 125-GeV Higgs mass. 
\item In case 2, ie. when the lighter CP-even scalar is SM-like in the right-sign region, the requirement of perturbative unitarity at the elctroweak scale, already favors lower non-standard scalar masses and consequently lower quartic couplings, facilitating high-scale validity.
\item In Scenario 2 (both cases 3 and 4), the non-standard scalar masses are on the lower side, as compared to Scenario 1. Therefore, here too, a major portion of the parameter space remains valid upto Planck scale. 
\item Case 1 is least favored among the four cases considered, when high-scale validity is demanded.   
\end{itemize}

\section{Conclusion}

\label{sec7}

   We have explored the high-scale validity of Type-X 2HDM,
   particularly in regions of the parameter space answering to  
   a low-mass neutral CP-odd spinless particle. Such a pseudoscalar is 
   not only consistent with all experimental limits so far but can 
   also help in explaining the observed discrepancy in $g_{\mu}-2$.
   The high-scale validity of the regions of the parameter space of this model, 
   where the above features of special interest are noticed, has been studied here.

   We have identified the regions in the parameter space, which are helpful in
   explaining $(g_{\mu}-2)$ including the most recent results. Other theoretical and 
   experimental constraints, starting from low-scale perturbative unitarity,
   vacuum stability etc., and all the way to the most recent LHC limits, have been
   used to filter out the surviving parameter regions. The two-loop running of 
   various couplings in such regions upto high scales has been studied thereafter,
   thus identifying regions where perturbative unitarity and vacuum stability
   are satisfied upto various high scales, ranging from $10^{4}$ GeV to the Planck 
   scale. Different benchmark points have been used, including both situations where
   the 125-GeV state is either the lighter or the heavier neutral CP-even scalar.   
   Scenarios with both right-and wrong-sign Yukawa couplings have also scanned across
   the parameter space.

   For regions in the parameter space having cut-off scales on the lower side,
   the aspiration for perturbative unification of the three SM gauge couplings 
   is found to necessitate UV completion of Type-X 2HDM 
   below the GUT scale. For regions with
   perturbative validity inching up to the Planck scale, on the other hand,
   the requirements for gauge coupling unification turn out to be similar
   to what they are for the standard model electroweak symmetry breaking sector.
   All this bears ample testimony to the Type-X 2HDM being a candidate theory that
   explains the observed value of $g_{\mu}-2$, keeping open a rich set of UV completion
   possibilities.

\section{Acknowledgements}

We thank Amitava Raychaudhuri for useful discussions. This work was partially supported by funding available from the Department of Atomic Energy, Government of India,
for the Regional Centre for Accelerator-based Particle Physics (RECAPP), Harish-Chandra Research Institute. BM would like to thank RECAPP, HRI, where significant part of the work was done. AD and JL would like to thank Indian Institute of Science Education and Research, Kolkata, where part of the work was done.

\bibliographystyle{jhep}
 \bibliography{ref1.bib}

\end{document}